\documentclass[twocolumn,showpacs,reprint,amsmath,amssymb, prd]{revtex4}

\usepackage{graphicx}% Include figure files
\usepackage{dcolumn}% Align table columns on decimal point
\usepackage{bm}% bold math
\usepackage{courier}

\def\BEq{\begin{equation}}
\def\EEq{\end{equation}}
\def\BEqA{\begin{eqnarray}}
\def\EEqA{\end{eqnarray}}
\def\BEn{\begin{enumerate}}
\def\EEn{\end{enumerate}}
\def\BWT{\begin{widetext}}
\def\EWT{\end{widetext}}
\def\a{\alpha}

\def\b{\beta}

\def\d{\delta}

\def\g{\gamma}
\def\G{\Gamma}

\def\l{\lambda}

\def\m{\mu}

\def\n{\nu}

\def\r{\rho}
\def\s{\sigma}

\begin{document}

%\preprint{APS/123-QED}

\title{Post-Newtonian celestial mechanics in scalar-tensor cosmology}

\author{Andrei Galiautdinov$^1$ and Sergei M.\ Kopeikin$^{2,3}$}
 \affiliation{$^1$Department of Physics and Astronomy, 
University of Georgia, Athens, Georgia 30602\\
$^2$ Department of Physics and Astronomy, University of Missouri--Columbia, Columbia, 
Missouri 65211\\
$^3$ Siberian State University of Geosystems and Technologies,
Plakhotny Street 10, Novosibirsk 630108, Russia
}

\date{\today}% It is always \today, today,

\begin{abstract}

Applying the recently developed dynamical perturbation formalism on cosmological
background to scalar-tensor theory, we provide a solid theoretical basis and a
rigorous justification for phenomenological models of orbital
dynamics that are currently used to interpret experimental measurements of the 
time-dependent gravitational
constant. We derive the field equations for the scalar-tensor perturbations and
study their gauge freedom associated with the cosmological expansion. We find a new
gauge eliminating a prohibitive number of gauge modes in the field equations and
significantly simplifying post-Newtonian equations of motion for localized
astronomical systems in the universe with time-dependent gravitational constant. We
identify several new post-Newtonian terms and calculate their effect on secular
cosmological evolution of the osculating orbital elements.

\end{abstract}


\maketitle

%\tableofcontents

\allowdisplaybreaks[1]

\section{Introduction}

Alternative theories of gravity --- the competitors to Einstein's general theory of 
relativity --- have been the subject of numerous investigations for almost a hundred 
years. One such theory is the scalar-tensor theory 
\cite{NORDTVEDT1970, DAMOUR-ESPOSITO1992-96,
DAMOUR-NORDTVEDT1993}, an outgrowth of theories by Jordan 
\cite{JORDAN1955} and Brans and Dicke \cite{BD1961}, in which, in addition 
to the metric tensor, the gravitational field is described by the fundamental scalar 
field, $\phi$. This so-called Brans-Dicke (BD) field has a broader meaning than 
the scalar field of standard general relativity, where it is present only in the 
stress-energy tensor and generates curvature via Einstein's field equations. 
In scalar-tensor theory, apart from having its own stress-energy tensor, the 
BD field appears explicitly in the Lagrangian through direct coupling to the 
curvature scalar, which makes Newton's gravitational constant variable, 
$G\propto 1/\phi$, and gives rise to additional terms in Einstein's field 
equations, presumably, with important observational consequences.

A typical approach to deriving such consequences is to assume that the 
cosmological evolution of the scalar field affects the value of the gravitational 
constant, making it time dependent, much as in the earlier proposals by 
Dirac \cite{DIRAC1937}. One then postulates a linear time dependence,
\BEq
 G(t)=G_0[1+(\dot{G}_0/G_0) (t-t_0)],
\EEq 
with $G_0$ representing the value of $G$ at reference epoch, say, 
$t_0= {\rm J2000}$ 
(which we set to zero, for convenience),  and uses it in the equation of 
motion for the gravitational probe 
moving in a spherically symmetric gravitational field of a point like mass $M$
\cite{NORTDTVEDT2001,MULLER2008},
\BEq
\label{eq:NewtonianSimpleModel}
\ddot{\bm r}=-\frac{G(t)M{\bm r}}{r^3}
+{\bm F}_{\rm Newtonian}+{\bm F}_{\rm relativistic},
\EEq
where the first term on the right-hand side of (\ref{eq:NewtonianSimpleModel})
represents what is known as the Gyld\'{e}n-Meshcherskii problem 
\cite{Meshcherskii1893,BERKOVIC1981}, the term
${\bm F}_{\rm Newtonian}$ includes additional Newtonian corrections, 
such as the influence of the other planets, and ${\bm F}_{\rm relativistic}$ 
includes the relativistic terms present in the Einstein-Infeld-Hoffman
equations \cite{EIH1938}. The so-called linear trend, $\dot{G}_0/G_0$, is 
then estimated from the astronomical observations. 
The Lunar Laser Ranging experiment based on the 44 years of data 
\cite{MULLER2014} and the Mars Reconnaissance Orbiter experiment
\cite{KONOPLIV2011} give the most stringent upper limits on the variability of $G$, 
$\dot{G}_0/G_0 = (1.4 \pm 1.5) \times 10^{-13} {\rm \; yr}^{-1}$
and $\dot{G}_0/G_0 = (0.1 \pm 1.6) \times 10^{-13} {\rm \; yr}^{-1}$,
correspondingly.

The above phenomenological approach can certainly be improved. 
We are particularly interested in determining the relativistic terms 
that come from a careful analysis of 
the scalar-tensor theory of an expanding universe. That can be done with 
the help of the dynamical perturbation theory of curved spacetime manifolds 
recently developed in Refs.\ \cite{SMK2012,SMK2013} which provides a 
rigorous method of calculating the gravitational fields of perturbations whose 
density contrast significantly exceeds the average density of the universe,
such as in the case of a localized gravitational system placed on cosmological 
background. The field variables of the theory are then naturally separated into two parts: 
the background part, whose dynamics is fully determined by the spherically symmetric
Freedman solution of the Brans-Dicke theory, and perturbations, whose evolution 
is governed by the field equations derived on the basis of the properly formulated 
variational procedure applied to the Brans-Dicke action functional. It is the perturbations 
of the background metric and the scalar field that we are interested in finding. They determine 
the effective gravitational force, from which the post-Newtonian terms in the equations 
of motion can be deduced, thus improving on Eq.\ (\ref{eq:NewtonianSimpleModel}).

\section{Lagrangian, field equations, and metric}

We take the localized system to be a planetary system or a binary pulsar
and the background manifold to be the spatially flat 
Friedman-Lemaitre-Robertson-Walker (FLRW) solution of the scalar-tensor 
theory. The background metric in the isotropic conformal coordinates, 
$x^\m\equiv (c\eta,x^i)$,  $\m=0, 1,2,3$, $i=1,2,3$, 
thus has the form
\BEq
\bar{g}_{\m\n}= a^2(\eta)\texttt{f}_{\m\n},
\quad
\texttt{f}_{\m\n} \equiv {\rm diag}(-1,1,1,1),
\EEq    
where $c$ is the speed of light, $a(\eta)$ is the cosmological scale factor, and $\eta$ 
is the conformal time that is related to the standard cosmological time, $t_{\rm H}$, 
measured by freely falling Hubble observers via $dt_{\rm H}=a(\eta)d\eta$. Additionally, 
in terms of $\eta$, the conformal Hubble constant is defined by
\BEq
{\cal H} \equiv (1/a)(da/d\eta).
\EEq
The full gravitational system is then described by the Lagrangian,
${\cal L}={\cal L}^{\rm st}+{\cal L}^{\rm bgc}+{\cal L}^{\rm p}$,
where
\BEqA
{\cal L}^{\rm st} = \frac{\sqrt{-g}c^3}{16\pi}
\left(-\phi R
+
\frac{\omega(\phi)}{\phi} g^{\a\b}\phi_{,\a}\phi_{,\b}
\right)
\EEqA
is the Lagrangian of the scalar-tensor theory, $R$ is the Ricci scalar curvature, 
$\phi$ is the BD field, $\phi_{,\a} \equiv  \partial_{\a} \phi \equiv 
\partial \phi/\partial x^\a$, $\omega$ 
is the BD coupling parameter (assumed to be a function of $\phi$), 
$g^{\a\b}$ is the (inverse) metric, $g\equiv {\rm det}(g_{\m\n})$,
 ${\cal L}^{\rm bgc}$ is the Lagrangian of the background content of the 
expanding universe (mainly dark matter and dark energy, but also includes 
background baryonic matter), and ${\cal L}^{\rm p}$ 
is the perturbing
Lagrangian of the localized system. The ${\cal L}^{\rm p}$ 
is assumed to be independent of 
$\phi$, which corresponds to the requirement that the geodesic motion of 
material objects is governed by the metric alone without any direct influence 
of the BD field.

In accordance with the dynamical perturbation formalism, we write the full 
metric and the BD field as the sums,
\BEq
\label{eq:metricDecomposed}
g_{\m\n}(x)=\bar{g}_{\m\n}(\eta)+\varkappa_{\m\n}(x),
\quad
\phi(x) = \bar{\phi}(\eta)+\varphi(x),
\EEq  
of their background parts, $\bar{g}_{\m\n}$ and $\bar{\phi}$, and 
perturbations, $\varkappa_{\m\n}$ and $\varphi$.
We also introduce the contravariant metric density, 
$\frak{g}^{\a\b}\equiv \sqrt{-g}g^{\a\b}$, its background value, 
$\frak{\bar{g}}^{\a\b}\equiv \sqrt{-\bar{g}}\bar{g}^{\a\b}$,
and the perturbation, 
$\frak{h}^{\a\b}\equiv \frak{g}^{\a\b}-\frak{\bar{g}}^{\a\b}$,
which is conveniently written in the form 
\BEq
\frak{h}^{\a\b}\equiv \sqrt{-\bar{g}}\,l^{\a\b}.
\EEq 
We then take $\frak{h}^{\a\b}$ and $\varphi$ to represent the dynamical 
variables of the theory and use the variational procedure of 
Refs.\ \cite{SMK2012,SMK2013} to write down
the linearized field equations for $\frak{h}^{\a\b}$ and $\varphi$,
\begin{align}
\label{eq:der21}
\frac{-16\pi}{c^3\sqrt{-\bar{g}}}
\frac{\d }{\d {\bar{\phi}}}
\left(
{\frak{h}}^{\r\s} \frac{\d \bar{\cal L}^{\rm st}}{\d \bar{\frak{g}}^{\r\s}} 
+\varphi \frac{\d \bar{\cal L}^{\rm st}}{\d \bar{\phi}}
\right)&=0,
\\
\label{eq:der01}
\frac{-16\pi}{c^3\sqrt{-\bar{g}}}
\frac{\d }{\d {\bar{g}}^{\m\n}}
\left(
{\frak{h}}^{\r\s} \frac{\d \bar{\cal L}^{\rm st}}{\d \bar{\frak{g}}^{\r\s}} 
+\varphi \frac{\d \bar{\cal L}^{\rm st}}{\d \bar{\phi}}
\right) &=\frac{8\pi}{c^4} \Lambda_{\m\n},
\end{align}
where $\d \bar{\cal L}^{\rm st}/\d \bar{\phi}$ 
and
$\d \bar{\cal L}^{\rm st}/\d \bar{\frak{g}}^{\r\s}$
stand for the variational derivatives of 
$\bar{\cal L}^{\rm st} \equiv {\cal L}^{\rm st} 
(\bar{\frak{g}}^{\a\b},  \bar{\phi})$ with respect to the background 
field $\bar{\phi}$ and metric density $\bar{\frak{g}}^{\r\s}$, 
and 
\BEq
\Lambda_{\m\n}\equiv 
\frac{2c}{\sqrt{-\bar{g}}}\frac{\d {\cal L}^{\rm p}}{\d {\bar{g}^{\m\n}}}
\EEq 
is the stress-energy tensor of the localized gravitational system. 
Once 
Eqs.\ (\ref{eq:der21}) and (\ref{eq:der01}) are worked out, we can 
find $l^{\a\b}$ and $\varphi$ by solving these equations, and then, via
\BEq
{{\varkappa}_{\m\n}}=-{l_{\m\n}}+(1/2)\bar{g}_{\m\n}{l},
\quad
l \equiv {l^\a}_{\a},
\EEq
find the full metric, $g_{\m\n}$. [Note that the background metric 
$\bar{g}_{\m\n}$ is used to raise and lower tensorial indices; covariant 
differentiation with respect to $\bar{g}_{\m\n}$ will be denoted with a vertical bar.] 

Applying (\ref{eq:der21}) and (\ref{eq:der01}) to $\bar{\cal L}^{\rm st}$,
and making the linearized Hubble approximation in which we ignore all terms 
containing ${\cal H}^2$, $d{\cal H}/d \eta$, $d^2 \bar{\phi}/d \eta^2$, 
etc., we get the system of differential equations for scalar-tensor perturbations,
\BWT
\begin{align}
\label{eq:varphi_perturbation_linearHubble1}
{\varphi^{|\a}}_{|\a}
+ 
\frac{2{\omega}'}{3+2\omega}
\bar{\phi}^{|\a}{\varphi}_{|\a}
+
A^{\a}\bar{\phi}_{|\a}
&=
\frac{8\pi \Lambda}{(3+2\omega)c^4},
\\
\label{eq:l_mn_perturbation_linearHubble1}
\left(
{{l_{\m\n}}^{|\a}}_{|\a}+
\bar{g}_{\m\n}{A^{\a}}_{|\a}-{A}_{\m|\n}-{A}_{\n|\m}
\right)
+
\frac{{\bar{\phi}}^{|\a}}{\bar{\phi}}
\left( 
l_{\m\n|\a}
- l_{\a\m|\n}- l_{\a\n|\m}
\right)
-
\bar{g}_{\m\n}\frac{\bar{\phi}^{|\a}}{\bar{\phi}}
\left(
\frac{1}{2}l_{|\a}-\frac{2\omega}{\bar{\phi}}\varphi_{|\a}
\right) 
&
\nonumber \\
+
\frac{\bar{\phi}_{|\m}}{\bar{\phi}}
\left(
\frac{1}{2}l_{|\n}-\frac{2\omega}{\bar{\phi}}\varphi_{|\n}
\right) 
+
\frac{\bar{\phi}_{|\n}}{\bar{\phi}}
\left(
\frac{1}{2}l_{|\m}-\frac{2\omega}{\bar{\phi}}\varphi_{|\m}
\right)
+ 
2\bar{g}_{\m\n} \frac{{\bar{\phi}}_{|\a}}{\bar{\phi}}A^{\a}
+ \frac{2}{\bar{\phi}}
\left( \bar{g}_{\m\n}{\varphi^{|\a}}_{|\a}-\varphi_{|\m\n} \right)
&=\frac{16\pi}{\bar{\phi} c^4} \Lambda_{\m\n} ,
\end{align}
where $A^{\a} \equiv {l^{\a\b}}_{|\b}$, 
$\omega \equiv\omega(\bar{\phi})$,
$\omega' \equiv d\omega(\bar{\phi})/d\bar{\phi}$,
and $\Lambda \equiv \bar{g}^{\m\n}\Lambda_{\m\n}$.
Equations (\ref{eq:varphi_perturbation_linearHubble1}) and 
(\ref{eq:l_mn_perturbation_linearHubble1})
admit an enormous number of gauge modes most of which can be 
eliminated if we impose the gauge condition 
[here, $\bar{u}^{\a} = (1/a,0,0,0)$ is the velocity of the Hubble flow],
\begin{align}
\label{eq:generalizedGauge}
& A^{\a} = 
-\frac{2\cal{H}}{ac}l^{\a\b}\bar{u}_{\b}
-l^{\a\b}\frac{\bar{\phi}_{|\b}}{\bar{\phi}}
+\frac{\bar{\phi}^{|\a}}{\bar{\phi}}
\left(
\frac{l}{2}
- \frac{2\omega}{\bar{\phi}} \varphi
\right)
-\frac{{\varphi}^{|\a}}{\bar{\phi}}
- \frac{2{\cal H}}{ac}\frac{\varphi}{\bar{\phi}}\bar{u}^{\a}
- \frac{\bar{\phi}^{|\a}\varphi}{\bar{\phi}^2}, 
\end{align}
\EWT
which generalizes the gauges used in Refs.\ \cite{BD1961} and 
\cite{SMK2012}.
Using (\ref{eq:generalizedGauge}) and rewriting everything in the 
isotropic conformal coordinates, we arrive at the wave 
equations for perturbations,
\begin{align}
\label{eq:varphi_perturbation_linearHubble2_simpler}
\Box {\varphi}
+ \frac{2}{c}
\left(
 -{\cal H}+\frac{{\cal F}}{2}-\frac{\omega'\bar{\phi}{\cal F}}{3+2\omega}
\right)
\varphi_{,0}
&=
\frac{8\pi \texttt{f}^{\a\b} \Lambda_{\a\b}}{(3+2\omega)c^4},
\\
\label{eq:l_mn_perturbation_linearHubble2_simpler}
\Box Q_{\m\n}
+ \frac{2}{c}\left({\cal H}-\frac{{\cal F}}{2}\right)Q_{\m\n,0}
&= \frac{16\pi a^2}{\bar{\phi}c^4}\Lambda_{\m\n},
\end{align}
where $\Box {\varphi} \equiv \texttt{f}^{\a\b} {\varphi}_{,{\a}{\b}}$
and $\Box Q_{\m\n} \equiv \texttt{f}^{\a\b} Q_{\m\n,{\a}{\b}}$. In the 
above, we introduced an auxiliary gravitational variable
\BEq
Q_{\m\n}\equiv l_{\m\n}+\bar{g}_{\m\n}\varphi/\bar{\phi},
\EEq
and defined
\BEq
{\cal F} \equiv (1/\bar{\phi})(d\bar{\phi}/{d\eta}).
\EEq

Equations (\ref{eq:varphi_perturbation_linearHubble2_simpler}) 
and (\ref{eq:l_mn_perturbation_linearHubble2_simpler}) have the general 
form
\BEq
\label{eq:generalPerturbationEquation_simpler}
\Box Q +(2/c){\cal B} Q_{,0} = 4\pi a^2 {\cal T},
\EEq
with
${\cal B}(\eta)\sim {\cal O}({\cal H})$, 
$d{\cal B}(\eta)/d\eta\sim {\cal O}({\cal H}^2)$.
This can be solved by introducing two new functions, $b=b(\eta)$ 
and $q = q(\eta, x^i)$, such that
$Q = b^2 q$,
with $d{b}/d\eta = {\cal B}b$.
Noticing that, in the linear Hubble approximation,
\BEq
\Box Q +(2/c){\cal B} Q_{,0} 
= b\Box(bq),
\EEq
we get the equation 
\BEq
\Box(bq) =  4\pi \frac{a^2 {\cal T}}{b},
\EEq
whose retarded solution is given by the volume integral,
\BEq
\label{eq:retardedSolution_q}
q(\eta,{\bf x})=-\frac{1}{ b(\eta)}\int \frac{a^2(\eta')}{b(\eta')}
\frac{{\cal T}(\eta',{\bf x}')}{|{\bf x}-{\bf x}'|}d^3x',
\EEq
with $\eta' = \eta - |{\bf x}-{\bf x}'|/c$ being the retarded time.
The corresponding solution to (\ref{eq:generalPerturbationEquation_simpler})
is then
\BEq
\label{eq:generalPerturbationEquationSolution_simpler}
Q(\eta,{\bf x})=- b(\eta)\int \frac{a^2(\eta')}{b(\eta')}
\frac{{\cal T}(\eta',{\bf x}')}{|{\bf x}-{\bf x}'|}d^3x'.
\EEq
Applying (\ref{eq:generalPerturbationEquationSolution_simpler}) to 
(\ref{eq:varphi_perturbation_linearHubble2_simpler}) 
and (\ref{eq:l_mn_perturbation_linearHubble2_simpler}) we get
\begin{align}
\label{eq:varphi_retardedSolutionRe-written}
 \varphi(\eta,{\bf x})
& =
- \frac{{b}_1(\eta)}{c^4}\int 
\frac{2\texttt{f}^{\a\b}\Lambda_{\a\b}(\eta', {\bf x}')d^3x'}
{{{b}_1}(\eta')\left[3+2\omega(\eta')\right]|{\bf x}-{\bf x}'|},
\\
\label{eq:Qmn_retardedSolutionRe-written}
Q_{\m\n}(\eta,{\bf x}) 
& =
-4 \frac{b_2(\eta)}{c^4}\int 
\frac{a^2(\eta')  \Lambda_{\m\n}(\eta', {\bf x}')  d^3x'}
{b_2(\eta')\bar{\phi}(\eta')|{\bf x}-{\bf x}'|},
\end{align}
with $b_{1}(\eta)$ and $b_{2}(\eta)$ satisfying the conditions
\begin{align}
\frac{1}{b_1}\frac{d{b}_1}{d\eta} 
&=
 -{\cal H}+\frac{{\cal F}}{2}-\frac{\omega'\bar{\phi}{\cal F}}{3+2\omega},
\\
\frac{1}{b_2}\frac{d{b}_2}{d\eta} 
&=
{\cal H}-\frac{{\cal F}}{2}.
\end{align}
Performing the near zone expansion \cite{POISSON&WILL2014} of  
(\ref{eq:varphi_retardedSolutionRe-written}) and
(\ref{eq:Qmn_retardedSolutionRe-written}) gives
\begin{align}
\varphi &= -\frac{2\texttt{f}^{\a\b}}{a(3+2\omega)c^4}
\left[
\Phi_{\a\b}+\frac{{\cal F}}{2c}
\left(1+\frac{2\omega'\bar{\phi}}{3+2\omega}\right)\Psi_{\a\b}
\right],
\end{align}
\begin{align}
Q_{\m\n} &= -\frac{4a}{\bar{\phi}c^4}
\left(
\Phi_{\m\n}+\frac{{\cal F}}{2c}\Psi_{\a\b}
\right),
\end{align}
where
\begin{align}
\label{eq:functionPhi}
\Phi_{\m\n}&= 
\int \frac{a\Lambda_{\m\n}({\bf x}')}{|{\bf x}-{\bf x}'|}d^3x'
- \frac{1}{c}\frac{d}{d\eta}\int a\Lambda_{\m\n}d^3x
\nonumber \\
&\quad
+\frac{1}{2c^2}\frac{d^2}{d\eta^2}
\int a\Lambda_{\m\n}({\bf x}')|{\bf x}-{\bf x}'|d^3x',
\\
\label{eq:functionPsi}
\Psi_{\m\n}&=
\int a\Lambda_{\m\n}d^3x
-\frac{1}{c}\frac{d}{d\eta}
\int a\Lambda_{\m\n}({\bf x}')|{\bf x}-{\bf x}'|d^3x',
\end{align}
from which the metric perturbation is found to be
\BWT
\begin{align}
\label{eq:varkappaGeneral}
{{\varkappa}_{\m\n}}
=
\frac{4a}{\bar{\phi}c^4}
\biggl\{
\Phi_{\m\n}
-\frac{\texttt{f}_{\m\n}}{2}\left(1-\frac{1}{3+2\omega}\right)
\texttt{f}^{\a\b}\Phi_{\a\b}
+
\frac{{\cal F}}{2c}
\biggl[
\Psi_{\m\n}-
\frac{\texttt{f}_{\m\n}}{2}
\biggl(1-\frac{1}{3+2\omega}
-\frac{2\omega'\bar{\phi}}{(3+2\omega)^2}\biggr)
\texttt{f}^{\a\b}\Psi_{\a\b}
\biggr]
\biggr\}.
\end{align}
\EWT

To simplify (\ref{eq:varkappaGeneral}), we appeal to the covariant continuity 
of the stress-energy tensor $\Lambda_{\m\n}$, which in zeroth order in 
perturbations can be expressed in the form
\begin{align}
\partial_{0}\left(a\Lambda_{00}\right) - \partial_{j}\left(a\Lambda_{0j}\right)
&=-({\cal H}/c)\left(a\Lambda_{kk}\right),
\\ 
\partial_{0}\left(a\Lambda_{i0}\right) - \partial_{j}\left(a\Lambda_{ij}\right)
&=-({\cal H}/c)\left(a\Lambda_{i0}\right),
\end{align} 
with $\Lambda_{kk}= \d^{ij} \Lambda_{ij}$,
or, upon volume integration,
\begin{align}
\int \partial_{0}\left(a\Lambda_{00}\right)d^3x 
&= -\frac{\cal H}{c}\int \left(a\Lambda_{kk}\right) d^3x,
\\
\int \partial_{0}\left(a\Lambda_{i0}\right)d^3x 
&= -\frac{\cal H}{c}\int \left(a\Lambda_{i0}\right)d^3x.
\end{align} 
We notice that for any three-vector $\psi_i(\eta, {\bf x})$, 
$\psi_i = \partial_{j}\left(\psi_{j}x^i\right) - \left(\partial_{j}\psi_{j}\right)x^i$,
and, upon integrating and discarding the divergence terms, 
$\int \psi_i d^3x= \int \left(-\partial_{j}\psi_{j}\right)x^i d^3x$.
Applying this to $\psi_i\equiv a\Lambda_{i0}$ gives the virial relation,
\begin{align}
\int \left(a\Lambda_{i0}\right)d^3x 
= 
-\frac{1}{c}\frac{d}{d\eta}\int \left( a\Lambda_{00}\right) x^id^3x 
- \frac{\cal H}{c} \int \left( a\Lambda_{kk}\right) x^id^3x.
\end{align} 
If we now define the mass, $M$, of the localized system by
\BEq
M \equiv (1/c^2)\int \left(a\Lambda_{00}\right)d^3x,
\EEq
as well as its momentum, $P^i \equiv -(1/c)\int \left(a\Lambda_{i0}\right)d^3x$, 
and dipole moment, $I^i \equiv (1/c^2)\int \left( a\Lambda_{00}\right) x^id^3x$, 
and restrict consideration to slowly moving sources only (that is, ignore all terms containing
$\Lambda_{ij}$), we find that in the linearized approximation the system's mass is conserved,
\BEq
dM/d\eta=0,
\EEq 
and that momentum and dipole moment are related to each other by
$P^i={dI^i}/{d\eta}$.
This allows us to introduce the system's rest frame in which the system's momentum vanishes,
$P^i=0$, provided the origin of coordinates is chosen at the system's center of mass,
$I^i=0$.
Combining this with the monopole approximation,
$|{\bf x}-{\bf x}'|\approx |{\bf x}|$, leads to 
$\Phi_{00}={M}/{|{\bf x}|}$,
$\Phi_{0i}=0$,
$\Phi_{ij}=0$,
$\Psi_{00}=M$,
$\Psi_{0i}=0$,
$\Psi_{ij}=0$,
and we get
\BWT
\begin{align}
\label{eq:metricPerturbationsFinal00}
{{\varkappa}_{00}}
 &= 
\frac{2aM}{\bar{\phi}c^2}
\biggl\{
\left(1+ \frac{1}{3+2\omega}\right)\frac{1}{|{\bf x}|}
+\frac{\cal F}{2c}
\left(1+ \frac{1}{3+2\omega}+\frac{2\omega'\bar{\phi}}{(3+2\omega)^2}\right)
\biggr\},
\\
\label{eq:metricPerturbationsFinali0}
{{\varkappa}_{i0}} 
&= 0,
\\
\label{eq:metricPerturbationsFinalij}
{{\varkappa}_{ij}} 
&= 
\frac{2aM}{\bar{\phi}c^2}
\biggl\{
\left(1 - \frac{1}{3+2\omega}\right)\frac{1}{|{\bf x}|}
+\frac{\cal F}{2c}
\left(1- \frac{1}{3+2\omega}-\frac{2\omega'\bar{\phi}}{(3+2\omega)^2}\right)
\biggr\}\d_{ij}.
\end{align}
\EWT
Introducing the post-Newtonian (PPN) parameters \cite{WILL1993},
\begin{align}
\gamma &
= \frac{1+\omega}{2+\omega}, 
\quad
G = \left(\frac{4+2\omega}{3+2\omega}\right)\frac{1}{\bar{\phi}}, 
\\
\label{eq:betaDefinition}
\beta &= 1+\frac{\omega' \bar{\phi}}{(3+2\omega)(4+2\omega)^2}, 
\end{align}
with $G$ having the meaning of the experimentally observable 
gravitational ``constant,'' brings $\varkappa_{\m\n}$ 
to a compact form,
\begin{align}
\label{eq:metricPerturbationsFinalPPN00}
{{\varkappa}_{00}}
 &= 
\frac{2aM}{c^2}
\left(\frac{G}{|{\bf x}|}-\frac{1}{2c}\frac{dG}{d\eta}\right),
 \\
\label{eq:metricPerturbationsFinalPPNi0}
{{\varkappa}_{i0}} 
&= 0,
 \\
\label{eq:metricPerturbationsFinalPPNij}
{{\varkappa}_{ij}} 
&= 
\frac{2aM}{c^2}
\left(
\frac{\gamma G}{|{\bf x}|}
-\frac{1}{2c}\frac{d(\gamma{G})}{d\eta}
\right)\d_{ij}.
\end{align}
In terms of the cosmological time, $t_{\rm H}$, the full metric 
is thus given by
\begin{align}
\label{eq:metricPerturbationsFinalPPN00g}
{g_{00}}
 &= 
-1+\frac{2M}{c^2}
\left(\frac{G}{a|{\bf x}|}-\frac{1}{2c}\frac{dG}{dt_{\rm H}}\right),
 \\
\label{eq:metricPerturbationsFinalPPNi0g}
{g_{i0}} 
&= 0,
 \\
\label{eq:metricPerturbationsFinalPPNijg}
{g_{ij}} 
&= 
\left[
1
+
\frac{2M}{c^2}
\left(
\frac{\gamma G}{a|{\bf x}|}
-\frac{1}{2c}\frac{d(\gamma{G})}{dt_{\rm H}}
\right)
\right] a^2\d_{ij},
\end{align}
which in the limit ${\cal H}$, ${\cal F} \rightarrow 0$ 
reproduces the standard BD result \cite{BD1961}.

\section{Equations of motion for gravitational probes}

To uncover the observational consequences of the found metric, we have to 
derive the equations of motion for point probes. For that, we introduce the 
local inertial coordinates, $(ct,X^i)$, associated with a freely falling Hubble 
observer,
\BEq
ct=ct_{\rm H} + a^2H \delta_{ij} x^{i} x^{j} /(2c), 
\quad 
X^i = ax^i,
\quad
H\equiv \dot{a}/a,
\EEq
where the overdot represents differentiation with respect to $t$, and $H$ 
stands for the usual Hubble constant. Denoting $r \equiv |{\bm X}|$, we find 
\begin{align}
\label{eq:metricPerturbationsFinalPPN00g}
{g_{00}}
 &= 
-1+\frac{2GM}{c^2 r}-\frac{M}{c^3}\frac{dG}{dt},
 \\
\label{eq:metricPerturbationsFinalPPNi0g}
{g_{i0}} 
&= - \frac{2(1+\g)GMHX^i}{c^3 r},
 \\
\label{eq:metricPerturbationsFinalPPNijg}
{g_{ij}} 
&= 
\left[
1
+
\frac{2\gamma GM}{c^2r}
-\frac{M}{c^3}\frac{d(\gamma{G})}{dt}
\right]\d_{ij},
\end{align}
with the linearized post-Newtonian connection coefficients being
\begin{align}
\G^{i}_{00}
&=
\frac{1}{c^2}\frac{GM X^i}{r^3},
\quad
\G^{i}_{j0}
=
\frac{1}{c^3}\frac{d(\gamma{G})}{dt}
\frac{M}{r}\d_{ij},
\nonumber \\
\G^{i}_{jk}
&=
-\frac{1}{c^2}
\frac{\gamma {G}M(\d_{ij}X^k+\d_{ik}X^j-\d_{jk}X^i)}{r^3},
\nonumber \\
\G^{0}_{00}&=
-\frac{1}{c^3}
\frac{dG}{dt}\frac{M}{r},
\;
\G^{0}_{j0}
=
\frac{1}{c^2}\frac{MG X^j}{r^3},
\;
\G^{0}_{jk} = 0.
\end{align}
These are substituted into the geodesic equation parametrized by the 
coordinate time,
\begin{align}
\frac{d^2X^i}{dt^2}&=
-c^2\G^{i}_{00}
-2c\G^{i}_{j0}\frac{dX^j}{dt}
-\G^{i}_{jk}\frac{dX^j}{dt}\frac{dX^k}{dt}
\nonumber \\
&\quad
+
\left(
c\G^{0}_{00}
+2\G^{0}_{j0}\frac{dX^j}{dt}
+\frac{1}{c}\G^{0}_{jk}\frac{dX^j}{dt}\frac{dX^k}{dt}
\right)
\frac{dX^i}{dt},
\end{align}
with the result (here, 
${\bm n}\equiv {\bm r}/{r}$,
${\bm v}\equiv \dot{\bm r}$,
$v \equiv |{\bm v}|$),
\BEq
\ddot{\bm r} = -G(t)M{\bm{n}}/r^2 + {\bm F},
\EEq
where the disturbing force per unit mass is given by
\begin{align}
\label{eq:disturbingForce}
{\bm F}&=
-\frac{{\gamma} GM}{c^2} \frac{v^2}{r^2} {\bm n}
+\frac{(2{\gamma}+2\beta) G^2M^2}{c^2r^3}{\bm n}
\nonumber\\
&\quad
+\frac{GM}{c^2}
\biggl\{
(2+2\g)\frac{\dot{r}}{r^2}
-\biggl[(1+2\gamma)\frac{\dot{G}}{G}+2\dot{\gamma}\biggr]
\frac{1}{r}
\biggr\}{\bm v}.
\end{align}
On the right-hand side of Eq.\ (\ref{eq:disturbingForce}) we have included the 
``standard'' first-order post-Newtonian (1PN) quadratic term (even though it does not formally 
follow from our linearized theory), which is expected on physical grounds.

We are particularly interested in the effect of Eq.\ (\ref{eq:disturbingForce}) 
on Keplerian orbits. It is immediately clear that the following result of 
standard general relativity, with 
${\dot{G}}/{G}=0$, $\dot{\gamma}=0$, holds: in the FLRW universe, in 
the linear Hubble approximation, planetary orbits do not change. The 
scalar-tensor theory, however, modifies that conclusion, as will be 
demonstrated below.

Because $d[{\bm r}\times{\bm v}]/dt\propto [{\bm r}\times{\bm v}]$,
the  motion is confined to a fixed orbital plane.
This allows us to simplify the description of post-Newtonian dynamics by 
taking the orbital plane to coincide with the $(X,Y)$ plane of the coordinate
system \cite{POISSON&WILL2014}. Introducing the orbital basis 
\cite{SMK2011, POISSON&WILL2014},
${\bm n}=[\cos f,\sin f, 0]$,
${\bm \l}=[-\sin f,\cos f, 0]$,
${\bm e}_z=[0,0,1]$,
in which  
${\bm v} = \dot{r}{\bm n}+r\dot{f}{\bm \l}$,
where $f$ is the true anomaly (the orbital angle measured relative 
to the pericenter),
brings ${\bm F}$ to the form
\BEq
{\bm F} = {\cal R}{\bm n} + {\cal S}{\bm \l},
\EEq
where 
\begin{align}
{\cal R}
&=
-\frac{GM}{c^2}
\biggl\{
\gamma \frac{v^2}{r^2} 
-(2+2\g)\frac{\dot{r}^2}{r^2}
-(2{\gamma}+2\beta) \frac{GM}{r^3}
\nonumber \\
&\quad\quad\quad\quad\quad
+\biggl[
(1+2\gamma)\frac{\dot{G}}{G}+2\dot{\gamma}
\biggr]\frac{\dot{r}}{r}
\biggr\},
\\
{\cal S} &=
+\frac{GM}{c^2}
\biggl\{
(2+2\g)\frac{\dot{r}\dot{f}}{r}
-
\biggl[
(1+2\gamma)\frac{\dot{G}}{G}+2\dot{\gamma}
\biggr]\dot{f}
\biggr\}.
\end{align}

\section{Secular evolution}

To find the secular changes of the orbital elements, 
$a$ (semimajor axis, not to be confused with the cosmological scale factor), $e$
(eccentricity), and $\varpi$ (longitude of pericenter), we use the osculating equations 
of the perturbed Gyld\'{e}n-Meshcherskii problem \cite{LUKYANOV2009, DUBOSHIN1968}
(also see \cite{VERAS2013}),
\BWT
\begin{align}
\label{eq:dadf}
\frac{da}{df}&
=
\biggl\{
-a
\biggl[
1+\frac{2e}{1-e^2}(e+\cos f)
\biggr]
\frac{\dot{G}}{G}
+
\frac{2}{n\sqrt{1-e^2}}
[e\sin f \; {\cal R} + (1+e\cos f)\, {\cal S}]
\biggr\}\frac{dt}{df},
\\
\label{eq:dedf}
\frac{de}{df}&
=
\biggl\{
-(e+\cos f)\frac{\dot{G}}{G}
+
\frac{\sqrt{1-e^2}}{n a}
\biggl[
\sin f \; {\cal R} 
+
\biggl(
\cos f + \frac{e+\cos f}{1+e\cos f}
\biggr){\cal S}
\biggr]
\biggr\}\frac{dt}{df},
\\
\label{eq:dwdf}
\frac{d\varpi}{df}&
=
\biggl\{
-\frac{\sin f}{e}\frac{\dot{G}}{G}
+
\frac{\sqrt{1-e^2}}{n ae}
\biggl[
-\cos f \;{\cal R} 
 + 
\frac{2+e\cos f}{1+e\cos f} \sin f \; {\cal S}
\biggr]
\biggr\}\frac{dt}{df},
\\
\label{eq:dtdf}
\frac{dt}{df}
&=
\biggl\{
\frac{n(1+e\cos f)^2}{(1-e^2)^{3/2}} 
+\frac{\sin f}{e}\frac{\dot{G}}{G}
-
\frac{\sqrt{1-e^2}}{n ae}
\biggl[
-\cos f \;{\cal R} 
 + 
\frac{2+e\cos f}{1+e\cos f} \sin f \; {\cal S}
\biggr]
\biggr\}^{-1},
\end{align}
\EWT
where 
\BEq
\label{eq:n(t)}
n(t)\equiv \sqrt{G(t)M/a^3(t)}=2\pi/P(t)
\EEq 
is the osculating mean motion, with $P(t)$ being the osculating orbital period. We write
\begin{align}
\label{eq:defGdotOverG}
G(t) &= G_0 + \dot{G}_0t, 
\quad 
\dot{G}_0/G_0\equiv {s_1} H,
\\
\label{eq:defgdot}
\gamma(t) &= \gamma_0 + \dot{\gamma}_0t, 
\quad
\dot{\gamma}_0\equiv {s_2} H,
\\
\label{eq:defbetadot}
\beta(t) &= \beta_0 + \dot{\beta}_0t, 
\quad
\dot{\beta}_0\equiv {s_3} H,
\end{align}
where ${s_1}$, ${s_2}$, and ${s_3}$ are the adjustable 
parameters to be fixed by observations.
Substituting the usual Keplerian relations for $r$, $\dot{r}$, $\dot{f}$, $v^2$ 
in ${\cal R}$ and ${\cal S}$, and using in Eqs.\ (\ref{eq:dadf}), (\ref{eq:dedf}), 
(\ref{eq:dwdf}), (\ref{eq:dtdf}) the zeroth-order orbital elements 
\cite{SUBBOTIN1974, POISSON&WILL2014} and the values of $G$, $\gamma$, 
and $\beta$ taken at the initial epoch,
we get, upon integrating each of Eqs.\ (\ref{eq:dadf}), (\ref{eq:dedf}), and 
(\ref{eq:dwdf}) with respect to $f$ from 0 to $2\pi$, the following 1PN
changes per anomalistic period:
\BWT
\begin{align}
\label{eq:dadfAVEfinal}
 \langle \dot{a}\rangle
&=
-a  H s_1
+
\frac{ H}{c^2}
\frac{ n^2 a^3 }{e^2 \sqrt{1-e^2}}
\biggl\{
\biggl[
2 \left(-3+e^2+3 \sqrt{1-e^2}\right)\b_0 
+
4 \left(-1+\sqrt{1-e^2}\right) (6+5\g_0)
\nonumber \\
&\quad\quad\quad\quad\quad\quad\quad\quad\quad\quad\quad\quad\quad\quad
+
e^2 \left(4+2 \sqrt{1-e^2}-4 \g_0+5 \sqrt{1-e^2} \g_0\right)
\biggr] s_1
+
4 e^2 \left(-2+\sqrt{1-e^2}\right) s_2
\biggr\},
\\
\label{eq:dedfAVEfinal}
 \langle \dot{e}\rangle
&=
\frac{ H}{c^2}
\frac{n^2 a^2 \sqrt{1-e^2}  }{e^3}
\biggl\{
\biggl[
2\left(-2+e^2+2 \sqrt{1-e^2}\right)\b_0
+
2 \left(-1+\sqrt{1-e^2}\right) (8+7 \g_0)
\nonumber \\
&\quad\quad\quad\quad\quad\quad\quad\quad\quad\quad\quad\quad\quad\quad
+
e^2 \left(6+2 \sqrt{1-e^2}+2 \g_0+5 \sqrt{1-e^2}\g_0\right)
\biggr] s_1
+
4 e^2 \left(-1+\sqrt{1-e^2}\right) s_2
\biggr\},
\\
\label{eq:dwdfAVEfinal}
 \langle \dot{\varpi}\rangle
&=
\frac{n^3 a^2 (2+2\g_0-\b_0)  }{c^2(1-e^2)}
+ \;
{\cal O}(H^2).
\end{align}
\EWT
Notice that in the limit $e\rightarrow 1$ our approximation breaks down, 
as $\langle \dot{a}\rangle$ increases without 
bound. For $e\rightarrow 0$, Eq.\ (\ref{eq:dedfAVEfinal}) gives 
$\langle \dot{e}\rangle \rightarrow 0$, as had to be expected. 
The leading contribution in (\ref{eq:dwdfAVEfinal}) is the standard 1PN 
general relativistic correction;  in the linearized Hubble approximation, 
there is no additional contribution to the advance of the pericenter coming 
from the scalar-tensor theory.

To find the anomalistic rate of change of the osculating period
we write (\ref{eq:n(t)}) in the variational form,
\BEq
\label{eq:Prate}
\d P = 2\pi \d \left(\sqrt{\frac{a^3}{GM}}\right)
=P \left(\frac{3}{2}\frac{\d a}{a}-\frac{1}{2}\frac{\d G}{G}\right).
\EEq
Taking $\d P$ to represent the anomalistic change of the osculating period
and denoting $\d P/P \equiv \langle \dot{P}\rangle$, we get upon substituting  
(\ref{eq:defGdotOverG}) and (\ref{eq:dadfAVEfinal}) in (\ref{eq:Prate})
\begin{align}
\label{eq:dotPoverP}
\frac{\langle \dot{P}\rangle}{P}
&=
-2\frac{\dot{G}}{{G}}
+
\frac{H}{c^2}
\frac{ 3n^2 a^2  }{2e^2 \sqrt{1-e^2}}
\biggl\{
\biggl[
2 \left(-3+e^2+3 \sqrt{1-e^2}\right)\b_0 
\nonumber \\
&\quad\quad
+
4 \left(-1+\sqrt{1-e^2}\right) (6+5\g_0)
\nonumber \\
&\quad\quad
+
e^2 \left(4+2 \sqrt{1-e^2}-4 \g_0+5 \sqrt{1-e^2} \g_0\right)
\biggr] s_1
\nonumber \\
&\quad\quad
+
4 e^2 \left(-2+\sqrt{1-e^2}\right) s_2
\biggr\}.
\end{align}
The first term on the right-hand side of (\ref{eq:dotPoverP}) reproduces the 
result of \cite{DAMOUR1988}; the term proportional to $H/c^2$ extends it to 
the post-Newtonian domain.

Next, with the definition of $\beta$ given in 
(\ref{eq:betaDefinition}), we see that
\begin{align}
\frac{\dot{G}}{G} &= 
-\biggl(1-\frac{4(\b-1)}{\g-1}\biggr){\cal F},
\\
\dot{\gamma} &= 
-\frac{4(\b-1)(1+\g)}{\g-1}{\cal F} ,
\end{align}
and thus $s_1$ and $s_2$ are related to each other via
\BEq
\label{eq:s2_vs_s1}
s_2 = \frac{4(\b-1)(1+\g)}{\g-1-4(\b-1)}s_1.
\EEq
The constraints \cite{WILL2014-lrr4, MULLER2014, BENNET2013},
\begin{align}
\gamma - 1 &= 2.3 \times 10^{-5},
\quad
\beta-1 = 8\times 10^{-5},
\\
\dot{G}_0/G_0 &= 1.4 \times 10^{-13} {\rm \; yr}^{-1}, 
\\
H &= 7 \times 10^{-11} {\rm \; yr^{-1}},
\end{align}
then give the estimates for $s_1$ and $s_2$, 
\BEq
s_1 \simeq 0.002,
\quad
s_2 \simeq -0.004,
\EEq
which, in turn, result in the following estimated secular changes 
per century for, say, the Hulse-Taylor binary,
\BEqA
(\Delta a)_{\rm 0PN} &\simeq& -0.027 {\rm \; m},
\\
(\Delta a)_{\rm 1PN} &\simeq& -7.3\times 10^{-7} {\rm \; m},
\\
(\Delta e)_{\rm 1PN} &\simeq& -6.3\times 10^{-17}.
\EEqA
These are too small to be detectable with presently available technology.

\section{Summary}

In conclusion, we performed post-Newtonian analysis of the equations 
of motion in the scalar-tensor theory of gravity for localized astronomical 
systems subjected to the time-dependent cosmological background. Several 
new cosmologically driven correction terms have been identified and their 
effects on the secular evolution of the orbital elements have been calculated. 
At the present level of observational astronomy, these contributions are 
negligible and cannot affect any realistic analysis of orbital motion based 
on Eq.\ (\ref{eq:NewtonianSimpleModel}) \cite{DAMOUR1988}. 
However, should experimental methods develop further,
the found corrections may prove helpful in establishing much stricter observational 
bounds on various PPN parameters as well as on the variability of the universal
gravitational constant. 

\begin{acknowledgments}

We thank the anonymous referee for carefully reading the manuscript and
for making suggestions that improved the presentation.
The work of S.\ K.\ has been supported by Grant No.\ 14-27-00068 
of the Russian Science Foundation.

\end{acknowledgments}

\end{document}

%\preprint{APS/123-QED}

\title{ Post-Newtonian celestial mechanics in scalar-tensor cosmology\\ 
(Supplemental materials: Derivation of the wave equations for perturbations)}

\author{Andrei Galiautdinov$^1$ and Sergei M.\ Kopeikin$^{2}$}
 \affiliation{$^1$Department of Physics and Astronomy, 
University of Georgia, Athens, Georgia 30602\\
$^2$Department of Physics and Astronomy, University of Missouri--Columbia, Columbia, 
Missouri 65211
}

\date{\today}% It is always \today, today,

\begin{abstract}
Detailed derivation of the wave equations for cosmological perturbations 
of the scalar-tensor theory used in [Phys.\ Rev.\ D {\bf 94}, 044015 (2016)] 
is provided.
\end{abstract}

%\pacs{96.50.S-, 98.62.En, 98.70.Sa}

\maketitle

\tableofcontents

\newpage

\allowdisplaybreaks[1]

\section{Notation}

\begin{itemize}

\item
$T$ and $X^i = \{X,Y,Z \}$ are the coordinate time and isotropic spatial coordinates on
the background manifold (in various parts of the manuscript other conventions may be used; 
e.\ g., in subsections of Section \ref{sec:BackgroundFriedmanCosmology});
\item
$X^\a = \{X^0, X^i\} =\{c\eta, X^i \}$ are the conformal coordinates with $\eta$ being 
the conformal time;
\item 
$x^\a = \{x^0, x^i\} = \{ct, x^i \}$ is an arbitrary coordinate chart on the background
manifold;
\item
 Greek indices $\a,\b\g, \dots, \m,\n, \dots$ run through values 0, 1, 2, 3, and label 
spacetime coordinates;
\item
 Roman indices $i, j, k, \dots$ take values 1, 2, 3, and label spatial coordinates;
\item
 Einstein summation convention for repeated (dummy) indices is always assumed, 
for example,
$P^{\a}Q_\a \equiv P^0Q_0+P^1Q_1+P^2Q_2+P^3Q_3$, and $P^i Q_i 
\equiv P^1Q_1+P^2Q_2+P^3Q_3$;
\item
 $g_{\a\b}$ is a full metric on the cosmological spacetime manifold;
\item
 $\bar{g}_{\a\b}$ is the Friedmann-Lemaitre-Robertson-Walker (FLRW) metric on 
the background spacetime manifold;
\item
$\frak{g}^{\a\b} = \sqrt{-g}g^{\a\b}$ is the (Gothic raised) metric tensor density 
of weight +1;
\item
$\frak{\bar g}^{\a\b} = \sqrt{-\bar{g}}\bar{g}^{\a\b}$ is the background metric 
tensor density of weight +1;
\item
$f_{\a\b}$ is the metric on the conformal spacetime manifold;
\item
$\eta_{\a\b} = {\rm diag}\{-1,+1,+1,+1\}$ is the Minkowski metric;
\item
$R = R(T)$, or $a = a(\eta) =
R[T (\eta)]$ is the scale factor of the FLRW metric;
\item
$H = R^{-1}dR/dT$ is the Hubble parameter;
\item
${\cal H} = a^{-1}da/d\eta$ is the conformal Hubble parameter;
\item
 a bar over a geometric object (as in ${\bar F}$), denotes the unperturbed value of 
$F$ on the
background manifold;
\item
 the tensor indices of geometric objects on the background manifold are raised and
lowered with the background metric $\bar{g}_{\a\b}$, for example 
$F_{\a\b} = \bar{g}_{\a\m}\bar{g}_{\b\n}F^{\m\n}$;
\item
 the tensor indices of geometric objects on the conformal spacetime are raised and
lowered with the conformal metric $f_{\a\b}$;
\item
 symmetrization of a geometric object with respect to two indices is denoted with 
the parenthesis, $F_{(\a\b)} \equiv (1/2)(F_{\a\b} + F_{\b\a})$;
\item
antisymmetrization of a geometric object with respect to two indices is denoted with 
the square brackets, $F_{[\a\b]} \equiv (1/2)(F_{\a\b} - F_{\b\a})$;
\item
 a prime, $W' = dW/d\phi$, denotes the derivative with respect to the scalar field $\phi$;
\item
 a dot, $\dot{F} = dF/d\eta$, denotes the total derivative with respect to the conformal
time $\eta$;
\item
$\partial_\a = \partial/\partial x^\a$ is a partial derivative with respect to 
coordinate $x^\a$;
\item
 a comma followed by an index, $F_{,\a} \equiv \partial_{\a}F$, indicates the
partial derivative with respect to coordinate $x^\a$, which is a convenient notation 
in some cases. When no confusion may arise, the comma as a symbol of
the partial derivative is omitted. For example, we may denote the partial derivatives of
the scalar field by $\varphi_{\a} \equiv \varphi_{,\a}$;
\item
 a vertical bar, $F_{|\a}$, denotes the covariant derivative associated with the background 
metric $\bar{g}_{\a\b}$. Covariant derivatives of scalar
fields coincide with their partial derivatives;
\item
 a semicolon, $F_{;\a}$ denotes the covariant derivative associated with the conformal 
metric $f_{\a\b}$
\item
 $\nabla_{\a}$ denotes the covariant derivative associated with the full metric ${g}_{\a\b}$;
\item
 $\phi$ is the fundamental scalar field of the Brans-Dicke theory;
\item
 $\omega$ is the Brans-Dicke parameter; in general, $\omega = \omega(\phi)$;
\item
$\bar{\phi}$ is the background value of the Brans-Dicke (scalar) field $\phi$;
\item
$\varphi = \phi - \bar{\phi}$ is the perturbation of $\phi$ from its background value 
$\bar{\phi}$. Fields $\phi$ and $\bar{\phi}$ refer to the same point on the spacetime 
manifold;
\item
 $\varkappa_{\a\b} \equiv g_{\a\b} - \bar{g}_{\a\b}$ is the metric tensor perturbation. 
Fields ${g}_{\a\b}$ and $\bar{g}_{\a\b}$ refer to the
same point on the spacetime manifold;
\item
$\frak{h}_{\a\b}\equiv \frak{g}_{\a\b} - \bar{\frak{g}}_{\a\b}$ is the perturbation of 
the metric density;
\item
$l^{\a\b} \equiv \frak{h}_{\a\b}/\sqrt{-\bar{g}}$. In the linear approximation, 
$l^{\a\b} = −\varkappa^{\a\b} + (1/2)\bar{g}^{\a\b}{\varkappa^{\a}}_{\a}$, 
where
${\varkappa^{\a}}_{\a} = \bar{g}^{\a\b}\varkappa_{\a\b}$;
\item
 the Christoffel symbols, 
${\G^{\a}}_{\b\g} = (1/2)g^{\a\kappa}
(g_{\kappa \g,\b}+g_{\kappa \b,\g}-g_{\b\g,\kappa})$;
\item
 the Riemann tensor, 
${R^{\a}}_{\b\m\n} = 
{\G^{\a}}_{\b\n,\m}-{\G^{\a}}_{\b\m,\n}
+{\G^{\a}}_{\m\kappa}{\G^{\kappa}}_{\b\n}
-{\G^{\a}}_{\n\kappa}{\G^{\kappa}}_{\b\mu}$;
\item
 the Ricci tensor, $R_{\a\b} = {R^{\m}}_{\a\m\b}$;
\item
 the Ricci scalar, $R = g^{\a\b}R_{\a\b}$.

\end{itemize}

%\newpage

\section{Brief review of dynamical perturbation theory}

In accordance with the dynamical perturbation theory of spacetime manifolds 
developed in Refs.\ \cite{SMKP2014, SMKP2017} we write the  
variables of the theory as the sums of their background values and 
the corresponding perturbations,
\begin{align}
\label{eq:metricDecomposed}
%g_{\m\n}(x)&=\bar{g}_{\m\n}(\eta)+\varkappa_{\m\n}(x),
%\nonumber\\
\Phi^A &= \bar{\Phi}^A +\varphi^A,
\end{align}
with $\Phi^A$ representing the generic multi-component 
field whose components are labeled by a generic index $A$. 
For example, $\Phi^A$ may collectively represent the metric density,
\begin{align}
\frak{g}^{\a\b}&=\frak{\bar{g}}^{\a\b} + \frak{h}^{\a\b},
\end{align}
and the scalar field,
\BEq
\phi = \bar{\phi}+\varphi,
\EEq 
of the Brans-Dicke theory.
Denoting by ${\cal L}$ the Lagrangian of the theory (regarded as a 
function of $\Phi^A$ and its derivatives of arbitrary, 
but finite, order), we first notice that the variational derivative of ${\cal L}$
obeys the rule
\begin{align}
\label{eq:useful}
\frac{\d {\cal L}}{\d {\varphi}^A}&=\frac{\d {\cal L}}{\d \bar{\Phi}^A},
\end{align}
which will be used in what follows. Expanding ${\cal L}$ in a Taylor series 
around $\bar{\Phi}^A$ gives
\BEq
\label{BriefReview_L}
{\cal L} = \bar{\cal L}+{\cal L}_1+{\cal L}^{\rm dyn}+{\cal L}^{\rm p},
\EEq
where $\bar{\cal L} \equiv {\cal L}(\bar{\Phi}^A)$ is the background Lagrangian,
\begin{align}
{\cal L}_1 &\equiv \varphi^A \frac{\d \bar{\cal L}}{\d \bar{\Phi}^A},
\end{align}
${\cal L}^{\rm dyn}$ is the infinite sum of the higher-order terms in $\varphi^A$ 
(in the linearized approximation, these are systematically discarded),
and ${\cal L}^{\rm p}$ is the Lagrangian of a localized gravitational source
(such as, e.\ g., a star or a planet), which is considered as a bare perturbation of the dynamical system). Because the barred variables satisfy 
the background field equations, we have
\begin{align}
\label{eq:on-shell}
\frac{\d \bar{\cal L}}{\d \bar{\Phi}^A}&=0,
\end{align}
which constitutes the so-called {\it on-shell} condition. 
The dynamical perturbation theory is then based on the assumption 
that the evolution of the field perturbations is governed by the 
variational equation (now ${\cal L}$ is formally regarded as the function 
of $\varphi^A$),
\BEq
\label{eq:varPrinciple}
\frac{\d {\cal L}}{\d \varphi^A}=0,
\EEq
subject to (\ref{eq:on-shell}).
Thus, applying (\ref{eq:useful}), (\ref{BriefReview_L}) and (\ref{eq:on-shell}) to (\ref{eq:varPrinciple})
gives
\begin{align}
\frac{\d {\cal L}}{\d \varphi^A }
&= \frac{\d }{\d \bar{\Phi}^A }
\left(\bar{\cal L}+{\cal L}_1 +{\cal L}^{\rm dyn}+{\cal L}^{\rm p} \right)
\nonumber \\
&= \frac{\d }{\d \bar{\Phi}^A }
\left({\cal L}_1+{\cal L}^{\rm dyn}+{\cal L}^{\rm p}\right)
= 0,
\end{align}
which results in the field equations for perturbations,
\begin{align}
\label{eq:forPerturbations}
\frac{2\kappa}{\sqrt{-\bar{g}}} \frac{\d }{\d \bar{\Phi}^A }
\left(
\varphi^B \frac{\d \bar{\cal L}}{\d \bar{\Phi}^B}
+{\cal L}^{\rm dyn}+{\cal L}^{\rm p}
\right)
= 0,
\end{align}
where the prefactor was inserted for future convenience, with $\kappa \equiv 8\pi$
being (dimensionless) Einstein's gravitational constant.
Notice that in (\ref{eq:forPerturbations}) the on-shell condition (\ref{eq:on-shell}) 
should not be imposed until after all the variational derivatives have been calculated.

%%%%%%%%%%%%%555

\section{Derivation of the wave equations for perturbations in the scalar-tensor theory}

Our main goal is to derive the wave equations for scalar 
field and metric perturbations of the scalar-tensor theory,
(\ref{eq:varphi_perturbation_linearHubble2_simpler}) and
(\ref{eq:l_mn_perturbation_linearHubble2_simpler}), 
which reproduce Eqs.\ (15) and (16) of Ref.\ \cite{AGSMK2016}.

\subsection{Lagrangian and stress-energy tensor}

We work with the Lagrangian of the form
\BEqA
{\cal L}^{G} &=& -\frac{1}{16\pi}\sqrt{-g}R\phi,
\\
{\cal L}^{BD} &=& \sqrt{-g}
\left[
\frac{1}{2}\tilde{\omega}(\phi) g^{\a\b}\phi_{,\a}\phi_{,\b}+W(\phi)
\right],
\EEqA
where
\BEq
\tilde{\omega}(\phi) = \frac{2}{16\pi}\frac{\omega}{\phi},
\quad
\tilde{\omega}'(\phi) = \frac{2}{16\pi}\frac{\omega}{\phi}
\left(\frac{\omega'}{\omega} - \frac{1}{{\phi}}\right),
\quad
W(\phi) = \frac{2}{16\pi}\lambda\phi,
\quad
W'(\phi) = \frac{2}{16\pi}(\lambda'\phi + \lambda).
\EEq
Notice that the associated stress-energy tensor of the scalar field is given by
\BEqA
T^{BD}_{\a\b} &=& \tilde{\omega}(\phi) \phi_{,\a}\phi_{,\b}
-g_{\a\b}
\left[
\frac{1}{2}\tilde{\omega}(\phi) g^{\r\s}\phi_{,\r}\phi_{,\s}+W(\phi)
\right].
\EEqA

\subsection{Background equations}

Upon direct variational calculation, we find the following background field equations
(here written in terms of $\tilde{\omega}$ and $\bar{W}$),
\BEqA
\label{eq:backgroundFieldEqs11}
\bar{R}_{\m\n}{\bar{\phi}}&=& {8\pi}
\left(
\bar{T}^{M}_{\m\n} -\frac{1}{2}\bar{g}_{\m\n}\bar{T}^{M}
+\bar{g}_{\m\n}\bar{W}+\tilde{\omega}{\bar{\phi}_{|\m}}{\bar{\phi}_{|\n}}
\right) +{\bar{\phi}_{|\m\n}}+\frac{1}{2}\bar{g}_{\m\n}{\bar{\phi}^{|\a}}_{|\a},
\\
\label{eq:backgroundFieldEqs12}
\bar{R}{\bar{\phi}}&=& {8\pi}
\left(
-\bar{T}^{M}+4\bar{W}+\tilde{\omega}{\bar{\phi}^{|\a}}{\bar{\phi}_{|\a}}
\right) +3{\bar{\phi}^{|\a}}_{|\a},
\\
\label{eq:backgroundFieldEqs13}
{\bar{\phi}^{|\a}}_{|\a}&=&\frac{8\pi}{3+16\pi\tilde{\omega}{\bar{\phi}}}
\left[
\bar{T}^{M}
-\left(
\tilde{\omega}+{\tilde{\omega}'}{\bar{\phi}}
\right){\bar{\phi}^{|\a}}{\bar{\phi}_{|\a}}
-4\bar{W}+2{\bar{W}'}\bar{\phi}
\right].
\EEqA
Alternatively, in terms of ${\omega}\equiv \bar{\omega}$ and $\bar{W}$,
\BEqA
\label{eq:backgroundFieldEqs31}
\left(\bar{R}_{\m\n}-\frac{1}{2}\bar{g}_{\m\n}\bar{R}\right)\bar{\phi}&=& {8\pi}
\left(
\bar{T}^{M}_{\m\n} -\bar{g}_{\m\n}\bar{W}
\right)
+\frac{\omega}{\bar{\phi}}
\left(
{\bar{\phi}_{|\m}}{\bar{\phi}_{|\n}}
-\frac{1}{2}\bar{g}_{\m\n}{\bar{\phi}^{|\a}}{\bar{\phi}_{|\a}}
\right)
 +{\bar{\phi}_{|\m\n}}-\bar{g}_{\m\n}{\bar{\phi}^{|\a}}_{|\a},
\\
\label{eq:backgroundFieldEqs33}
{\bar{\phi}^{|\a}}_{|\a}&=&\frac{1}{3+2{\omega}}
\left(
8\pi\bar{T}^{M}
-{\omega}'{\bar{\phi}^{|\a}}{\bar{\phi}_{|\a}}
-4\bar{W}+2{\bar{W}'}\bar{\phi}
\right),
\EEqA
which immediately shows that
\BEq
\label{eq:TandWpropH2}
\bar{T}^{M}_{\m\n} \sim \bar{W} \sim {\cal O}({\cal H}^2).
\EEq
Also, in terms of ${\omega}$ and $\lambda$,
\BEqA
\label{eq:backgroundFieldEqs21}
\bar{R}_{\m\n}&=& \frac{8\pi}{\bar{\phi}}
\left(
\bar{T}^{M}_{\m\n} -\frac{1+\omega}{3+2\omega}\bar{g}_{\m\n}\bar{T}^{M}
\right)
+\frac{\bar{g}_{\m\n}}{3+2\omega}
\left[
2(1+\omega)\lambda + {\lambda'} \bar{\phi}
-\frac{1}{2}\frac{\omega'}{\bar \phi}{\bar{\phi}^{|\a}}{\bar{\phi}_{|\a}}
\right]
 +\frac{1}{\bar \phi}{\bar{\phi}_{|\m\n}}+\frac{\omega}{\bar{\phi}^2}
\bar{\phi}_{|\m}\bar{\phi}_{|\n},
\\
\label{eq:backgroundFieldEqs22}
\bar{R}&=& -\frac{8\pi}{\bar{\phi}}\frac{2\omega}{3+2\omega}\bar{T}^{M}
+\frac{2(3+4\omega)\lambda }{3+2\omega}
+ \frac{6{\lambda'} \bar{\phi}}{3+2\omega}
+\left(\frac{\omega}{\bar{\phi}}-\frac{3{\omega'}}{3+2\omega}\right)
\frac{\bar{\phi}^{|\a}\bar{\phi}_{|\a}}{\bar{\phi}},
\\
\label{eq:backgroundFieldEqs23}
{\bar{\phi}^{|\a}}_{|\a}&=&\frac{1}{3+2\omega}
\left(
8\pi\bar{T}^{M}
-2\lambda{\bar{\phi}} +2{\lambda'}\bar{\phi}^2
-{\omega'}{\bar{\phi}^{|\a}}{\bar{\phi}_{|\a}}
\right).
\EEqA
In the above,
\BEqA
\label{eq:background_phi_mn}
{\bar{\phi}_{|\m\n}}
&=&{\bar{\phi}_{,\m\n}}-\bar{\Gamma}^{\r}_{\m\n}{\bar{\phi}_{,\r}},
\\
\label{eq:background_phi_aa}
{\bar{\phi}^{|\a}}_{|\a} &=& 
\bar{g}^{\a\b}\left({\bar{\phi}_{,\a\b}}
-\bar{\Gamma}^{\r}_{\a\b}{\bar{\phi}_{,\r}}\right),
\\
\label{eq:background_phi_a_phi_a}
\bar{\phi}^{|\a}\bar{\phi}_{|\a} 
&=& \bar{g}^{\a\b}{\bar{\phi}_{,\a}}{\bar{\phi}_{,\b}}.
\EEqA

%%%%%%%%%%%%%%%

\subsection{Equation for $l_{\m\n}$ perturbation}

In accordance with the dynamical perturbation theory of spacetime manifolds developed in
Ref.\ \cite{SMKP2014}, the field equations for metric perturbations, 
Eqs.\ (\ref{eq:forPerturbations}), are
\BEq
\label{eq:der01}
F^{G}_{\m\n}+F^{BD}_{\m\n}=8\pi \Lambda_{\m\n},
\EEq
where
\BEq
\label{eq:der02}
F^{G}_{\m\n}
= 
\frac{-16\pi}{\sqrt{-\bar{g}}}
\frac{\d }{\d {\bar{g}}^{\m\n}}
\left(
{\frak{h}}^{\r\s} \frac{\d \bar{\cal L}^{G}}{\d \bar{\frak{g}}^{\r\s}} 
+\varphi \frac{\d \bar{\cal L}^{G}}{\d \bar{\phi}}
\right)
\EEq
\BEq
\label{eq:der03}
F^{BD}_{\m\n}
= 
\frac{-16\pi}{\sqrt{-\bar{g}}}
\frac{\d }{\d {\bar{g}}^{\m\n}}
\left(
{\frak{h}}^{\r\s} \frac{\d \bar{\cal L}^{BD}}{\d \bar{\frak{g}}^{\r\s}} 
+\varphi \frac{\d \bar{\cal L}^{BD}}{\d \bar{\phi}}
\right),
\EEq
and $\Lambda_{\m\n}$ is the stress-energy tensor of the localized source.
Notice that in deriving (\ref{eq:der01}) we defined the stress-energy tensor of the source via
\begin{align}
\Lambda_{\m\n}
& \equiv + \frac{2}{\sqrt{-\bar{g}}} \frac{\d {\cal L}^{\rm p}}{\d \bar{g}^{\m\n} },
\end{align}
treated ${\cal L}^{\rm p}$ as being of first order of smallness, and used the chain rule,
\begin{align}
\frac{\d}{\d \bar{\frak{g}}^{\a\b}}
&=\frac{\d \bar{{g}}^{\r\s}}{\d \bar{\frak{g}}^{\a\b}}
\frac{\d}{\d \bar{{g}}^{\r\s}}
=\frac{1}{2\sqrt{-\bar{g}}}(\d^{\r}_{\a}\d^{\s}_{\b}+\d^{\r}_{\b}\d^{\s}_{\a}
-\bar{g}_{\a\b}\bar{g}^{\r\s})
\frac{\d}{\d \bar{{g}}^{\r\s}}.
\end{align}

\subsubsection{Derivation of $F^{G}_{\m\n}$}

We have,
\BEqA
\label{eq:der04}
-16\pi \frac{\d \bar{\cal L}^{G}}{\d \bar{\frak{g}}^{\r\s}} 
&=&
-16\pi 
\frac{\partial {\bar{g}}^{\m\n}}{\partial \bar{\frak{g}}^{\r\s}} 
\frac{\d \bar{\cal L}^{G}}{\d {\bar{g}}^{\m\n}} 
\nonumber \\
&=&
-16\pi 
\frac{\partial {\bar{g}}^{\m\n}}{\partial \bar{\frak{g}}^{\r\s}} 
\frac{\partial {\bar{g}}_{\a\b}}{\partial \bar{g}^{\m\n}} 
\frac{\d \bar{\cal L}^{G}}{\d {\bar{g}}_{\a\b}} 
\nonumber \\
&=&
-16\pi 
\frac{\partial {\bar{g}}^{\m\n}}{\partial \bar{\frak{g}}^{\r\s}} 
\left(- \bar{g}_{\m\a} \bar{g}_{\n\b} \right)
\frac{\d \bar{\cal L}^{G}}{\d {\bar{g}}_{\a\b}} 
\nonumber \\
&=& 
\frac{1}{2\sqrt{-\bar{g}}}
\left(
\d^{\m}_{\r}\d^{\n}_{\s} + \d^{\m}_{\s}\d^{\n}_{\r} 
- \bar{g}^{\m\n} \bar{g}_{\r\s} 
\right) 
\left(
- \bar{g}_{\m\a} \bar{g}_{\n\b} 
\right)
\left(-16\pi \frac{\d \bar{\cal L}^{G}}{\d {\bar{g}}_{\a\b}} \right).
\EEqA
Now, using (\ref{eq:extra3}),
\BEqA
\label{eq:der05}
-16\pi \frac{\d \bar{\cal L}^{G}}{\d {\bar{g}}_{\m\n}} 
&=&
\frac{\d}{\d \bar{g}_{\m\n}}\left(\sqrt{-\bar{g}}\bar{\phi}\bar{R}\right)
\nonumber \\
&=&
\frac{\d}{\d \bar{g}_{\m\n}}
\left(
\sqrt{-\bar{g}}\bar{\phi} \bar{g}^{\l\k}\d^{\g}_{\r}\bar{R}^{\r}_{\l\g\k}
\right)
\nonumber \\
&=&
\frac{\partial\left(\sqrt{-\bar{g}}\bar{g}^{\l\k}\right)}{\partial \bar{g}_{\m\n}}
\bar{\phi}\d^{\g}_{\r}\bar{R}^{\r}_{\l\g\k}
+
\left[
\sqrt{-\bar{g}}\bar{\phi} \bar{g}^{\l\k}\d^{\g}_{\r}
\left(
\bar{g}^{\s\n}\frac{\partial \bar{R}^{\r}_{\l\g\k}}{\partial \bar{R}^{\s}_{\m\b\a}}
+
\bar{g}^{\s\m}\frac{\partial \bar{R}^{\r}_{\l\g\k}}{\partial \bar{R}^{\s}_{\a\b\n}}
-
\bar{g}^{\s\a}\frac{\partial \bar{R}^{\r}_{\l\g\k}}{\partial \bar{R}^{\s}_{\m\b\n}}
\right)
\right]_{|\b\a}
\nonumber \\
&=&
\left[
\frac{\partial\left(\sqrt{-\bar{g}}\right)}{\partial \bar{g}_{\m\n}}
\bar{g}^{\l\k}
+
\sqrt{-\bar{g}}
\frac{\partial \bar{g}^{\l\k} }{\partial \bar{g}_{\m\n}}
\right]
\bar{\phi}\d^{\g}_{\r}\bar{R}^{\r}_{\l\g\k}
+
\left[
\sqrt{-\bar{g}}\bar{\phi} \bar{g}^{\l\k}\d^{\g}_{\r}
\left(
\bar{g}^{\s\n}\frac{\partial \bar{R}^{\r}_{\l\g\k}}{\partial \bar{R}^{\s}_{\m\b\a}}
+
\bar{g}^{\s\m}\frac{\partial \bar{R}^{\r}_{\l\g\k}}{\partial \bar{R}^{\s}_{\a\b\n}}
-
\bar{g}^{\s\a}\frac{\partial \bar{R}^{\r}_{\l\g\k}}{\partial \bar{R}^{\s}_{\m\b\n}}
\right)
\right]_{|\b\a}
\nonumber \\
&=&
\left[
\left(+\frac{1}{2}\sqrt{-\bar{g}} \bar{g}^{\m\n}\right)
\bar{g}^{\l\k}
+
\sqrt{-\bar{g}}
\left(-\bar{g}^{\l\m} \bar{g}^{\k\n}\right)
\right]
\bar{\phi}\bar{R}_{\l\k}
+
\left[
\sqrt{-\bar{g}}\bar{\phi} \bar{g}^{\l\k}\d^{\g}_{\r}
\left(
\bar{g}^{\s\n}\frac{\partial \bar{R}^{\r}_{\l\g\k}}{\partial \bar{R}^{\s}_{\m\b\a}}
+
\bar{g}^{\s\m}\frac{\partial \bar{R}^{\r}_{\l\g\k}}{\partial \bar{R}^{\s}_{\a\b\n}}
-
\bar{g}^{\s\a}\frac{\partial \bar{R}^{\r}_{\l\g\k}}{\partial \bar{R}^{\s}_{\m\b\n}}
\right)
\right]_{|\b\a}
\nonumber \\
&=&
\sqrt{-\bar{g}}
\left(
\frac{1}{2} \bar{g}^{\m\n}\bar{R} - \bar{R}^{\m\n}
\right)
\bar{\phi}
+
\left[
\sqrt{-\bar{g}}\bar{\phi} \bar{g}^{\l\k}\d^{\g}_{\r}
\left(
\bar{g}^{\s\n}\frac{\partial \bar{R}^{\r}_{\l\g\k}}{\partial \bar{R}^{\s}_{\m\b\a}}
+
\bar{g}^{\s\m}\frac{\partial \bar{R}^{\r}_{\l\g\k}}{\partial \bar{R}^{\s}_{\a\b\n}}
-
\bar{g}^{\s\a}\frac{\partial \bar{R}^{\r}_{\l\g\k}}{\partial \bar{R}^{\s}_{\m\b\n}}
\right)
\right]_{|\b\a}
\nonumber \\
&=&
%%%%%%%%%%%%%%%%%%%%
\sqrt{-\bar{g}}
\left(
\frac{1}{2} \bar{g}^{\m\n}\bar{R} - \bar{R}^{\m\n}
\right)
\bar{\phi}
\nonumber \\
&&
+
\left[
\sqrt{-\bar{g}}\bar{\phi} \bar{g}^{\l\k}\d^{\g}_{\r}
\left(
\bar{g}^{\s\n} 
\d^{\r}_{\s}\d^{\m}_{\l}
\frac{\d^{\b}_{\g}\d^{\a}_{\k}-\d^{\b}_{\k}\d^{\a}_{\g}}{2}
+
\bar{g}^{\s\m}
\d^{\r}_{\s}\d^{\a}_{\l}
\frac{\d^{\b}_{\g}\d^{\n}_{\k}-\d^{\b}_{\k}\d^{\n}_{\g}}{2}
-
\bar{g}^{\s\a}
\d^{\r}_{\s}\d^{\m}_{\l}
\frac{\d^{\b}_{\g}\d^{\n}_{\k}-\d^{\b}_{\k}\d^{\n}_{\g}}{2}
\right)
\right]_{|\b\a}
\nonumber \\
&=&
%%%%%%%%%%%%%%%%%%%%
\sqrt{-\bar{g}}
\left(
\frac{1}{2} \bar{g}^{\m\n}\bar{R} - \bar{R}^{\m\n}
\right)
\bar{\phi}
\nonumber \\
&&
+
\frac{\sqrt{-\bar{g}}}{2}
\left[
\bar{\phi} 
\left(
\bar{g}^{\l\k}\d^{\g}_{\r}
\bar{g}^{\s\n} 
\d^{\r}_{\s}\d^{\m}_{\l}
\left(\d^{\b}_{\g}\d^{\a}_{\k}-\d^{\b}_{\k}\d^{\a}_{\g}\right)
+
\bar{g}^{\l\k}\d^{\g}_{\r}
\bar{g}^{\s\m}
\d^{\r}_{\s}\d^{\a}_{\l}
\left(\d^{\b}_{\g}\d^{\n}_{\k}-\d^{\b}_{\k}\d^{\n}_{\g}\right)
-
\bar{g}^{\l\k}\d^{\g}_{\r}
\bar{g}^{\s\a}
\d^{\r}_{\s}\d^{\m}_{\l}
\left(\d^{\b}_{\g}\d^{\n}_{\k}-\d^{\b}_{\k}\d^{\n}_{\g}\right)
\right)
\right]_{|\b\a}
\nonumber \\
&=&
%%%%%%%%%%%%%%%%%%%%
\sqrt{-\bar{g}}
\left(
\frac{1}{2} \bar{g}^{\m\n}\bar{R} - \bar{R}^{\m\n}
\right)
\bar{\phi}
+
\frac{\sqrt{-\bar{g}}}{2}
\left[
\bar{\phi} 
\left(
\left(\bar{g}^{\m\a}\bar{g}^{\b\n}
-\bar{g}^{\m\b}\bar{g}^{\a\n}\right)
+
\left(\bar{g}^{\a\n}\bar{g}^{\b\m}
-\bar{g}^{\a\b}\bar{g}^{\m\n}\right)
-
\left(\bar{g}^{\m\n}\bar{g}^{\b\a}
-\bar{g}^{\m\b}\bar{g}^{\n\a}\right)
\right)
\right]_{|\b\a}
\nonumber \\
&=&
%%%%%%%%%%%%%%%%%%%%
\sqrt{-\bar{g}}
\left(
\frac{1}{2} \bar{g}^{\m\n}\bar{R} - \bar{R}^{\m\n}
\right)
\bar{\phi}
+
\frac{\sqrt{-\bar{g}}}{2}
\left(
\bar{g}^{\a\m}\bar{g}^{\b\n}
+
\bar{g}^{\a\n}\bar{g}^{\b\m}
-2\bar{g}^{\a\b}\bar{g}^{\m\n}
\right)
\bar{\phi}_{|\b\a}
\nonumber \\
&=&
%%%%%%%%%%%%%%%%%%%%
\sqrt{-\bar{g}}
\left[
\left(
\frac{1}{2} \bar{g}^{\m\n}\bar{R} - \bar{R}^{\m\n}
\right)
\bar{\phi}
+
\frac{\bar{\phi}^{|\m\n} + \bar{\phi}^{|\n\m}}{2}
-\bar{g}^{\m\n}{\bar{\phi}^{|\a}}_{|\a}
\right]
\nonumber \\
&=&
-\sqrt{-\bar{g}}
\left[
\left(
\bar{R}^{\m\n}-\frac{1}{2} \bar{g}^{\m\n}\bar{R} 
\right)
\bar{\phi}
+\bar{g}^{\m\n}{\bar{\phi}^{|\a}}_{|\a}
-\bar{\phi}^{|\m\n}
\right].
\EEqA
Substituting (\ref{eq:der05}) in (\ref{eq:der04}) gives
\BEqA
\label{eq:der06}
-16\pi \frac{\d \bar{\cal L}^{G}}{\d \bar{\frak{g}}^{\r\s}} 
&=& 
\frac{1}{2}
\left(
\d^{\m}_{\r}\d^{\n}_{\s} + \d^{\m}_{\s}\d^{\n}_{\r} 
- \bar{g}^{\m\n} \bar{g}_{\r\s} 
\right) 
 \bar{g}_{\m\a} \bar{g}_{\n\b} 
\left[
\left(
\bar{R}^{\a\b}-\frac{1}{2} \bar{g}^{\a\b}\bar{R} 
\right)
\bar{\phi}
+\bar{g}^{\a\b}{\bar{\phi}^{|\k}}_{|\k}
-\bar{\phi}^{|\a\b}
\right]
\nonumber \\
&=& 
\frac{1}{2}
\left(
\d^{\m}_{\r}\d^{\n}_{\s} + \d^{\m}_{\s}\d^{\n}_{\r} 
- \bar{g}^{\m\n} \bar{g}_{\r\s} 
\right) 
\left[
\left(
\bar{R}_{\m\n}-\frac{1}{2} \bar{g}_{\m\n}\bar{R} 
\right)
\bar{\phi}
+\bar{g}_{\m\n}{\bar{\phi}^{|\k}}_{|\k}
-\bar{\phi}_{|\m\n}
\right]
\nonumber \\
&=& 
\left(
\bar{R}_{\r\s}-\frac{1}{2} \bar{g}_{\r\s}\bar{R} 
\right)
\bar{\phi}
+\bar{g}_{\r\s}{\bar{\phi}^{|\k}}_{|\k}
-\bar{\phi}_{|\r\s}
- \frac{1}{2}
 \bar{g}_{\r\s} 
\left[
\left(
\bar{R}-2\bar{R} 
\right)
\bar{\phi}
+4{\bar{\phi}^{|\k}}_{|\k}
-\bar{g}^{\m\n}\bar{\phi}_{|\m\n}
\right]
\nonumber \\
&=& 
\bar{R}_{\r\s}\bar{\phi}
- \frac{1}{2} \bar{g}_{\r\s} {\bar{\phi}^{|\k}}_{|\k}
-\bar{\phi}_{|\r\s}.
\EEqA
Additionally,
\BEq
\label{eq:der07}
 -16\pi\frac{\d \bar{\cal L}^{G}}{\d \bar{\phi}} = \sqrt{-\bar{g}}\bar{R}.
\EEq
Thus, from (\ref{eq:der06}) and (\ref{eq:der07}),
\BEq
\label{eq:der08}
-16\pi
\left(
{\frak{h}}^{\r\s} \frac{\d \bar{\cal L}^{G}}{\d \bar{\frak{g}}^{\r\s}} 
+\varphi \frac{\d \bar{\cal L}^{G}}{\d \bar{\phi}}
\right)
=
{\frak{h}}^{\r\s}
\left(
\bar{R}_{\r\s}\bar{\phi}
- \frac{1}{2} \bar{g}_{\r\s} {\bar{\phi}^{|\k}}_{|\k}
-\bar{\phi}_{|\r\s}
\right)
+
\varphi\sqrt{-\bar{g}}\bar{R}.
\EEq

To get (\ref{eq:der02}) we still have to take the variational derivative of 
(\ref{eq:der08}) with respect to ${\bar{g}}^{\m\n}$. 

%%%%%%%%%%%%%%%%%%%%%%%%%%
First, by analogy with (\ref{eq:der05}), and taking into account 
an extra minus sign due to the derivative being with respect to the {\it raised} 
metric,
we have
\BEq
\label{eq:der10}
\frac{1}{\sqrt{-\bar{g}}}\frac{\d}{\d \bar{g}^{\m\n}}
\left(\varphi\sqrt{-\bar{g}}\bar{R}\right)
=
\left(
\bar{R}_{\m\n}-\frac{1}{2} \bar{g}_{\m\n}\bar{R} 
\right)
\varphi
+\bar{g}_{\m\n}{\varphi^{|\a}}_{|\a}
-{\varphi}_{|\m\n}.
\EEq
Next, by analogy with general relativity,
\BEqA
\label{eq:der09}
\frac{1}{\sqrt{-\bar{g}}}\frac{\d}{\d \bar{g}^{\m\n}}
\left({\frak{h}}^{\r\s}\bar{R}_{\r\s}\bar{\phi}\right)
&=&
\frac{1}{2}
\left[
{\left(\bar{\phi}l_{\m\n}\right)^{|\a}}_{|\a}
+\bar{g}_{\m\n}\left(\bar{\phi}l^{\a\b}\right)_{|\a\b}
-\left(\bar{\phi}{l^{\a}}_{\m}\right)_{|\n\a}
-\left(\bar{\phi}{l^{\a}}_{\n}\right)_{|\m\a}
\right]
%%%%%%%%%%%
\nonumber \\
&=&
\frac{1}{2}
\left(
{\bar{\phi}^{|\a}}_{|\a} l_{\m\n}
+ 2\bar{\phi}^{|\a} l_{\m\n|\a}
+ \bar{\phi}{{l_{\m\n}}^{|\a}}_{|\a}
\right)
%%%%%%%%%%%
\nonumber \\
&&+
\frac{1}{2}\bar{g}_{\m\n}
\left(
\bar{\phi}_{|\a\b}l^{\a\b}
+2\bar{\phi}_{|\a}{l^{\a\b}}_{|\b}
+\bar{\phi}{l^{\a\b}}_{|\a\b}
\right)
\nonumber \\
&&-
\frac{1}{2}
\left(
\bar{\phi}_{|\n\a}{l^{\a}}_{\m}
+ \bar{\phi}_{|\n}{l^{\a}}_{\m |\a}
+ \bar{\phi}_{|\a}{l^{\a}}_{\m |\n}
+ \bar{\phi}{l^{\a}}_{\m|\n\a}
\right)
\nonumber \\
&&-\frac{1}{2}
\left(
\bar{\phi}_{|\m\a}{l^{\a}}_{\n}
+ \bar{\phi}_{|\m}{l^{\a}}_{\n|\a}
+ \bar{\phi}_{|\a}{l^{\a}}_{\n|\m}
+ \bar{\phi}{{l^{\a}}_{\n}}_{|\m\a}
\right).
\EEqA
Now,
\BEqA
\label{eq:der11}
&&
\frac{1}{\sqrt{-\bar{g}}}\frac{\d}{\d \bar{g}^{\m\n}}
\left(
\frac{1}{2}{\frak{h}}^{\r\s}
 \bar{g}_{\r\s}\bar{g}^{\a\b} \bar{\phi}_{|\a\b}
\right)
=
-\frac{\bar{g}_{\m\k}\bar{g}_{\n\l} }{2\sqrt{-\bar{g}}}
\frac{\d}{\d \bar{g}_{\k\l}}
\left[
{\frak{h}}^{\r\s}
 \bar{g}_{\r\s}\bar{g}^{\a\b} 
\left(
\bar{\phi}_{,\a\b} - \bar\G^{\g}_{\a\b}\bar{\phi}_{,\g}
\right)
\right]
\nonumber \\
&&=
\frac{-\bar{g}_{\m\k}\bar{g}_{\n\l} }{2\sqrt{-\bar{g}}}
\left\{
{\frak{h}}^{\k\l}\bar{g}^{\a\b} \bar{\phi}_{|\a\b}
-
{\frak{h}}^{\r\s}
 \bar{g}_{\r\s}\bar{g}^{\a\k} \bar{g}^{\b\l} \bar{\phi}_{|\a\b}
-
\frac{1}{2}
\left[
\left(
{g}^{\chi\k}\frac{\partial }{\partial {\bar\G}^{\chi}_{\l\pi}}
+
{g}^{\chi\l}\frac{\partial }{\partial {\bar\G}^{\chi}_{\k\pi}}
-
{g}^{\chi\pi}\frac{\partial }{\partial {\bar\G}^{\chi}_{\k\l}}
\right)
{\frak{h}}^{\r\s}
 \bar{g}_{\r\s}\bar{g}^{\a\b} 
\left(
\bar{\phi}_{,\a\b} - \bar\G^{\g}_{\a\b}\bar{\phi}_{,\g}
\right)
\right]_{|\pi}
\right\}
%%%%%%%%%%%%%%%%%%%%%%%%%
\nonumber \\
&&=
-\frac{1}{2}
l_{\m\n}{\bar{\phi}^{|\a}}_{|\a}
+
\frac{1 }{2}
l\bar{\phi}_{|\m\n}
-
\frac{ 1}{4}\bar{g}_{\m\k}\bar{g}_{\n\l}\bar{g}^{\a\b}
\left[
\left(
\bar{g}^{\chi\k}\frac{\partial  \bar\G^{\g}_{\a\b}}{\partial {\bar\G}^{\chi}_{\l\pi}}
+
\bar{g}^{\chi\l}\frac{\partial  \bar\G^{\g}_{\a\b}}{\partial {\bar\G}^{\chi}_{\k\pi}}
-
\bar{g}^{\chi\pi}\frac{\partial  \bar\G^{\g}_{\a\b}}{\partial {\bar\G}^{\chi}_{\k\l}}
\right) 
l\bar{\phi}_{,\g}
\right]_{|\pi}
%%%%%%%%%%%%%%%%%%%%%%%%%
\nonumber \\
&&=
-\frac{1}{2}
l_{\m\n}{\bar{\phi}^{|\a}}_{|\a}
+
\frac{1 }{2}
l\bar{\phi}_{|\m\n}
-
\frac{ 1}{4}\bar{g}_{\m\k}\bar{g}_{\n\l}\bar{g}^{\a\b}
\left[
\left(
\bar{g}^{\chi\k}
\d^{\g}_{\chi}
\frac{\d^{\l}_{\a}\d^{\pi}_{\b}+\d^{\l}_{\b}\d^{\pi}_{\a}}{2}
+
\bar{g}^{\chi\l}
\d^{\g}_{\chi}
\frac{\d^{\k}_{\a}\d^{\pi}_{\b}+\d^{\k}_{\b}\d^{\pi}_{\a}}{2}
-
\bar{g}^{\chi\pi}
\d^{\g}_{\chi}
\frac{\d^{\k}_{\a}\d^{\l}_{\b}+\d^{\k}_{\b}\d^{\l}_{\a}}{2}
\right) 
l\bar{\phi}_{,\g}
\right]_{|\pi}
%%%%%%%%%%%%%%%%%%%%%%%%%
\nonumber \\
&&=
-\frac{1}{2}
l_{\m\n}{\bar{\phi}^{|\a}}_{|\a}
+
\frac{1 }{2}
l\bar{\phi}_{|\m\n}
-
\frac{ 1}{4}\bar{g}_{\m\k}\bar{g}_{\n\l}\bar{g}^{\a\b}
\left[
\left(
\bar{g}^{\g\k}
\frac{\d^{\l}_{\a}\d^{\pi}_{\b}+\d^{\l}_{\b}\d^{\pi}_{\a}}{2}
+
\bar{g}^{\g\l}
\frac{\d^{\k}_{\a}\d^{\pi}_{\b}+\d^{\k}_{\b}\d^{\pi}_{\a}}{2}
-
\bar{g}^{\g\pi}
\frac{\d^{\k}_{\a}\d^{\l}_{\b}+\d^{\k}_{\b}\d^{\l}_{\a}}{2}
\right) 
l\bar{\phi}_{|\g}
\right]_{|\pi}
%%%%%%%%%%%%%%%%%%%%%%%%%
\nonumber \\
&&=
-\frac{1}{2}
l_{\m\n}{\bar{\phi}^{|\a}}_{|\a}
+
\frac{1 }{2}
l\bar{\phi}_{|\m\n}
-
\frac{ 1}{4}
\left(
\d^{\g}_{\m}\d^{\p}_{\n}
+
\d^{\p}_{\m}\d^{\g}_{\n}
-
\bar{g}_{\m\n}\bar{g}^{\g\pi}
\right) 
\left(
l\bar{\phi}_{|\g}\right)_{|\pi}
%%%%%%%%%%%%%%%%%%%%%%%%%
\nonumber \\
&&=
-\frac{1}{2}
l_{\m\n}{\bar{\phi}^{|\a}}_{|\a}
+
\frac{1 }{2}
l\bar{\phi}_{|\m\n}
-
\frac{ 1}{4}
\left[
\left(
l\bar{\phi}_{|\m}\right)_{|\n}
+
\left(
l\bar{\phi}_{|\n}\right)_{|\m}
-
\bar{g}_{\m\n}
\left(
l\bar{\phi}_{|\a}\right)^{|\a}
\right] 
%%%%%%%%%%%%%%%%%%%%%%%%%
\nonumber \\
&&=
-\frac{1}{2}
l_{\m\n}{\bar{\phi}^{|\a}}_{|\a}
+
\frac{1 }{2}
l\bar{\phi}_{|\m\n}
-
\frac{ 1}{4}
\left[
l_{|\n}\bar{\phi}_{|\m}
+
l\bar{\phi}_{|\m\n}
+
l_{|\m}\bar{\phi}_{|\n}
+
l\bar{\phi}_{|\n\m}
-
\bar{g}_{\m\n}
l^{|\a}\bar{\phi}_{|\a}
-
\bar{g}_{\m\n}
l{\bar{\phi}_{|\a}}^{|\a}
\right] 
%%%%%%%%%%%%%%%%%%%%%%%%%
\nonumber \\
&&=
-\frac{1}{2}
l_{\m\n}{\bar{\phi}^{|\a}}_{|\a}
+
\frac{1 }{2}
l\bar{\phi}_{|\m\n}
-
\frac{ 1}{4}
\left[
l_{|\n}\bar{\phi}_{|\m}
+
l_{|\m}\bar{\phi}_{|\n}
+
2l\bar{\phi}_{|\n\m}
-
\bar{g}_{\m\n}
l^{|\a}\bar{\phi}_{|\a}
-
\bar{g}_{\m\n}
l{\bar{\phi}_{|\a}}^{|\a}
\right] 
%%%%%%%%%%%%%%%%%%%%%%%%%
\nonumber \\
&&=
-\frac{1}{2}
l_{\m\n}{\bar{\phi}^{|\a}}_{|\a}
-
\frac{ 1}{4}
\left(
l_{|\m}\bar{\phi}_{|\n} + l_{|\n}\bar{\phi}_{|\m}
\right)
+\frac{ 1}{4}
\bar{g}_{\m\n}
\left(
l^{|\a}\bar{\phi}_{|\a}
+
l{\bar{\phi}_{|\a}}^{|\a}
\right).
\EEqA
Also,
\BEqA
\label{eq:der12}
\frac{1}{\sqrt{-\bar{g}}}\frac{\d}{\d \bar{g}^{\m\n}}
\left({\frak{h}}^{\r\s}\bar{\phi}_{|\r\s}\right)
&=&
-\frac{\bar{g}_{\m\k}\bar{g}_{\n\l} }{\sqrt{-\bar{g}}}
\frac{\d}{\d \bar{g}_{\k\l}}
\left[
{\frak{h}}^{\r\s}
\left(
\bar{\phi}_{,\r\s} - \bar\G^{\g}_{\r\s}\bar{\phi}_{,\g}
\right)
\right]
\nonumber \\
&=&
\frac{-\bar{g}_{\m\k}\bar{g}_{\n\l} }{2\sqrt{-\bar{g}}}
\left[
\left(
{g}^{\chi\k}\frac{\partial }{\partial {\bar\G}^{\chi}_{\l\pi}}
+
{g}^{\chi\l}\frac{\partial }{\partial {\bar\G}^{\chi}_{\k\pi}}
-
{g}^{\chi\pi}\frac{\partial }{\partial {\bar\G}^{\chi}_{\k\l}}
\right)
{\frak{h}}^{\r\s}
\bar\G^{\g}_{\r\s}\bar{\phi}_{,\g}
\right]_{|\pi}
%%%%%%%%%%%%%%%%%%%%%%%%%
\nonumber \\
&=&
\frac{-\bar{g}_{\m\k}\bar{g}_{\n\l} }{2}
\left[
\left(
\bar{g}^{\chi\k}
\d^{\g}_{\chi}
\frac{\d^{\l}_{\r}\d^{\pi}_{\s}+\d^{\l}_{\s}\d^{\pi}_{\r}}{2}
+
\bar{g}^{\chi\l}
\d^{\g}_{\chi}
\frac{\d^{\k}_{\r}\d^{\pi}_{\s}+\d^{\k}_{\s}\d^{\pi}_{\r}}{2}
-
\bar{g}^{\chi\pi}
\d^{\g}_{\chi}
\frac{\d^{\k}_{\r}\d^{\l}_{\s}+\d^{\k}_{\s}\d^{\l}_{\r}}{2}
\right) 
l^{\r\s}
\bar{\phi}_{|\g}
\right]_{|\pi}
%%%%%%%%%%%%%%%%%%%%%%%%%
\nonumber \\
&=&
\frac{-\bar{g}_{\m\k}\bar{g}_{\n\l} }{2}
\left[
\left(
\bar{g}^{\g\k}
\frac{\d^{\l}_{\r}\d^{\pi}_{\s}+\d^{\l}_{\s}\d^{\pi}_{\r}}{2}
+
\bar{g}^{\g\l}
\frac{\d^{\k}_{\r}\d^{\pi}_{\s}+\d^{\k}_{\s}\d^{\pi}_{\r}}{2}
-
\bar{g}^{\g\pi}
\frac{\d^{\k}_{\r}\d^{\l}_{\s}+\d^{\k}_{\s}\d^{\l}_{\r}}{2}
\right) 
l^{\r\s}
\bar{\phi}_{|\g}
\right]_{|\pi}
%%%%%%%%%%%%%%%%%%%%%%%%%
\nonumber \\
&=&
\frac{-\bar{g}_{\m\k}\bar{g}_{\n\l} }{2}
\left[
\left(
\bar{g}^{\g\k}l^{\l\p}
+
\bar{g}^{\g\l}l^{\k\p}
-
\bar{g}^{\g\pi}l^{\k\l}
\right) 
\bar{\phi}_{|\g}
\right]_{|\pi}
%%%%%%%%%%%%%%%%%%%%%%%%%
\nonumber \\
&=&
-\frac{1 }{2}
\left(
{l_{\n}}^{\p}\bar{\phi}_{|\m}
+
{l_{\m}}^{\p}\bar{\phi}_{|\n}
-
l_{\m\n}\bar{\phi}^{|\p}
\right)_{|\pi}
%%%%%%%%%%%%%%%%%%%%%%%%%
\nonumber \\
&=&
-\frac{1 }{2}
\left[
\left(
\bar{\phi}_{|\m}{l_{\n}}^{\a}
\right)_{|\a}
+
\left(
\bar{\phi}_{|\n}{l_{\m}}^{\a}
\right)_{|\a}
-
\left(
\bar{\phi}^{|\a}l_{\m\n}
\right)_{|\a}
\right]
%%%%%%%%%%%%%%%%%%%%%%%%%
\nonumber \\
&=&
-\frac{1 }{2}
\left[
\bar{\phi}_{|\m\a}{l_{\n}}^{\a}
+
\bar{\phi}_{|\n\a}{l_{\m}}^{\a}
-
{\bar{\phi}^{|\a}}_{|\a}l_{\m\n}
+
\bar{\phi}_{|\m}{{l_{\n}}^{\a}}_{|\a}
+
\bar{\phi}_{|\n}{{l_{\m}}^{\a}}_{|\a}
-
\bar{\phi}^{|\a}l_{\m\n|\a}
\right].
\EEqA

Combining (\ref{eq:der09}), (\ref{eq:der11}), and (\ref{eq:der12}) gives
\BEqA
\label{eq:der13}
\frac{1}{\sqrt{-\bar{g}}}\frac{\d}{\d \bar{g}^{\m\n}}
\left[
{\frak{h}}^{\r\s}
\left(
\bar{R}_{\r\s}\bar{\phi}
- \frac{1}{2} \bar{g}_{\r\s} {\bar{\phi}^{|\k}}_{|\k}
-\bar{\phi}_{|\r\s}
\right)
\right]
&=&
\frac{1}{2}
\left(
\underbrace{
{\bar{\phi}^{|\a}}_{|\a} l_{\m\n}
}_{5}
+ 
\underbrace{
2\bar{\phi}^{|\a} l_{\m\n|\a}
}_{[6]}
+ \bar{\phi}{{l_{\m\n}}^{|\a}}_{|\a}
\right)
\nonumber \\
&&+
\frac{1}{2}\bar{g}_{\m\n}
\left(
\bar{\phi}_{|\a\b}l^{\a\b}
+2\bar{\phi}_{|\a}{l^{\a\b}}_{|\b}
+\bar{\phi}{l^{\a\b}}_{|\a\b}
\right)
\nonumber \\
&&-
\frac{1}{2}
\left(
\underbrace{
\bar{\phi}_{|\n\a}{l^{\a}}_{\m}
}_{2}
+ 
\underbrace{
\bar{\phi}_{|\n}{l^{\a}}_{\m |\a}
}_{4}
+ \bar{\phi}_{|\a}{l^{\a}}_{\m |\n}
+ \bar{\phi}{l^{\a}}_{\m|\n\a}
\right)
\nonumber \\
&&-\frac{1}{2}
\left(
\underbrace{
\bar{\phi}_{|\m\a}{l^{\a}}_{\n}
}_{1}
+ 
\underbrace{
\bar{\phi}_{|\m}{l^{\a}}_{\n|\a}
}_{3}
+ \bar{\phi}_{|\a}{l^{\a}}_{\n|\m}
+ \bar{\phi}{{l^{\a}}_{\n}}_{|\m\a}
\right)
\nonumber \\
&&
+\frac{1}{2}
l_{\m\n}{\bar{\phi}^{|\a}}_{|\a}
+
\frac{ 1}{4}
\left(
l_{|\m}\bar{\phi}_{|\n} + l_{|\n}\bar{\phi}_{|\m}
\right)
-\frac{ 1}{4}
\bar{g}_{\m\n}
\left(
l^{|\a}\bar{\phi}_{|\a}
+
l{\bar{\phi}_{|\a}}^{|\a}
\right)
\nonumber \\
&&
+
\frac{1}{2}
\left(
\underbrace{
\bar{\phi}_{|\m\a}{l_{\n}}^{\a}
}_{1}
+
\underbrace{
\bar{\phi}_{|\n\a}{l_{\m}}^{\a}
}_{2}
-
\underbrace{
{\bar{\phi}^{|\a}}_{|\a}l_{\m\n}
}_{5}
+
\underbrace{
\bar{\phi}_{|\m}{{l_{\n}}^{\a}}_{|\a}
}_{3}
+
\underbrace{
\bar{\phi}_{|\n}{{l_{\m}}^{\a}}_{|\a}
}_{4}
-
\underbrace{
\bar{\phi}^{|\a}l_{\m\n|\a}
}_{[6]}
\right)
%%%%%%%%%%%
\nonumber \\
&=&
\frac{1}{2}
\left(
\bar{\phi}^{|\a} l_{\m\n|\a}
+ \bar{\phi}{{l_{\m\n}}^{|\a}}_{|\a}
\right)
\nonumber \\
&&+
\frac{1}{2}\bar{g}_{\m\n}
\left(
\bar{\phi}_{|\a\b}l^{\a\b}
+2\bar{\phi}_{|\a}{l^{\a\b}}_{|\b}
+\bar{\phi}{l^{\a\b}}_{|\a\b}
\right)
\nonumber \\
&&-
\frac{1}{2}
\left(
\bar{\phi}_{|\a}{l^{\a}}_{\m |\n}
+ \bar{\phi}{l^{\a}}_{\m|\n\a}
\right)
-\frac{1}{2}
\left(
\bar{\phi}_{|\a}{l^{\a}}_{\n|\m}
+ \bar{\phi}{{l^{\a}}_{\n}}_{|\m\a}
\right)
\nonumber \\
&&
+\frac{1}{2}
l_{\m\n}{\bar{\phi}^{|\a}}_{|\a}
+
\frac{ 1}{4}
\left(
l_{|\m}\bar{\phi}_{|\n} + l_{|\n}\bar{\phi}_{|\m}
\right)
-\frac{ 1}{4}
\bar{g}_{\m\n}
\left(
l^{|\a}\bar{\phi}_{|\a}
+
l{\bar{\phi}_{|\a}}^{|\a}
\right)
%%%%%%%%%%%
\nonumber \\
&=&
 \frac{1}{2}\bar{\phi}
\left(
{{l_{\m\n}}^{|\a}}_{|\a}
+\bar{g}_{\m\n}{l^{\a\b}}_{|\a\b}
- {l^{\a}}_{\m|\n\a}
- {{l^{\a}}_{\n}}_{|\m\a}
\right)
\nonumber \\
&&+
\frac{1}{2}\bar{g}_{\m\n}
\left(
\bar{\phi}_{|\a\b}l^{\a\b}
+2\bar{\phi}_{|\a}{l^{\a\b}}_{|\b}
-\frac{ 1}{2}\bar{\phi}_{|\a}l^{|\a}
\right)
\nonumber \\
&&
+ \frac{1}{2}\bar{\phi}^{|\a} 
\left(
l_{\m\n|\a}-l_{\a\m |\n}-l_{\a\n|\m}
\right)
\nonumber \\
&&
+\frac{1}{2}{\bar{\phi}^{|\a}}_{|\a}\left(l_{\m\n}-\frac{1}{2}\bar{g}_{\m\n}l\right)
+
\frac{ 1}{4}
\left(
l_{|\m}\bar{\phi}_{|\n} + l_{|\n}\bar{\phi}_{|\m}
\right).
\EEqA
Combining (\ref{eq:der13}) with (\ref{eq:der10}) gives
\BEqA
\label{eq:der14}
F^{G}_{\m\n}
&=& 
\frac{1}{2}
\bar{\phi}
\underbrace{
\left(
{{l_{\m\n}}^{|\a}}_{|\a}+\bar{g}_{\m\n}{A^{\a}}_{|\a}
-{A}_{\m|\n}-{A}_{\n|\m}
-{\bar{R}^{\a}}_{\n}l_{\m\a}-{\bar{R}^{\a}}_{\m}l_{\n\a}
-2\bar{R}_{\m\a\b\n}l^{\a\b}
\right)
}_{
\equiv 
{{l_{\m\n}}^{|\a}}_{|\a}+\bar{g}_{\m\n}{A^{\a}}_{|\a}
-{l^{\a}}_{\m|\n\a}-{l^{\a}}_{\n|\m\a}
}
\nonumber \\
&&+
\frac{1}{2}
{\bar{\phi}}^{|\a}
\left( l_{\m\n|\a}- l_{\a\m|\n}- l_{\a\n|\m} \right)
+
\frac{1}{2}
{{\bar{\phi}}^{|\a}}_{|\a}\left(l_{\m\n}-\frac{1}{2}\bar{g}_{\m\n}l\right)
\nonumber \\
&&
+ 
\frac{1}{2}\bar{g}_{\m\n}
\left(
\bar{\phi}_{|\a\b}l^{\a\b}+2\bar{\phi}_{|\a}A^{\a}-\frac{1}{2}\bar{\phi}_{|\a}l^{|\a}
\right)
+
\frac{1}{4}
\left( l_{|\m}{\bar{\phi}}_{|\n}+l_{|\n}{\bar{\phi}}_{|\m}\right)
\nonumber \\
&&+
\left(\bar{R}_{\m\n}-\frac{1}{2}\bar{g}_{\m\n}\bar{R}\right)\varphi 
+\bar{g}_{\m\n}{\varphi^{|\a}}_{|\a}-\varphi_{|\m\n}.
\EEqA

%%%%%%%%%%%%%%%%%%%%%%%%

\subsubsection{Derivation of $F^{BD}_{\m\n}$}

We have,
\BEqA
\label{eq:der15}
\frac{\d \bar{\cal L}^{BD}}{\d \bar{\frak{g}}^{\r\s}} 
&=&
\frac{\partial {\bar{g}}^{\m\n}}{\partial \bar{\frak{g}}^{\r\s}} 
\frac{\d \bar{\cal L}^{BD}}{\d {\bar{g}}^{\m\n}} 
\nonumber \\
&=& 
\frac{1}{2\sqrt{-\bar{g}}}
\left(
\d^{\m}_{\r}\d^{\n}_{\s} + \d^{\m}_{\s}\d^{\n}_{\r} 
- \bar{g}^{\m\n} \bar{g}_{\r\s} 
\right) 
 \frac{\d \bar{\cal L}^{BD}}{\d {\bar{g}}^{\m\n}}.
\EEqA
Now,
\BEqA
\label{eq:der16}
 \frac{\d \bar{\cal L}^{BD}}{\d {\bar{g}}^{\m\n}}
&=&
 \frac{\d }{\d {\bar{g}}^{\m\n}}
\left[
\sqrt{-\bar{g}}
\left(
\frac{1}{2}\tilde{\omega}(\bar{\phi}) 
\bar{g}^{\a\b}\bar{\phi}_{,\a}\bar{\phi}_{,\b}+W(\bar{\phi})
\right)
\right]
\nonumber \\
&=&
-\frac{1}{2}\sqrt{-\bar{g}} \bar{g}_{\m\n}
\left(
\frac{1}{2}\tilde{\omega}(\bar{\phi}) 
\bar{g}^{\a\b}\bar{\phi}_{,\a}\bar{\phi}_{,\b}+W(\bar{\phi})
\right)
+
\frac{1}{2}\sqrt{-\bar{g}}\tilde{\omega}(\bar{\phi}) \bar{\phi}_{,\m}\bar{\phi}_{,\n}
\nonumber \\
&=&
\frac{1}{2}\sqrt{-\bar{g}}
\underbrace{
\left[
\tilde{\omega}(\bar{\phi}) \bar{\phi}_{,\m}\bar{\phi}_{,\n}
- \bar{g}_{\m\n}
\left(
\frac{1}{2}\tilde{\omega}(\bar{\phi}) 
\bar{g}^{\a\b}\bar{\phi}_{,\a}\bar{\phi}_{,\b}+W(\bar{\phi})
\right)
\right]
}_{\equiv \bar{T}^{BD}_{\m\n}}.
\EEqA
Substituting (\ref{eq:der16}) in (\ref{eq:der15}) gives
\BEq
\label{eq:der17}
 \frac{\d \bar{\cal L}^{BD}}{\d \bar{\frak{g}}^{\r\s}} 
=
\frac{1}{2}
\left[
\tilde{\omega}(\bar{\phi}) \bar{\phi}_{,\r}\bar{\phi}_{,\s}
+ \bar{g}_{\r\s} W(\bar{\phi})
\right].
\EEq
Also, using (\ref{eq:extra2}),
\BEqA
\label{eq:der18}
 \frac{\d \bar{\cal L}^{BD}}{\d \bar{\phi}}
&=&
\frac{\partial \bar{\cal L}^{BD}}{\partial \bar{\phi}}
-\left[\frac{\partial \bar{\cal L}^{BD}}{\partial \bar{\phi}_{|\r}}\right]_{|\r}
\nonumber \\
&=&
\frac{\partial }{\partial \bar{\phi}}
\left[
\sqrt{-\bar{g}}
\left(
\frac{1}{2}\tilde{\omega}(\bar{\phi}) 
\bar{g}^{\a\b}\bar{\phi}_{,\a}\bar{\phi}_{,\b}+W(\bar{\phi})
\right)
\right]
-
\left\{
\frac{\partial }{\partial \bar{\phi}_{|\r}}
\left[
\sqrt{-\bar{g}}
\left(
\frac{1}{2}\tilde{\omega}(\bar{\phi}) 
\bar{g}^{\a\b}\bar{\phi}_{,\a}\bar{\phi}_{,\b}+W(\bar{\phi})
\right)
\right]
\right\}_{|\r}
\nonumber \\
&=&
\sqrt{-\bar{g}}
\left(
\frac{1}{2}\tilde{\omega}'(\bar{\phi}) 
\bar{g}^{\a\b}\bar{\phi}_{,\a}\bar{\phi}_{,\b}+W'(\bar{\phi})
\right)
-
\left(
\sqrt{-\bar{g}}
\tilde{\omega}(\bar{\phi}) \bar{\phi}^{|\r}
\right)_{|\r}.
\EEqA
Thus, from (\ref{eq:der17}) and (\ref{eq:der18}),
\BEqA
\label{eq:der19}
{\frak{h}}^{\r\s} \frac{\d \bar{\cal L}^{BD}}{\d \bar{\frak{g}}^{\r\s}} 
+\varphi \frac{\d \bar{\cal L}^{BD}}{\d \bar{\phi}}
&=& 
{\frak{h}}^{\r\s}
\frac{1}{2}
\left[
\tilde{\omega}(\bar{\phi}) \bar{\phi}_{,\r}\bar{\phi}_{,\s}
+ \bar{g}_{\r\s} W(\bar{\phi})
\right]
+
\varphi
\left[
\sqrt{-\bar{g}}
\left(
\frac{1}{2}\tilde{\omega}'(\bar{\phi}) 
\bar{g}^{\a\b}\bar{\phi}_{,\a}\bar{\phi}_{,\b}+W'(\bar{\phi})
\right)
-
\left(
\sqrt{-\bar{g}}
\tilde{\omega}(\bar{\phi}) \bar{\phi}^{|\r}
\right)_{|\r}
\right]
\nonumber \\
&=& 
{\frak{h}}^{\r\s}
\frac{1}{2}
\left[
\tilde{\omega}(\bar{\phi}) \bar{\phi}_{,\r}\bar{\phi}_{,\s}
+ \bar{g}_{\r\s} W(\bar{\phi})
\right]
\nonumber\\
&&+
\sqrt{-\bar{g}}
\left(
\frac{1}{2}\tilde{\omega}'(\bar{\phi}) 
\bar{g}^{\a\b}\bar{\phi}_{,\a}\bar{\phi}_{,\b}+W'(\bar{\phi})
\right)\varphi
+
\sqrt{-\bar{g}}
\tilde{\omega}(\bar{\phi}) \bar{\phi}^{|\r}
\varphi_{|\r}
-
\underbrace{
\left(
\varphi
\sqrt{-\bar{g}}
\tilde{\omega}(\bar{\phi}) \bar{\phi}^{|\r}
\right)_{|\r}
}_{\rm cov.\;divergence}
\nonumber \\
&=& 
{\frak{h}}^{\r\s}
\frac{1}{2}
\left[
\tilde{\omega}(\bar{\phi}) \bar{\phi}_{,\r}\bar{\phi}_{,\s}
+ \bar{g}_{\r\s} W(\bar{\phi})
\right]
\nonumber\\
&&+
\sqrt{-\bar{g}}
\left[
\left(
\frac{1}{2}\tilde{\omega}'(\bar{\phi}) 
\bar{g}^{\a\b}\bar{\phi}_{,\a}\bar{\phi}_{,\b}+W'(\bar{\phi})
\right)\varphi
+
\tilde{\omega}(\bar{\phi})\bar{g}^{\r\s} \bar{\phi}_{,\s}
\varphi_{,\r}
\right]
-
\underbrace{
\left(
\varphi
\sqrt{-\bar{g}}
\tilde{\omega}(\bar{\phi}) \bar{\phi}^{|\r}
\right)_{|\r}
}_{\rm cov.\;divergence}.
\EEqA
Taking the variational derivative of (\ref{eq:der19}) with respect to $\bar{g}^{\m\n}$,
dropping the covariant divergence term, and noticing that ${\frak{h}}^{\r\s}$
is independent of $\bar{g}^{\m\n}$ 
(even though we formally write ${\frak{h}}^{\r\s}\equiv {\sqrt{-\bar{g}}}l^{\r\s}$), we get
\BEqA
\label{eq:der20}
F^{BD}_{\m\n}
&=&
\frac{-16\pi}{{\sqrt{-\bar{g}}}}\frac{\d}{\d \bar{g}^{\m\n}}
\left(
{\frak{h}}^{\r\s} \frac{\d \bar{\cal L}^{BD}}{\d \bar{\frak{g}}^{\r\s}} 
+\varphi \frac{\d \bar{\cal L}^{BD}}{\d \bar{\phi}}
\right)
\nonumber \\
&=& 
8\pi
\left[
{l}^{\m\n} \bar{W}
+\bar{g}_{\m\n}
\left(
\frac{1}{2}\tilde{\omega}' \bar{\phi}^{|\a}\bar{\phi}_{|\a}\varphi
+\bar{W}'\varphi
+
\tilde{\omega} \bar{\phi}^{|\a}\varphi_{|\a}
\right)
-
\tilde{\omega}'\bar{\phi}_{|\m}\bar{\phi}_{|\n}\varphi
-\tilde{\omega}
\left(
\bar{\phi}_{|\m}\varphi_{|\n}+\bar{\phi}_{|\n}\varphi_{|\m}
\right)
\right].
\EEqA

%%%%%%%%%%%%%%%%%%%%%%%%%

\subsubsection{Final result}

Thus,
\BEq
\label{eq:l_mn_perturbation}
F^{G}_{\m\n}+F^{BD}_{\m\n}=8\pi \Lambda_{\m\n},
\EEq
where
\BEqA
\label{eq:der10_consequence1}
F^{G}_{\m\n}
&=& 
\frac{1}{2}
\bar{\phi}
\underbrace{
\left(
{{l_{\m\n}}^{|\a}}_{|\a}+\bar{g}_{\m\n}{A^{\a}}_{|\a}
-{A}_{\m|\n}-{A}_{\n|\m}
-{\bar{R}^{\a}}_{\n}l_{\m\a}-{\bar{R}^{\a}}_{\m}l_{\n\a}
-2\bar{R}_{\m\a\b\n}l^{\a\b}
\right)
}_{
\equiv 
{{l_{\m\n}}^{|\a}}_{|\a}+\bar{g}_{\m\n}{A^{\a}}_{|\a}
-{l^{\a}}_{\m|\n\a}-{l^{\a}}_{\n|\m\a}
}
\nonumber \\
&&+
\frac{1}{2}
{\bar{\phi}}^{|\a}
\left( l_{\m\n|\a}- l_{\a\m|\n}- l_{\a\n|\m} \right)
+
\frac{1}{2}
{{\bar{\phi}}^{|\a}}_{|\a}\left(l_{\m\n}-\frac{1}{2}\bar{g}_{\m\n}l\right)
\nonumber \\
&&
+ 
\frac{1}{2}\bar{g}_{\m\n}
\left(
\bar{\phi}_{|\a\b}l^{\a\b}+2\bar{\phi}_{|\a}A^{\a}
-\frac{1}{2}\bar{\phi}_{|\a}l^{|\a}
\right)
+
\frac{1}{4}
\left( l_{|\m}{\bar{\phi}}_{|\n}+l_{|\n}{\bar{\phi}}_{|\m}\right)
\nonumber \\
&&+
\left(\bar{R}_{\m\n}-\frac{1}{2}\bar{g}_{\m\n}\bar{R}\right)\varphi 
+\bar{g}_{\m\n}{\varphi^{|\a}}_{|\a}-\varphi_{|\m\n},
\EEqA
\BEq
F^{BD}_{\m\n}
= 
8\pi 
\left[
l_{\m\n}\bar{W}+
\bar{g}_{\m\n}
\left(
\frac{\tilde{\omega}'}{2}\bar{\phi}^{|\a}\bar{\phi}_{|\a}\varphi
+\bar{W}'\varphi
+\tilde{\omega}\bar{\phi}^{|\a}{\varphi}_{|\a}
\right)
-\tilde{\omega}'\bar{\phi}_{|\m}\bar{\phi}_{|\n}\varphi
-\tilde{\omega}
\left(
\bar{\phi}_{|\m}{\varphi}_{|\n}+\bar{\phi}_{|\n}{\varphi}_{|\m}
\right)
\right],
\EEq
\BEq
A^{\a} \equiv {l^{\a\b}}_{|\b},
\EEq
and
\BEq
\tilde{\omega}= \tilde{\omega}(\bar{\phi}),
\quad
\tilde{\omega}'= \frac{d\tilde{\omega}(\bar{\phi})}{d\bar{\phi}},
\quad
\bar{W}= {W}(\bar{\phi}),
\quad
\bar{W}'= \frac{dW(\bar{\phi})}{d\bar{\phi}}.
\EEq

%%%%%%%%%%%%%%%%%%%%%%%%%

\subsection{Equation for $l$ perturbation}

Taking the trace of (\ref{eq:l_mn_perturbation}) gives
\BEq
\label{eq:l_perturbation}
\bar{\phi}
\left(
\frac{1}{2}{l^{|\a}}_{|\a}+{A^{\a}}_{|\a}
\right)
-\frac{1}{2}{\bar{\phi}^{|\a}}_{|\a} l
+2 \bar{\phi}_{|\a\b}l^{\a\b}
+3\bar{\phi}_{|\a}A^{\a}
-\bar{R}\varphi 
+3{\varphi^{|\a}}_{|\a}
+ 8\pi 
\left[
\bar{W} l
+ \tilde{\omega}'\bar{\phi}^{|\a}\bar{\phi}_{|\a}\varphi
+ 4\bar{W}'\varphi
+ 2\tilde{\omega}\bar{\phi}^{|\a}{\varphi}_{|\a}
\right]
=
8\pi \Lambda.
\EEq

\subsection{Equation for $\varphi$ perturbation}

\subsubsection{Derivation}

From general theory of Ref.\ \cite{SMKP2014},
\BEq
\label{eq:der21}
F^{G}_{A}+F^{BD}_{A}=0.
\EEq
Using (\ref{eq:forPerturbations}) and (\ref{eq:extra2}) we get
\BEqA
\label{eq:der22}
F^{G}_{A}
&=& 
\frac{-16\pi}{\sqrt{-\bar{g}}}
\frac{\d }{\d {\bar{\phi}}}
\left(
{\frak{h}}^{\r\s} \frac{\d \bar{\cal L}^{G}}{\d \bar{\frak{g}}^{\r\s}} 
+\varphi \frac{\d \bar{\cal L}^{G}}{\d \bar{\phi}}
\right)
\nonumber \\
&=&
\frac{1}{\sqrt{-\bar{g}}}
\frac{\d }{\d {\bar{\phi}}}
\left[
{\frak{h}}^{\r\s}
\left(
\bar{R}_{\r\s}\bar{\phi}
- \frac{1}{2} \bar{g}_{\r\s}  \bar{g}^{\a\b}\bar{\phi}_{|\a\b}
-\bar{\phi}_{|\r\s}
\right)
+
\varphi\sqrt{-\bar{g}}\bar{R}
\right]
\nonumber \\
&=&
{l}^{\r\s}
\bar{R}_{\r\s}
-
\frac{1}{2} \bar{g}^{\a\b} {l}_{|\a\b}
- {{l}^{\a\b}}_{|\a\b}.
\EEqA
and
\BEqA
\label{eq:der23}
F^{BD}_{A}
&=& 
\frac{-16\pi}{\sqrt{-\bar{g}}}
\frac{\d }{\d \bar{\phi}}
\left(
{\frak{h}}^{\r\s} \frac{\d \bar{\cal L}^{BD}}{\d \bar{\frak{g}}^{\r\s}} 
+\varphi \frac{\d \bar{\cal L}^{BD}}{\d \bar{\phi}}
\right)
\nonumber \\
&=&
\frac{-16\pi}{\sqrt{-\bar{g}}}
\frac{\d }{\d {\bar{\phi}}}
\left\{
{\frak{h}}^{\r\s}
\frac{1}{2}
\left[
\tilde{\omega}(\bar{\phi}) \bar{\phi}_{,\r}\bar{\phi}_{,\s}
+ \bar{g}_{\r\s} W(\bar{\phi})
\right]
+
\sqrt{-\bar{g}}
\left[
\left(
\frac{1}{2}\tilde{\omega}'(\bar{\phi}) 
\bar{g}^{\a\b}\bar{\phi}_{,\a}\bar{\phi}_{,\b}+W'(\bar{\phi})
\right)\varphi
+
\tilde{\omega}(\bar{\phi})\bar{g}^{\r\s} \bar{\phi}_{,\s}
\varphi_{,\r}
\right]
\right\}
\nonumber \\
&=&
-16\pi
\left\{
{l}^{\r\s}
\frac{1}{2}
\left(
\tilde{\omega}' \bar{\phi}_{|\r}\bar{\phi}_{|\s}
+ \bar{g}_{\r\s} \bar{W}'
\right)
+
\left(
\frac{1}{2}\tilde{\omega}''
\bar{g}^{\a\b}\bar{\phi}_{|\a}\bar{\phi}_{|\b}+\bar{W}''
\right)\varphi
+
\tilde{\omega}'\bar{g}^{\r\s} \bar{\phi}_{|\s}
\varphi_{|\r}
\right\}
\nonumber \\
&&
+
16\pi
\left\{
\left(
{l}^{\r\s}\tilde{\omega} \bar{\phi}_{|\r}
\right)_{|\s}
+
\left(
\tilde{\omega}'
\bar{g}^{\a\b}\bar{\phi}_{|\b}
\varphi
\right)_{|\a}
+
\left(
\tilde{\omega}
\varphi^{|\r}
\right)_{|\r}
\right\}
\nonumber \\
&=&
-16\pi
\left\{
{l}^{\r\s}
\frac{1}{2}
\left(
\tilde{\omega}' \bar{\phi}_{|\r}\bar{\phi}_{|\s}
+ \bar{g}_{\r\s} \bar{W}'
\right)
+
\left(
\frac{1}{2}\tilde{\omega}''
\bar{g}^{\a\b}\bar{\phi}_{|\a}\bar{\phi}_{|\b}+\bar{W}''
\right)\varphi
+
\tilde{\omega}'\bar{g}^{\r\s} \bar{\phi}_{|\s}
\varphi_{|\r}
\right\}
\nonumber \\
&&
+
16\pi
\left\{
\tilde{\omega} {{l}^{\r\s}}_{|\s}\bar{\phi}_{|\r}
+
\tilde{\omega}'{l}^{\r\s} \bar{\phi}_{|\r} \bar{\phi}_{|\s}
+
\tilde{\omega}{l}^{\r\s} \bar{\phi}_{|\r\s}
\right\}
%%%%%%%%%%%%%%
\nonumber \\
&&
+
16\pi
\left\{
\tilde{\omega}''
\bar{g}^{\a\b}\bar{\phi}_{|\b}\bar{\phi}_{|\a}
\varphi
+
\tilde{\omega}'
\bar{g}^{\a\b}\bar{\phi}_{|\b\a}
\varphi
+
\tilde{\omega}'
\bar{g}^{\a\b}\bar{\phi}_{|\b}
\varphi_{|\a}
\right\}
%%%%%%%%%%%%%%
\nonumber \\
&&
+
16\pi
\left\{
\tilde{\omega}'\bar{\phi}_{|\a}
\varphi^{|\a}
+
\tilde{\omega}
{\varphi^{|\a}}_{|\a}
\right\}
\nonumber \\
&=&
16\pi
\left\{
\tilde{\omega}{\varphi^{|\a}}_{|\a}
+\tilde{\omega}'\bar{\phi}_{|\a}\varphi^{|\a}
+ \frac{1}{2}\tilde{\omega}''\bar{\phi}^{|\a}\bar{\phi}_{|\a}\varphi
+ \tilde{\omega}'{\bar{\phi}^{|\a}}_{|\a}\varphi
-\bar{W}''\varphi
+\frac{1}{2}\tilde{\omega}' {l}^{\r\s}\bar{\phi}_{|\r}\bar{\phi}_{|\s}
+\tilde{\omega}{{l}^{\r\s}}_{|\s} \bar{\phi}_{|\r}
+\tilde{\omega}{l}^{\r\s} \bar{\phi}_{|\r\s}
-\frac{1}{2}\bar{W}'l
\right\}.
\nonumber \\
\EEqA
Combining (\ref{eq:der22}) and (\ref{eq:der23}) gives
\BEqA
&&
16\pi
\left\{
\tilde{\omega}{\varphi^{|\a}}_{|\a}
+\tilde{\omega}'\bar{\phi}_{|\a}\varphi^{|\a}
+ \frac{1}{2}\tilde{\omega}''\bar{\phi}^{|\a}\bar{\phi}_{|\a}\varphi
+ \tilde{\omega}'{\bar{\phi}^{|\a}}_{|\a}\varphi
-\bar{W}''\varphi
+\frac{1}{2}\tilde{\omega}' {l}^{\r\s}\bar{\phi}_{|\r}\bar{\phi}_{|\s}
+\tilde{\omega}{{l}^{\r\s}}_{|\s} \bar{\phi}_{|\r}
+\tilde{\omega}{l}^{\r\s} \bar{\phi}_{|\r\s}
-\frac{1}{2}\bar{W}'l
\right\}
\nonumber \\
&&+
{l}^{\r\s}
\bar{R}_{\r\s}
-
\frac{1}{2} \bar{g}^{\a\b} {l}_{|\a\b}
- {{l}^{\a\b}}_{|\a\b}
=0.
\EEqA

%%%%%%%%%%%%%%%%%%%%%%%%%%%%%%%%

\subsubsection{Final result}

\BEqA
\label{eq:varphi_perturbation}
&&
{\varphi^{|\a}}_{|\a}
+\frac{\tilde{\omega}'}{\tilde{\omega}}
\bar{\phi}^{|\a}\varphi_{|\a} 
+ 
\frac{1}{\tilde{\omega}}
\left(
\tilde{\omega}'{\bar{\phi}^{|\a}}_{|\a} 
+\frac{1}{2}\tilde{\omega}''\bar{\phi}^{|\a}\bar{\phi}_{|\a} 
-\bar{W}''
\right)\varphi
\nonumber \\
&& \quad
+ A^{\a}{\bar{\phi}}_{|\a}
+ l^{\a\b}{\bar{\phi}}_{|\a\b}
+ \frac{1}{2}\frac{{\tilde{\omega}'}}{\tilde{\omega}}
l^{\a\b}{\bar{\phi}}_{|\a}{\bar{\phi}}_{|\b}
-\frac{1}{2}\frac{\bar{W}'}{\tilde \omega}l
+ 
\frac{1}{16\pi}\frac{1}{\tilde \omega}
\left(
\bar{R}_{\a\b}l^{\a\b}-\frac{1}{2}{l^{|\a}}_{|\a}-{A^{\a}}_{|\a}
\right)
=0.
\EEqA

%%%%%%%%%%%%%%%%%%%%%%%%%%%%%%

\section{Linear Hubble approximation in the scalar-tensor theory}

In the linear Hubble approximation for perturbations $\varphi$ and $l_{\m\n}$ 
we ignore the terms containing 
${\cal H}^2$, $\dot{\cal H}$, and $\ddot{\bar{\phi}}/\bar{\phi}$. 
(The overdot, we recall, represents the derivative with respect to the conformal 
time, $d/d\eta$.) To keep track of the time dependence of the gravitational ``constant'', 
despite of (\ref{eq:twoSmallParametersInequality}),
we must retain all terms containing $\dot{\bar{\phi}}/\bar{\phi}$.

Additionally, we will use the following formula valid in isotropic conformal coordinates:
\BEqA
\label{eq:scalarBox}
{\varphi^{|\a}}_{|\a} &=& \frac{1}{a^2}\left(\Box \varphi -2{\cal H}\varphi_{,0}\right).
\EEqA

\subsection{Field equations}

To arrive at the linear Hubble approximation,
we first drop the ``obvious'' terms proportional to
${\bar{\phi}^{|\m}}_{|\n}$, $\bar{\phi}_{|\m}\bar{\phi}_{|\n}$, 
$\bar{R}_{\m\n}$, $\bar{W}$, $\bar{W}'$, in Eqs.\ (\ref{eq:l_perturbation}), 
(\ref{eq:varphi_perturbation}), (\ref{eq:l_mn_perturbation}), and get the following 
three equations,
\BEq
\label{eq:l_perturbation_RECALL}
\bar{\phi}
\left(
\frac{1}{2}{l^{|\a}}_{|\a}+{A^{\a}}_{|\a}
\right)
+3\bar{\phi}_{|\a}A^{\a}
+3{\varphi^{|\a}}_{|\a}
+ \frac{2{\omega}}{\bar{\phi}}\bar{\phi}^{|\a}{\varphi}_{|\a}
=
8\pi \Lambda,
\EEq
\BEq
\label{eq:trace_perturbation_linearHubble1}
\frac{1}{2}{l^{|\a}}_{|\a}
-\frac{2\omega}{\bar{\phi}}{\varphi^{|\a}}_{|\a}
-\frac{2\omega}{\bar{\phi}}\left(\frac{\omega'}{\omega}-\frac{1}{\bar{\phi}}\right)
\bar{\phi}^{|\a}\varphi_{|\a} 
-\frac{2\omega}{\bar{\phi}}A^{\a}\bar{\phi}_{|\a}
+ {A^{\a}}_{|\a}
=0,
\EEq
and
\BEqA
\label{eq:l_mn_perturbation_linearHubble1}
&&
\left(
{{l_{\m\n}}^{|\a}}_{|\a}+
\bar{g}_{\m\n}{A^{\a}}_{|\a}-{A}_{\m|\n}-{A}_{\n|\m}
\right)
+
\frac{{\bar{\phi}}^{|\a}}{\bar{\phi}}
\left( 
l_{\m\n|\a}
- l_{\a\m|\n}- l_{\a\n|\m}
\right)
\nonumber \\
&& \quad
-
\bar{g}_{\m\n}\frac{\bar{\phi}^{|\a}}{\bar{\phi}}
\left(
\frac{1}{2}l_{|\a}-\frac{2\omega}{\bar{\phi}}\varphi_{|\a}
\right) 
+
\frac{\bar{\phi}_{|\m}}{\bar{\phi}}
\left(
\frac{1}{2}l_{|\n}-\frac{2\omega}{\bar{\phi}}\varphi_{|\n}
\right) 
+
\frac{\bar{\phi}_{|\n}}{\bar{\phi}}
\left(
\frac{1}{2}l_{|\m}-\frac{2\omega}{\bar{\phi}}\varphi_{|\m}
\right)
\nonumber \\
&& \quad
+ 
2\bar{g}_{\m\n} \frac{{\bar{\phi}}_{|\a}}{\bar{\phi}}A^{\a}
+ \frac{2}{\bar{\phi}}\left( \bar{g}_{\m\n}{\varphi^{|\a}}_{|\a}-\varphi_{|\m\n} \right)
=\frac{16\pi}{\bar{\phi}} \Lambda_{\m\n} ,
\EEqA
where, we recall, $A^{\a} \equiv {l^{\a\b}}_{|\b}$, $\omega \equiv\omega(\bar{\phi})$,
$\omega' \equiv d\omega/d\bar{\phi}$.
Instead of (\ref{eq:l_perturbation_RECALL}), by combining 
(\ref{eq:l_perturbation_RECALL}) and (\ref{eq:trace_perturbation_linearHubble1}), we get
\BEq
\label{eq:varphi_perturbation_linearHubble1}
{\varphi^{|\a}}_{|\a}
+ 
\frac{2{\omega}'}{3+2\omega}
\bar{\phi}^{|\a}{\varphi}_{|\a}
+
A^{\a}\bar{\phi}_{|\a}
=
\frac{8\pi}{3+2\omega}\Lambda,
\EEq
which will be used in what follows.
Thus, we have the system of equations
(\ref{eq:varphi_perturbation_linearHubble1}) and (\ref{eq:l_mn_perturbation_linearHubble1}).
Substituting
\BEqA
-\frac{2\omega}{\bar{\phi}}\varphi_{|\a} 
= 
\left(-\frac{2\omega}{\bar{\phi}}\varphi\right)_{|\a} + {\cal O}({\cal H})
\EEqA
in (\ref{eq:l_mn_perturbation_linearHubble1}), we get to order ${\cal O}({\cal H})$,
\BEqA
\label{eq:l_mn_perturbation_linearHubble101}
&&
\left(
{{l_{\m\n}}^{|\a}}_{|\a}+
\underbrace{
\bar{g}_{\m\n}{A^{\a}}_{|\a}-{A}_{\m|\n}-{A}_{\n|\m}
}_{\rm gauge}
\right)
+
\frac{{\bar{\phi}}^{|\a}}{\bar{\phi}}
\left( 
l_{\m\n|\a}
\underbrace{
- l_{\a\m|\n}- l_{\a\n|\m}
}_{{\rm will\;cancel\;by\;}C^{\a}{\rm \; gauge}}
\right)
\nonumber \\
&&\quad\quad
\underbrace{
-
\bar{g}_{\m\n}\frac{\bar{\phi}^{|\a}}{\bar{\phi}}
\left(
\frac{1}{2}l-\frac{2\omega}{\bar{\phi}}\varphi
\right) _{|\a}
+
\frac{\bar{\phi}_{|\m}}{\bar{\phi}}
\left(
\frac{1}{2}l-\frac{2\omega}{\bar{\phi}}\varphi
\right) _{|\n}
+
\frac{\bar{\phi}_{|\n}}{\bar{\phi}}
\left(
\frac{1}{2}l-\frac{2\omega}{\bar{\phi}}\varphi
\right) _{|\m}
}_{{\rm will\;cancel\;by\;}C^{\a}{\rm \; gauge}}
\nonumber \\
&& \quad\quad\quad
+ 
2\bar{g}_{\m\n} \frac{{\bar{\phi}}_{|\a}}{\bar{\phi}}A^{\a}
+ \frac{2}{\bar{\phi}}\left( \bar{g}_{\m\n}{\varphi^{|\a}}_{|\a}-\varphi_{|\m\n} \right)
=\frac{16\pi}{\bar{\phi}} \Lambda_{\m\n} .
\EEqA

We now introduce the gauge
\BEq
\label{eq:generalizedGauge}
A^{\a} = B^{\a} + C^{\a} + D^{\a},
\EEq
where we define
\BEqA
B^{\a} &\equiv&  
\underbrace{
-\frac{2\cal{H}}{a}l^{\a\b}\bar{u}_{\b}
}_{{\rm "Celestial..."}},
\\
C^{\a} &\equiv&
-\frac{\bar{\phi}^{|\b}}{\bar{\phi}}{l_{\b}}^{\a}
+\frac{\bar{\phi}^{|\a}}{\bar{\phi}}
\left(
\frac{1}{2}l
- \frac{2\omega}{\bar{\phi}} \varphi
\right),
\\
D^{\a} &\equiv& 
\underbrace{
-\frac{{\varphi}^{|\a}}{\bar{\phi}}
}_{{\rm Brans-Dicke}} 
- \frac{2{\cal H}}{a}\bar{u}^{\a}\frac{\varphi}{\bar{\phi}}
- \frac{\bar{\phi}^{|\a}}{\bar{\phi}^2}\varphi.
\EEqA
This gauge generalizes both the gauge used in the ``Celestial ephemerides'' paper 
\cite{SMK2012}, and the gauge used in the original Brans-Dicke paper \cite{BD1961}.
We notice that to order $O({\cal H})$,
\BEqA
\label{eq:BgaugeCombination}
{{l_{\m\n}}^{|\a}}_{|\a}+\bar{g}_{\m\n}{B^{\a}}_{|\a}-{B}_{\m|\n}-{B}_{\n|\m} 
&=&
\bar{g}^{\a\b}l_{\m\n,\a\b}
+\frac{2\cal{H}}{a}\bar{u}^{\a} \partial_{\a}l_{\m\n}
+\frac{2{\cal H}}{a}
\left( 
\bar{g}_{\m\n} \bar{u}^{\a}\frac{\varphi_{|\a}}{\bar{\phi}}
-\bar{u}_{\m} \frac{\varphi_{|\n}}{\bar{\phi}}-\bar{u}_{\n} \frac{\varphi_{|\m}}{\bar{\phi}}
\right),
\\
\bar{g}_{\m\n}{C^{\a}}_{|\a}-{C}_{\m|\n}-{C}_{\n|\m}
&=&
\bar{g}_{\m\n}\frac{\bar{\phi}^{|\a}}{\bar{\phi}}
\left(
\frac{1}{2}l-\frac{2\omega}{\bar{\phi}}\varphi
\right) _{|\a}
-
\frac{\bar{\phi}_{|\m}}{\bar{\phi}}
\left(
\frac{1}{2}l-\frac{2\omega}{\bar{\phi}}\varphi
\right) _{|\n}
-
\frac{\bar{\phi}_{|\n}}{\bar{\phi}}
\left(
\frac{1}{2}l-\frac{2\omega}{\bar{\phi}}\varphi
\right) _{|\m}
\nonumber \\
&& 
+
\frac{{\bar{\phi}}^{|\a}}{\bar{\phi}}\left( l_{\a\m|\n}+ l_{\a\n|\m} \right)
+
\bar{g}_{\m\n} \frac{{\bar{\phi}}^{|\a}\varphi_{|\a}}{\bar{\phi}^2},
\\
\bar{g}_{\m\n}{D^{\a}}_{|\a}-{D}_{\m|\n}-{D}_{\n|\m}
&=&
- \bar{g}_{\m\n}\frac{{\varphi^{|\a}}_{|\a}}{\bar{\phi}}
+
\frac{2\varphi_{|\m\n}}{\bar{\phi}}
 - \frac{2{\cal H}}{a}
\left( 
\bar{g}_{\m\n} \bar{u}^{\a}\frac{\varphi_{|\a}}{\bar{\phi}}
-\bar{u}_{\m} \frac{\varphi_{|\n}}{\bar{\phi}}-\bar{u}_{\n} \frac{\varphi_{|\m}}{\bar{\phi}}
\right),
\EEqA
where in (\ref{eq:BgaugeCombination}) we used Eq.\ (\ref{eq:workingOutBgauge}) 
of Sec.\ \ref{sec:Laplace-Beltrami}. This gives the system,
\BEqA
\label{eq:varphi_perturbation_linearHubble102}
{\varphi^{|\a}}_{|\a}
+ 
\left(\frac{2{\omega}'\bar{\phi}}{3+2\omega}-1\right)
\frac{\bar{\phi}^{|\a}}{\bar{\phi}}{\varphi}_{|\a}
&=&
\frac{8\pi}{3+2\omega} \Lambda,
\nonumber \\
\label{eq:l_mn_perturbation_linearHubble102}
\bar{g}^{\a\b}l_{\m\n,\a\b}
+
\left(
\frac{2\cal{H}}{a}\bar{u}^{\a}
+\frac{{\bar{\phi}}^{|\a}}{\bar{\phi}}
\right) 
l_{\m\n,\a}
+\bar{g}_{\m\n}\frac{{\varphi^{|\a}}_{|\a}}{\bar{\phi}}
-\bar{g}_{\m\n}\frac{\bar{\phi}^{|\a}\varphi_{|\a}}{\bar{\phi}^2}
&=&\frac{16\pi}{\bar{\phi}}\Lambda_{\m\n}.
\EEqA
Re-writing everything in the Hubble conformal coordinates with the help of
(\ref{eq:scalarBox}),
and taking into account that, to order $O({\cal H})$,
\BEqA
\Box\left(\frac{a^2\varphi}{\bar{\phi}}\right)
= 
\frac{a^2}{\bar{\phi}}\Box\varphi
-
2(2{\cal H}-{\cal F})\left(\frac{a^2\varphi}{\bar{\phi}}\right)_{,0}
\;,
\EEqA
and
\BEqA
\bar{g}_{\m\n}\frac{{\varphi^{|\a}}_{|\a}}{\bar{\phi}} 
&=&  
\frac{a^2f_{\m\n}}{\bar{\phi}}\frac{1}{a^2}
\left(\Box\varphi -2{\cal H}\varphi_{,0}\right)
\nonumber\\
&=&
\frac{f_{\m\n}}{a^2}
\left[
\frac{a^2}{\bar{\phi}}
\Box \varphi 
-2{\cal H}\left(\frac{a^2\varphi}{\bar{\phi}}\right)_{,0}
\right]
\nonumber \\
&=& \frac{f_{\m\n}}{a^2}
\left[
\Box\left(\frac{a^2\varphi}{\bar{\phi}}\right)
+2(2{\cal H}-{\cal F})\left(\frac{a^2\varphi}{\bar{\phi}}\right)_{,0}
-2{\cal H}\left(\frac{a^2\varphi}{\bar{\phi}}\right)_{,0}
\right]
\nonumber \\
&=& \frac{1}{a^2}
\left[
\Box\left(\frac{a^2f_{\m\n}\varphi}{\bar{\phi}}\right)
+(2{\cal H}-{\cal F})\left(\frac{a^2f_{\m\n}\varphi}{\bar{\phi}}\right)_{,0}
-{\cal F}\left(\frac{a^2f_{\m\n}\varphi}{\bar{\phi}}\right)_{,0}
\right]
\nonumber \\
&=& \frac{1}{a^2}
\left[
\Box\left(\frac{a^2f_{\m\n}\varphi}{\bar{\phi}}\right)
+(2{\cal H}-{\cal F})\left(\frac{a^2f_{\m\n}\varphi}{\bar{\phi}}\right)_{,0}
\right]
+a^2f_{\m\n}\left(-\frac{1}{a^2}\right){\cal F}\frac{{\varphi}_{,0}}{\bar{\phi}}
\nonumber \\
%%%%
&=& \frac{1}{a^2}
\left[
\Box\left(\frac{a^2f_{\m\n}\varphi}{\bar{\phi}}\right)
+(2{\cal H}-{\cal F})\left(\frac{a^2f_{\m\n}\varphi}{\bar{\phi}}\right)_{,0}
\right]
+ \bar{g}_{\m\n} \frac{{\bar{\phi}}^{|\a}\varphi_{|\a}}{\bar{\phi}^2},
\EEqA
we get
\BEqA
\label{eq:varphi_perturbation_linearHubble2_simpler}
\Box {\varphi}
- 
2\left(
{\cal H}-\frac{{\cal F}}{2}+\frac{{\omega}'\bar{\phi}}{3+2\omega}{\cal F}
\right)\varphi_{,0}
&=&
\frac{8\pi}{3+2\omega}f^{\a\b} \Lambda_{\a\b},
\\
\label{eq:l_mn_perturbation_linearHubble2_simpler}
\Box Q_{\m\n}
+ 2\left({\cal H}-\frac{{\cal F}}{2}\right)Q_{\m\n,0}
&=& \frac{16\pi a^2}{\bar{\phi}}\Lambda_{\m\n},
\EEqA
which reproduces Eqs.\ (15) and (16) of Ref.\ \cite{AGSMK2016}.
In the above, we defined
\BEq
{\cal F} \equiv \frac{\bar{\phi}_{,0}}{\bar{\phi}}
=\frac{1}{\bar{\phi}}\frac{d\bar{\phi}}{d\eta},
\EEq
and introduced a new gravitational variable $Q_{\m\n}$ (a direct analogue of
the variable $\a_{ij}$ that appears in Eq.\ (23) of the original 
Brans-Dicke paper \cite{BD1961}),
\BEq
\label{eq:newGravitationalVariable}
Q_{\m\n}\equiv l_{\m\n}+\bar{g}_{\m\n}\frac{\varphi}{\bar{\phi}}.
\EEq

Additionally, and this will turn out to be important for checking the gauge condition
in Section \ref{sec:CheckingGauge}, substituting
\BEqA
-\frac{2\omega}{\bar{\phi}}{\varphi^{|\a}}_{|\a} 
= 
{\left(-\frac{2\omega}{\bar{\phi}}\varphi\right)^{|\a}}_{|\a} 
+ {\cal O}({\cal H}),
\EEqA
in (\ref{eq:trace_perturbation_linearHubble1}),
we get to order ${\cal O}({\cal H})$,
\BEq
\label{eq:trace_perturbation_linearHubble101}
{\left( \frac{1}{2}l -\frac{2\omega}{\bar{\phi}}\varphi\right)^{|\a}}_{|\a} 
+ {A^{\a}}_{|\a} + {\cal O}({\cal H})
=0.
\EEq
Using (\ref{eq:generalizedGauge}) gives
\BEq
{A^{\a}}_{|\a}=
-\frac{{{\varphi}^{|\a}}_{|\a}}{\bar{\phi}} 
+{\cal O}({\cal H}),
\EEq
and thus, from (\ref{eq:trace_perturbation_linearHubble101}),
\BEq
\label{eq:trace_perturbation_linearHubble102}
{\left( \frac{1}{2}l -\frac{2\omega}{\bar{\phi}}\varphi\right)^{|\a}}_{|\a} 
=\frac{{{\varphi}^{|\a}}_{|\a}}{\bar{\phi}} 
+{\cal O}({\cal H}),
\EEq
which shows that the field perturbations satisfy the constraint
\BEq
\label{eq:trace_perturbation_linearHubble1020}
 \frac{1}{2}l -\frac{2\omega}{\bar{\phi}}\varphi = \frac{{\varphi}}{\bar{\phi}} 
+ {\cal O}({\cal H}),
\EEq
and thus the actual form of the gauge satisfied by the field perturbations is not
(\ref{eq:generalizedGauge}), but a somewhat simpler,
\BEq
\label{eq:generalizedGaugeActual}
A^{\a} = 
-\frac{2\cal{H}}{a}l^{\a\b}\bar{u}_{\b}
-\frac{\bar{\phi}^{|\b}}{\bar{\phi}}{l_{\b}}^{\a}
- \frac{2{\cal H}}{a}\bar{u}^{\a}\frac{\varphi}{\bar{\phi}}
-\frac{{\varphi}^{|\a}}{\bar{\phi}}.
\EEq

\subsection{Solving the wave equations}

\label{sec:solvingTheWaveEq}

Eqs.\ (\ref{eq:varphi_perturbation_linearHubble2_simpler})
and (\ref{eq:l_mn_perturbation_linearHubble2_simpler})
have the general form
\BEq
\label{eq:generalPerturbationEquation_simpler}
\Box Q +2{\cal B} Q_{,0} = 4\pi a^2 {\cal T},
\quad 
{\cal B}\sim {\cal O}({\cal H}),
\quad 
\dot{\cal B}\sim {\cal O}({\cal H}^2).
\EEq
This can be solved by introducing two new functions, $b=b(\eta)$ and $q = q(\eta, x^i)$, 
such that
\BEq
Q = b^2 q,
\EEq
where $b=b(\eta)$ is {\it defined} by
\BEq
\dot{b} = {\cal B}b, \quad \dot{b} \sim {\cal O}({\cal H}),
\quad \ddot{b} \sim {\cal O}({\cal H}^2).
\EEq
Noticing that, in the linear ${\cal O}({\cal H})$ approximation,
\BEq
\Box Q +2{\cal B} Q_{,0} 
= b\Box(bq),
\EEq
we get the equation 
\BEq
\Box(bq) =  4\pi \frac{a^2 {\cal T}}{b},
\EEq
whose retarded solution is given by
\BEq
\label{eq:retardedSolution_q}
q(\eta,{\bf x})=-\frac{1}{ b(\eta)}\int \frac{a^2(\eta')}{b(\eta')}
\frac{{\cal T}(\eta',{\bf x}')}{|{\bf x}-{\bf x}'|}d^3x',
\quad
\eta' = \eta - |{\bf x}-{\bf x}'|.
\EEq
The corresponding solution to (\ref{eq:generalPerturbationEquation_simpler})
is then given by
\BEq
\label{eq:generalPerturbationEquationSolution_simpler}
Q(\eta,{\bf x})=- b(\eta)\int \frac{a^2(\eta')}{b(\eta')}
\frac{{\cal T}(\eta',{\bf x}')}{|{\bf x}-{\bf x}'|}d^3x',
\quad
\eta' = \eta - |{\bf x}-{\bf x}'|,
\EEq
where, we recall, $\dot{b} = {\cal B}b$.

For example, when applied to $Q=b^2q_{\m\n} = \varphi$, the retarded solution
(\ref{eq:generalPerturbationEquationSolution_simpler}) takes the form
\BEqA
\label{eq:varphi_retarded_solution}
\varphi(\eta,{\bf x})
&=&
- \tilde{b}(\eta)\int 
\frac{a^2(\eta')}{\tilde{b}(\eta')}
\frac{2}{3+2\omega(\eta')}
\frac{\Lambda(\eta',{\bf x}')}{|{\bf x}-{\bf x}'|}d^3x'
\nonumber \\
&=&
- \tilde{b}(\eta)\int 
\frac{1}{\tilde{b}(\eta')}
\frac{2}{3+2\omega(\eta')}
\frac{f^{\a\b}\Lambda_{\a\b}(\eta',{\bf x}')}{|{\bf x}-{\bf x}'|}d^3x',
\quad
\tilde{\cal B} \equiv \frac{\dot{\tilde{b}}}{\tilde{b}} 
= -{\cal H}+\frac{{\cal F}}{2}-\frac{\omega'\bar{\phi}}{3+2\omega}{\cal F},
\quad
\omega' \equiv \frac{d\omega}{d\bar{\phi}}.
\EEqA
When applied to $Q_{\m\n}=b^2q_{\m\n} = l_{\m\n}+ \bar{g}_{\m\n}\varphi/\bar{\phi}$ 
introduced in (\ref{eq:newGravitationalVariable}) and  
(\ref{eq:l_mn_perturbation_linearHubble2_simpler}), the retarded solution becomes
\BEq
\label{eq:l_mn_retarded_solution}
 l_{\m\n}(\eta,{\bf x}) = 
\underbrace{
b(\eta)\int 
\frac{{\cal S}_{\m\n}(\eta',{\bf x}')}{|{\bf x}-{\bf x}'|}d^3x'
}_{\equiv Q_{\m\n}}
-\bar{g}_{\m\n}\frac{\varphi(\eta,{\bf x})}{\bar{\phi}(\eta)},
\quad
{\cal S}_{\m\n} = 
\frac{-4a^2\Lambda_{\m\n}}
{b\bar{\phi}},
\quad
{\cal B} \equiv \frac{\dot{b}}{b} = {\cal H}-\frac{{\cal F}}{2}.
\EEq
We will use this form of $ l_{\m\n}$ in Sec.\ \ref{sec:checkingTheGaugeCondition_OURS} 
to check the gauge condition.

%%%%%%%%%%%%%%%%%%%%%%%%%%%%%%%%%%

%\newpage

\section{APPENDIX: Some useful formulas}

Given
\BEq
S = \int {\cal F}(Q) d^4 x,
\EEq
the variational derivative of ${\cal F}$ with respect to the variable $Q$ is defined by
\BEq
\label{eq:extra1}
\frac{\d {\cal F}}{\d Q}
=
\frac{\partial {\cal F}}{\partial Q}
-\frac{\partial }{\partial x^\a}\frac{\partial {\cal F}}{\partial Q_{,\a}}
+\frac{\partial^2 }{\partial x^\a \partial x^\b}\frac{\partial {\cal F}}{\partial Q_{,\a\b}}.
\EEq
It can then be shown that
\BEq
\label{eq:extra2}
\frac{\d {\cal F}}{\d Q}
=
\frac{\partial {\cal F}}{\partial Q}
-
\left[
\frac{\partial {\cal F}}{\partial Q_{;\a}}
\right]_{;\a}
+
\left[
\frac{\partial {\cal F}}{\partial Q_{;\a\b}}
\right]_{;\b\a},
\EEq
and, in the case of $Q = g_{\m\n}$ and 
${\cal F}={\cal F}(g_{\m\n}, {\G}^{\a}_{\m\n}, {R}^{\a}_{\b\m\n})$,
\BEq
\label{eq:extra3}
\frac{\d {\cal F}}{\d g_{\m\n}}
=
\frac{\partial {\cal F}}{\partial g_{\m\n}}
-
\frac{1}{2}
\left(
{g}^{\s\n}\frac{\partial {\cal F}}{\partial {\G}^{\s}_{\m\a}}
+
{g}^{\s\m}\frac{\partial {\cal F}}{\partial {\G}^{\s}_{\a\n}}
-
{g}^{\s\a}\frac{\partial {\cal F}}{\partial {\G}^{\s}_{\m\n}}
\right)_{;\a}
+
\left(
{g}^{\s\n}\frac{\partial {\cal F}}{\partial {R}^{\s}_{\m\b\a}}
+
{g}^{\s\m}\frac{\partial {\cal F}}{\partial {R}^{\s}_{\a\b\n}}
-
{g}^{\s\a}\frac{\partial {\cal F}}{\partial {R}^{\s}_{\m\b\n}}
\right)_{;\b\a}.
\EEq

Now, the full metric is
\BEq
g_{\m\n} = \bar{g}_{\m\n}+ {\varkappa}_{\m\n},
\quad
g^{\a\b} = \bar{g}^{\a\b}+ \d g^{\a\b},
\quad
\d g^{\a\b}\approx -{\varkappa}^{\a\b}+{{\varkappa}^{\a}}_{\n}{\varkappa}^{\n\b},
\quad
{{\varkappa}^{\a}}_{\n} = \bar{g}^{\a\m}{\varkappa}_{\m\n},
\quad
{\varkappa}^{\n\b} = \bar{g}^{\n\m}{\varkappa}_{\m\s}\bar{g}^{\s\b}.
\EEq
The raised Gothic metric is
\BEq
\frak{g}^{\a\b}\equiv \sqrt{-g}g^{\a\b},
\quad
\bar{\frak{g}}^{\a\b}\equiv \sqrt{-\bar{g}}\bar{g}^{\a\b},
\quad
\frak{g}^{\a\b} = \bar{\frak{g}}^{\a\b} + \frak{h}^{\a\b},
\quad
\frak{h}^{\a\b} = \sqrt{-\bar{g}}l^{\a\b}.
\EEq
Given
\BEq
\frak{g}^{\a\b}\equiv \sqrt{-g}g^{\a\b},
\EEq
the relationship between the determinants of $g_{\a\b}$ and 
$\frak{g}_{\a\b}=g_{\a\b}/\sqrt{-g}$,
\BEq
g = \frak{g}^{-1}, \quad \frak{g} \equiv {\rm det} \left[\frak{g}_{\a\b}\right],
\EEq
and using
\BEq
\frac{\partial\left(\sqrt{-{g}}\right)}{\partial {g}_{\m\n}}
=
+\frac{1}{2}\sqrt{-{g}}{g}^{\m\n},
\quad
\frac{\partial\left(\sqrt{-{g}}\right)}{\partial {g}^{\m\n}}
=
-\frac{1}{2}\sqrt{-{g}} {g}_{\m\n},
\quad
\frac{\partial {g}^{\l\k} }{\partial{g}_{\m\n}}
=-{g}^{\l\m}{g}^{\k\n},
\EEq
we have,
\BEqA
\frac{\partial g^{\a\b}}{\partial \frak{g}^{\m\n}} 
&=&
\frac{\partial \left(
\frac{1}{\sqrt{-g}}\frak{g}^{\a\b}\right)}{\partial \frak{g}^{\m\n}} 
\nonumber \\
&=&
\frac{\partial \left(
\sqrt{-\frak{g}}\frak{g}^{\a\b}\right)}{\partial \frak{g}^{\m\n}} 
\nonumber \\
&=& -\frac{1}{2}\sqrt{-\frak{g}} \frak{g}_{\m\n} \frak{g}^{\a\b}
+ \sqrt{-\frak{g}}\frac{\partial \left(
\frak{g}^{\a\b}\right)}{\partial \frak{g}^{\m\n}} 
\nonumber \\
&=& -\frac{1}{2}\sqrt{-\frak{g}} \frak{g}_{\m\n} \frak{g}^{\a\b}
+ \frac{1}{2}\sqrt{-\frak{g}}
\left(
\d^{\a}_{\m}\d^{\b\n} + \d^{\a}_{\n}\d^{\b\m}
\right)
\nonumber \\
&=& \frac{1}{2\sqrt{-g}} 
\left(
\d^{\a}_{\m}\d^{\b}_{\n} + \d^{\a}_{\n}\d^{\b}_{\m}
-{g}^{\a\b}{g}_{\m\n}
\right),
\EEqA
and, similarly,
\BEq
\frac{\partial g_{\a\b}}{\partial \frak{g}^{\m\n}} 
= -\frac{1}{2\sqrt{-g}} 
\left(
g_{\a\m}g_{\b\n} + g_{\a\n}g_{\b\m}-{g}_{\a\b}{g}_{\m\n}
\right).
\EEq
It then follows that
\BEq
\frac{\partial }{\partial \frak{g}^{\m\n}}\left(g_{\a\b}g_{\r\s}\right) 
= -\frac{1}{2\sqrt{-g}} 
\left[
g_{\a\b}
\left(
g_{\r\m}g_{\s\n} + g_{\r\n}g_{\s\m}-{g}_{\r\s}{g}_{\m\n}
\right)
+
\left(
g_{\a\m}g_{\b\n} + g_{\a\n}g_{\b\m}-{g}_{\a\b}{g}_{\m\n}
\right)g_{\r\s}
\right].
\EEq
Then,
\BEq
l^{\a\b} 
= 
-{\varkappa}^{\a\b}+\frac{1}{2} \bar{g}^{\a\b}\varkappa
+
{\varkappa}^{\m(\a}{\varkappa^{\b)}}_{\m}
-\frac{1}{2}{\varkappa}^{\a\b}{\varkappa}
-\frac{1}{4}\bar{g}^{\a\b}
\left({\varkappa}^{\m\n}{\varkappa}_{\m\n}-\frac{1}{2}{\varkappa}^2\right),
\EEq
and
\BEq
\varkappa_{\a\b} 
= 
-{l}_{\a\b}+\frac{1}{2} \bar{g}_{\a\b}l
+{l^{\m}}_{(\a}l_{\b)\m}
-\frac{1}{2}{l}_{\a\b}{l}
-\frac{1}{4}\bar{g}_{\a\b}
\left({l}^{\m\n}{l}_{\m\n}-\frac{1}{2}{l}^2\right),
\EEq
where
\BEq
\varkappa \equiv {{\varkappa}^{\a}}_{\a} = \bar{g}^{\a\b} {{\varkappa}}_{\a\b},
\EEq
so in linear order in ${\varkappa}^{\a\b}$  we get
\BEq
l = {\varkappa},
\quad
l_{\m\n} = -{\varkappa}_{\m\n}+\frac{1}{2} \bar{g}_{\m\n}\varkappa,
\quad
{\varkappa}_{\m\n} = -l_{\m\n}+\frac{1}{2} \bar{g}_{\m\n}{l},
\quad
{\varkappa}^{\a\b} = -l^{\a\b}+\frac{1}{2} \bar{g}^{\a\b}{l}.
\EEq

For additional details the reader may consult Ref.\ \cite{SMKP2014}.

%%%%%%%%%%%%%%%%%%%%%%%%%%%%%%%%%%

%\newpage

\section{APPENDIX: Working out ${{l_{\m\n}}^{|\a}}_{|\a}$ to linear order in $\cal H$}

\label{sec:Laplace-Beltrami}

We first notice that for
\BEq
B^{\a} =-\frac{2\cal{H}}{a}l^{\a\b}\bar{u}_{\b},
\EEq
we have
\BEq
\label{eq:B^a_|a}
{B^{\a}}_{|\a}= 
-\frac{2{\cal H}}{a}{l^{\a\b}}_{|\a}\bar{u}_{\b} + O({\cal H}^2)
= -\frac{2{\cal H}}{a}\bar{u}^{\a}A_{\a} + O({\cal H}^2)
= +\frac{2{\cal H}}{a}\bar{u}^{\a}\frac{\varphi_{|\a}}{\bar{\phi}} + O({\cal H}^2),
\EEq
where the gauge condition (\ref{eq:generalizedGauge}) has been used.
Additionally, we have:
\BEqA
{{l_{\a\b}}^{|\m}}_{|\m}&=& \bar{g}^{\m\n}l_{\a\b|\m\n}
\nonumber \\
&=&
\bar{g}^{\m\n}
\left[
\left(
l_{\a\b|\m}
\right)_{,\nu}
-\bar{\Gamma}^{\rho}_{\a\nu} l_{\rho\b|\m}
-\bar{\Gamma}^{\rho}_{\b\nu} l_{\a\rho|\m}
-\bar{\Gamma}^{\rho}_{\m\n} l_{\a\b|\rho}
\right]
\nonumber \\
&=&
\bar{g}^{\m\n}
\left[
\left(
l_{\a\b,\m}
-\bar{\Gamma}^{\rho}_{\a\mu} l_{\rho\b}
-\bar{\Gamma}^{\rho}_{\b\mu} l_{\a\rho}
\right)_{,\nu}
-\bar{\Gamma}^{\rho}_{\a\nu} l_{\rho\b|\m}
-\bar{\Gamma}^{\rho}_{\b\nu} l_{\a\rho|\m}
-\bar{\Gamma}^{\rho}_{\m\n} l_{\a\b|\rho}
\right]
\nonumber \\
&=&
\bar{g}^{\m\n}
l_{\a\b,\m\n}
+
\bar{g}^{\m\n}
\left[
\frac{\cal{H}}{a}
\left(
\d^{\rho}_{\a}\bar{u}_{\m} + \d^{\rho}_{\m}\bar{u}_{\a} 
- \bar{u}^{\rho}\bar{g}_{\a\m}
\right) l_{\rho\b}
+
\frac{\cal{H}}{a}
\left(
\d^{\rho}_{\b}\bar{u}_{\m} + \d^{\rho}_{\m}\bar{u}_{\b} 
- \bar{u}^{\rho}\bar{g}_{\b\m}
\right) l_{\a\rho}
\right]_{,\nu}
\nonumber \\
&&+
\frac{\cal{H}}{a}\bar{g}^{\m\n}
\left[
\left(
\d^{\rho}_{\a}\bar{u}_{\n} + \d^{\rho}_{\n}\bar{u}_{\a} 
- \bar{u}^{\rho}\bar{g}_{\a\n}
\right) l_{\rho\b|\m}
+
\left(
\d^{\rho}_{\b}\bar{u}_{\n} + \d^{\rho}_{\n}\bar{u}_{\b} 
- \bar{u}^{\rho}\bar{g}_{\b\n}
\right) l_{\a\rho|\m}
+
\left(
\d^{\rho}_{\m}\bar{u}_{\n} + \d^{\rho}_{\n}\bar{u}_{\m} 
- \bar{u}^{\rho}\bar{g}_{\m\n}
\right) l_{\a\b|\rho}
\right]
\nonumber \\
&\approx&
\bar{g}^{\m\n}l_{\a\b,\m\n}
+
\frac{2\cal{H}}{a}
\left[
\bar{u}^{\m} l_{\a\b,\m}
-  \bar{u}^{\m}l_{\m\b,\a}
- \bar{u}^{\m}l_{\a\m,\b}
+ \bar{g}^{\m\n} \bar{u}_{\a} l_{\m\b,\n}
+ \bar{g}^{\m\n}\bar{u}_{\b} l_{\a\m,\n}
\right]
\nonumber \\
&\approx&
\bar{g}^{\m\n}l_{\a\b,\m\n}
+
\frac{2\cal{H}}{a}
\left[
\bar{u}^{\m} l_{\a\b,\m}
-  \bar{u}^{\m}l_{\m\b|\a}
- \bar{u}^{\m}l_{\a\m|\b}
+ \bar{u}_{\a} {l_{\b\m}}^{|\m}
+ \bar{u}_{\b} {l_{\a\m}}^{|\m}
\right]
\nonumber \\
\nonumber \\
%%%%%%%%%%%%
&\approx&
\bar{g}^{\m\n}l_{\a\b,\m\n}
+
\frac{2\cal{H}}{a}
\bar{u}^{\m} \partial_{\m}l_{\a\b}
-  \left(\frac{2\cal{H}}{a}\bar{u}^{\m}l_{\m\b}\right)_{|\a}
- \left(\frac{2\cal{H}}{a}\bar{u}^{\m}l_{\m\a}\right)_{|\b}
+ 
\frac{2\cal{H}}{a}\bar{u}_{\a} {l_{\b\m}}^{|\m}
+ \frac{2\cal{H}}{a}\bar{u}_{\b} {l_{\a\m}}^{|\m}
\nonumber \\
&=&
\bar{g}^{\m\n}l_{\a\b,\m\n}
+
\frac{2\cal{H}}{a}
\bar{u}^{\m} \partial_{\m}l_{\a\b}
-  \left(\frac{2\cal{H}}{a}\bar{u}^{\m}l_{\m\b}\right)_{|\a}
- \left(\frac{2\cal{H}}{a}\bar{u}^{\m}l_{\m\a}\right)_{|\b}
+ \frac{2\cal{H}}{a}\bar{u}_{\a} A_{\b}
+ \frac{2\cal{H}}{a}\bar{u}_{\b} A_{\a}
\nonumber \\
&=&
\bar{g}^{\m\n}l_{\a\b,\m\n}
+
\frac{2\cal{H}}{a}
\bar{u}^{\m} \partial_{\m}l_{\a\b}
+B_{\b|\a}+B_{\a|\b}
- \frac{2\cal{H}}{a}\bar{u}_{\a} \frac{\varphi_{|\b}}{\bar{\phi}}
- \frac{2\cal{H}}{a}\bar{u}_{\b} \frac{\varphi_{|\a}}{\bar{\phi}}.
\EEqA
Then, using (\ref{eq:B^a_|a}), we get
\BEqA
\label{eq:workingOutBgauge}
{{l_{\a\b}}^{|\m}}_{|\m}+\bar{g}_{\a\b}{B^{\m}}_{|\m}-{B}_{\a|\b}-{B}_{\b|\a} 
&=&
\bar{g}^{\m\n}l_{\a\b,\m\n}
+\frac{2\cal{H}}{a}\bar{u}^{\m} \partial_{\m}l_{\a\b}
+ \bar{g}_{\a\b}\frac{2{\cal H}}{a}\bar{u}^{\m}\frac{\varphi_{|\m}}{\bar{\phi}}
- \frac{2\cal{H}}{a}
\left(\bar{u}_{\a} \frac{\varphi_{|\b}}{\bar{\phi}}
+\bar{u}_{\b} \frac{\varphi_{|\a}}{\bar{\phi}}\right)
\nonumber \\
&=&
\frac{1}{a^2}
\left(
\Box l_{\a\b}
+ 2{\cal H}l_{\a\b,0}
\right) 
+ \frac{2{\cal H}}{a}
\left(
\bar{g}_{\a\b}\bar{u}^{\m}\frac{\varphi_{|\m}}{\bar{\phi}}
-\bar{u}_{\a} \frac{\varphi_{|\b}}{\bar{\phi}}
-\bar{u}_{\b} \frac{\varphi_{|\a}}{\bar{\phi}}
\right).
\EEqA

%%%%%%%%%%%%%%%%%%%%%%%%%%%%%%%%%%%%

%\newpage

\section{APPENDIX: Checking the gauge condition}
\label{sec:CheckingGauge}

\subsection{Classical electrodynamics}

As a warm-up exercise, let us recall how the gauge condition is checked in 
classical electrodynamics. In that case, the retarded solution for the vector 
potential $A^{\m}$ is
given by
\BEq
A^{\m}(t,{\bf x}) = \int 
\frac{{j}^{\m}(t',{\bf x}')}{|{\bf x}-{\bf x}'|}d^3x',
\quad
t' = t - |{\bf x}-{\bf x}'|.
\EEq
Using the notation
\BEq
{j'}^{\m} \equiv {j}^{\m}(t',{\bf x}'),
\quad
t' = t - R,
\quad
R \equiv |{\bf x}-{\bf x}'|,
\quad
\partial'_{\mu} \equiv \partial/\partial{{x'}^{\m}},
\EEq
we get
\BEqA
\label{eq:ED_gaugeCheck1}
\partial_{\m}A^{\m}(t,{\bf x}) 
&=& 
\partial_{\m}
\left(
\int 
\frac{{j'}^{\m}}{R}d^3x'
\right)
=\int
\partial_{\m}
\left( 
\frac{{j'}^{\m}}{R}\right)d^3x'
\nonumber \\
&=&
\int
\left[ 
\frac{1}{R}\partial_{\m}{j'}^{\m}
+ {j'}^{\m}\partial_{\m} \left(\frac{1}{R}\right)
\right]d^3x'
\nonumber \\
&=&
\int
\left[ 
\frac{1}{R}
\left(
\partial_{0}{j'}^{0}
+\partial_{k}{j'}^{k}
\right)
+ {j'}^{k}\partial_{k} \left(\frac{1}{R}\right)
\right]d^3x'
\nonumber \\
&=&
\int
\left\{ 
\frac{1}{R}
\left[
\partial'_{0}{j'}^{0}
+\left(\partial'_{0}{j'}^{k}\right)\left( \partial_{k}t'\right)
\right]
+ {j'}^{k}\partial_{k} \left(\frac{1}{R}\right)
\right\}d^3x'
\nonumber \\
&=&
\int
\left\{ 
\frac{1}{R}
\left[
\partial'_{0}{j'}^{0}
+\left(\partial'_{0}{j'}^{k}\right)\left( -\partial_{k}R\right)
\right]
- {j'}^{k}\partial'_{k} \left(\frac{1}{R}\right)
\right\}d^3x'
\nonumber \\
&=&
\int
\left\{ 
\frac{1}{R}
\left[
\partial'_{0}{j'}^{0}
+\left(\partial'_{0}{j'}^{k}\right)\left( \partial'_{k}R\right)
\right]
- {j'}^{k}\partial'_{k} \left(\frac{1}{R}\right)
\right\}d^3x'.
\EEqA
We now notice that
\BEq
\partial'_{k}{j'}^{k} = \left(\partial'_{k}{j'}^{k}\right)_{t' = {\rm const}}
+ \left(\partial'_{0}{j'}^{k}\right)\left( -\partial'_{k}R\right),
\EEq
and thus
\BEq
\label{eq:ED_gaugeCheck2}
\left(\partial'_{0}{j'}^{k}\right)\left(\partial'_{k}R\right) 
= 
\left(\partial'_{k}{j'}^{k}\right)_{t' = {\rm const}}-\partial'_{k}{j'}^{k}.
\EEq
Substituting (\ref{eq:ED_gaugeCheck2}) into (\ref{eq:ED_gaugeCheck1})
finally gives
\BEqA
\label{eq:ED_gaugeCheck3}
\partial_{\m}A^{\m}(t,{\bf x}) 
&=& 
\int
\left\{ 
\frac{1}{R}
\left[
\partial'_{0}{j'}^{0}
+
\left(\partial'_{k}{j'}^{k}\right)_{t' = {\rm const}}
\right]
-
\left[
\frac{1}{R}\partial'_{k}{j'}^{k}
+ {j'}^{k}\partial'_{k} \left(\frac{1}{R}\right)
\right]
\right\}d^3x'
\nonumber \\
&=&
\int
\frac{1}{R}
\underbrace{
\left[
\partial'_{0}{j'}^{0}
+
\left(\partial'_{k}{j'}^{k}\right)_{t' = {\rm const}}
\right]
}_{=0, {\rm \; continuity}}
d^3x'
-
\underbrace{
\int
\partial'_{k}\left(\frac{{j'}^{k}}{R}\right)d^3x'
}_{= 0, {\rm \; divergence}}
\nonumber \\
&=&0,
\EEqA
as expected.

%% \newpage

\subsection{``Celestial ephemerides" solution}

Now let us verify the gauge,
\BEq
B^{\a} \equiv  -\frac{2\cal{H}}{a}l^{\a\b}\bar{u}_{\b},
\EEq
used in the ``Celestial ephemerides" paper \cite{SMK2012}.

First, notice that in conformal coordinates, for any symmetric tensor $l^{\m\n}$,
\BEqA
\label{eq:covDerOf_lmn_generalResultProof}
{l^{\m\n}}_{|\n}&=&
\bar{g}^{\m\a}\bar{g}^{\n\b}
l_{\a\b|\n}
\nonumber \\
&=&
\bar{g}^{\m\a}\bar{g}^{\n\b}
\left(
l_{\a\b,\n}
-\G^{\kappa}_{\a \n}l_{\kappa\b}
-\G^{\kappa}_{\b \n}l_{\a\kappa}
\right)
\nonumber \\
&=&
\bar{g}^{\m\a}\bar{g}^{\n\b}l_{\a\b,\n}
+\bar{g}^{\m\a}
\frac{\cal{H}}{a}
\left(
\d^{\kappa}_{\a}\bar{u}_{\nu} + \d^{\kappa}_{\nu}\bar{u}_{\a} 
- \bar{u}^{\kappa}\bar{g}_{\a\nu}
\right)
{l_{\kappa}}^{\n}
+\bar{g}^{\n\b}\frac{\cal{H}}{a}
\left(
\d^{\kappa}_{\b}\bar{u}_{\nu} + \d^{\kappa}_{\nu}\bar{u}_{\b} 
- \bar{u}^{\kappa}\bar{g}_{\b\nu}
\right)
{l^{\m}}_{\kappa}
\nonumber \\
&=&
\bar{g}^{\m\a}\bar{g}^{\n\b}l_{\a\b,\n}
+
\frac{\cal H}{a}
\left(
\bar{g}^{\m\kappa}\bar{u}_{\nu} + \d^{\kappa}_{\nu}\bar{u}^{\m} 
- \bar{u}^{\kappa}\d^{\m}_{\nu}
\right)
{l_{\kappa}}^{\n}
+\frac{\cal H}{a}
\left(
\bar{g}^{\n\kappa}\bar{u}_{\nu} + \bar{u}^{\kappa} 
- 4\bar{u}^{\kappa}
\right)
{l^{\m}}_{\kappa}
\nonumber \\
&=&
\bar{g}^{\m\a}\bar{g}^{\n\b}l_{\a\b,\n}
+ \frac{\cal H}{a}l\bar{u}^{\m}
-\frac{2{\cal H}}{a}{l}^{\m\n}\bar{u}_{\nu}  
\nonumber \\
&=&
\bar{g}^{\m\a}\bar{g}^{\n\b}l_{\a\b,\n}
+ \frac{\cal H}{a^2}l\d^{\m}_{0}
+2{\cal H}{l}^{\m0}
\nonumber \\
&=&
\bar{g}^{\m\a}\bar{g}^{\n\b}l_{\a\b,\n}
- {\cal H}l\bar{g}^{\m0}+2{\cal H}{l}^{\m0}.
\EEqA
Next, the retarded solution, $l^{\m\n}$, used in the ``Celestial ephemerides'' paper is
\BEq
l_{\m\n}(t,{\bf x}) 
= -4a(\eta)
\int 
\frac{a(\eta'){\cal T}_{\m\n}(\eta',{\bf x}')}{|{\bf x}-{\bf x}'|}d^3x',
\quad
\eta' = \eta - |{\bf x}-{\bf x}'|,
\EEq 
where ${\cal T}_{\m\n}$ is the stress-energy tensor of the matter perturbation.
We have,
\BEqA
{l^{\m\n}}_{|\n}
&=&
\bar{g}^{\m\a}\bar{g}^{\n\b}l_{\a\b,\n}
- {\cal H}l\bar{g}^{\m0}+2{\cal H}{l}^{\m0}
\nonumber \\
&=&
\bar{g}^{\m\a}\bar{g}^{\n\b}
\partial_{\n}
\left(
-4a
\int 
\frac{a'{\cal T}'_{\a\b}}{R}d^3x'
\right)
- {\cal H}l\bar{g}^{\m0}+2{\cal H}{l}^{\m0}
\nonumber \\
&=&
\bar{g}^{\m\a}\bar{g}^{\n\b}
(\partial_{\n}a)
\left(
-4
\int 
\frac{a'{\cal T}'_{\a\b}}{R}d^3x'
\right)
+\bar{g}^{\m\a}\bar{g}^{\n\b}
(-4a)\int 
\partial_{\n}
\left(
\frac{a'{\cal T}'_{\a\b}}{R}\right)
d^3x'
- {\cal H}l\bar{g}^{\m0}+2{\cal H}{l}^{\m0}
\nonumber \\
%%%%%%%%%
&=&
\bar{g}^{\m\a}\bar{g}^{0\b}
{\cal H}{l}_{\a\b}
+\bar{g}^{\m\a}\bar{g}^{\n\b}
(-4a)\int 
\partial_{\n}
\left(
\frac{a'{\cal T}'_{\a\b}}{R}\right)
d^3x'
- {\cal H}l\bar{g}^{\m0}+2{\cal H}{l}^{\m0}
\nonumber \\
%%%%%%%%%
&=&
{\cal H}{l}^{\m0}
+\bar{g}^{\m\a}\bar{g}^{\n\b}
(-4a)\int 
\partial_{\n}
\left(
\frac{a'{\cal T}'_{\a\b}}{R}\right)
d^3x'
- {\cal H}l\bar{g}^{\m0}+2{\cal H}{l}^{\m0}
%%%
\nonumber \\
&=&
\bar{g}^{\m\a}\bar{g}^{\n\b}
(-4a)\int 
\partial_{\n}
\left(
\frac{a'{\cal T}'_{\a\b}}{R}\right)
d^3x'
- {\cal H}l\bar{g}^{\m0}+3{\cal H}{l}^{\m0}
%%%
\nonumber \\
&=&
\frac{1}{a^4} (-4a)
\int 
\partial_{\n}
\left(
\frac{a'f^{\m\a}f^{\n\b}{\cal T}'_{\a\b}}{R}\right)
d^3x'
- {\cal H}l\bar{g}^{\m0}+3{\cal H}{l}^{\m0}.
\EEqA
Now performing a few steps analogous to those in electrodynamics 
(see Eq.\ (\ref{eq:ED_gaugeCheck1})) gives
\BEqA
{l^{\m\n}}_{|\n}&=&
- {\cal H}l\bar{g}^{\m0}+3{\cal H}{l}^{\m0}
\nonumber \\
&&
+\frac{1}{a^4} (-4a)
\int
\frac{1}{R}
\left\{
\partial'_{0}\left(a' f^{\m\a}f^{0\b}{\cal T}'_{\a\b} \right)
+
\left[\partial'_{k}\left(a'f^{\m\a}f^{k\b}{\cal T}'_{\a\b}\right)\right]_{t' = {\rm const}}
\right\}
d^3x'
-\frac{1}{a^4} (-4a)
\underbrace{
\int
\partial'_{k}\left(\frac{a'f^{\m\a}f^{k\b}{\cal T}'_{\a\b}}{R}\right)d^3x'
}_{= 0, {\rm \; divergence}}
\nonumber \\
&=&
- {\cal H}l\bar{g}^{\m0}+3{\cal H}{l}^{\m0}
+\frac{1}{a^4} (-4a)
\int
\frac{1}{R}
\underbrace{
\left\{
\partial'_{0}\left(a' f^{\m\a}f^{0\b}{\cal T}'_{\a\b} \right)
+
\left[\partial'_{k}\left(a'f^{\m\a}f^{k\b}{\cal T}'_{\a\b}\right)\right]_{t' = {\rm const}}
\right\}
}_{\equiv I'^{\mu}}
d^3x'.
\EEqA
Conservation of ${\cal T}^{\m\n}$,
\BEq
\partial_{\b}\left(\sqrt{-\bar{g}}{\cal T}^{\a\b}\right)
+\sqrt{-\bar{g}}{\G^{\a}}_{\b\g}{\cal T}^{\b\g}
=0,
\EEq
gives, in conformal coordinates,
\BEqA
\partial_{0}\left(a{\cal T}_{00}\right) - \partial_{j}\left(a{\cal T}_{0j}\right)
&=&-{\cal H}\left(a{\cal T}_{kk}\right),
\\
\partial_{0}\left(a{\cal T}_{i0}\right) - \partial_{j}\left(a{\cal T}_{ij}\right)
&=&-{\cal H}\left(a{\cal T}_{i0}\right).
\EEqA
Then for the $0$-th component we get
\BEqA
I^{0}
&=&
\partial_{0}\left(af^{0\a}f^{0\b}{\cal T}_{\a\b}\right)
+\partial_{k}\left(af^{0\a}f^{k\b}{\cal T}_{\a\b}\right)
\nonumber \\
&=&
\partial_{0}\left(af^{00}f^{00}{\cal T}_{00}\right)
+\partial_{k}\left(af^{00}f^{km}{\cal T}_{0m}\right)
\nonumber \\
&=&
\partial_{0}\left(a{{\cal T}}_{00}\right)
-\partial_{k}\left(a{{\cal T}}_{{0}k}\right)
\nonumber \\
&=&
-{\cal H}\left(a{\cal T}_{kk}\right),
\nonumber \\
&=&
-{\cal H}\left(af^{ik}{\cal T}_{ik}\right),
\EEqA
and thus,
\BEqA
{l^{0\n}}_{|\n}&=&
- {\cal H}l\bar{g}^{00}+3{\cal H}{l}^{00}
-\frac{f^{ik}}{a^4} (-4a) 
\underbrace{
\int
\frac{{\cal H}'}{R}\left(a'{\cal T}'_{ik}\right)
d^3x'
}_{{\rm treat\;}{\cal H}'{\rm \; as\; constant}}
\nonumber \\
&\approx&
- {\cal H}l\bar{g}^{00}+3{\cal H}{l}^{00}
-{\cal H}\frac{f^{ik}}{a^4} (-4a) 
\int
\frac{1}{R}\left(a'{\cal T'}_{ik}\right)
d^3x'
\nonumber \\
&=&
- {\cal H}l\bar{g}^{00}+3{\cal H}{l}^{00}
-{\cal H}\frac{\bar{g}^{ik}}{a^2}
(-4a) 
\int
\frac{1}{R}\left(a'{\cal T'}_{ik}\right)
d^3x'
\nonumber \\
&=&
- {\cal H}l\bar{g}^{00}+3{\cal H}{l}^{00}
-{\cal H}\frac{\bar{g}^{ik}}{a^2}{l}_{ik}
\nonumber \\
&=&
- {\cal H}l\bar{g}^{00}+3{\cal H}{l}^{00}
+{\cal H}\bar{g}^{ik}{l}_{ik}\bar{g}^{00}
\nonumber \\
&=&
- {\cal H}l\bar{g}^{00}+3{\cal H}{l}^{00}
+{\cal H}\bar{g}^{ik}{l}_{ik}\bar{g}^{00}
+\left(
{\cal H}\bar{g}_{00}{l}^{00}\bar{g}^{00}
-{\cal H}\bar{g}_{00}{l}^{00}\bar{g}^{00}
\right)
\nonumber \\
&=&
- {\cal H}l\bar{g}^{00}+3{\cal H}{l}^{00}
+{\cal H}l\bar{g}^{00}
-{\cal H}\bar{g}_{00}{l}^{00}\bar{g}^{00}
\nonumber \\
&=&2{\cal H}{l}^{00}.
\EEqA
For the {\it i}-th component,
\BEqA
I^{i}
&=&
\partial_{0}\left(af^{i\a}f^{0\b}{\cal T}_{\a\b}\right)
+\partial_{k}\left(af^{i\a}f^{k\b}{\cal T}_{\a\b}\right)
\nonumber \\
&=&
\partial_{0}\left(af^{ij}f^{00}{\cal T}_{j0}\right)
+\partial_{k}\left(af^{ij}f^{km}{\cal T}_{jm}\right)
\nonumber \\
&=&
-\partial_{0}\left(a{{\cal T}}_{i0}\right)
+\partial_{k}\left(a{{\cal T}}_{ik}\right)
\nonumber \\
&=&
+{\cal H}\left(a{\cal T}_{i0}\right),
\EEqA
and thus
\BEqA
{l^{i\n}}_{|\n}&=&
3{\cal H}{l}^{i0}
+\frac{1}{a^4} (-4a) 
\underbrace{
\int
\frac{{\cal H}'}{R}\left(a'{\cal T}'_{i0}\right)
d^3x'
}_{{\rm treat\;}{\cal H}'{\rm \; as\; constant}}
\nonumber \\
&\approx&
3{\cal H}{l}^{i0}
+{\cal H}\frac{1}{a^4} (-4a) 
\int
\frac{1}{R}\left(a'{\cal T'}_{i0}\right)
d^3x'
\nonumber \\
&=&
3{\cal H}{l}^{i0}
+{\cal H}\frac{1}{a^4}{l}_{i0}
\nonumber \\
&=&
3{\cal H}{l}^{i0}
+{\cal H}\frac{1}{a^4}\bar{g}_{i\m}\bar{g}_{0\n}{l}^{\m\n}
\nonumber \\
&=&
3{\cal H}{l}^{i0}
+{\cal H}f_{i\m}f_{0\n}{l}^{\m\n}
\nonumber \\
&=&
3{\cal H}{l}^{i0}
+{\cal H}f_{ik}f_{00}{l}^{k0}
\nonumber \\
&=&
3{\cal H}{l}^{i0}
-{\cal H}{l}^{i0}
\nonumber \\
&=&2{\cal H}{l}^{i0}.
\EEqA
This shows that in conformal coordinates
\BEqA
{l^{\m\n}}_{|\n}&=&2{\cal H}{l}^{\m0},
\EEqA
as required.

%%%%%%%%%%%%%%%%%%%%%%%%

%\newpage

\subsection{Our scalar-tensor theory solution}

\label{sec:checkingTheGaugeCondition_OURS}

We will again use the primed notation,
\BEq
\label{eq:primed_notation_OURS}
{\cal S}'_{\m\n} \equiv {\cal S}_{\m\n}(t',{\bf x}'),
\quad
t' = t - R,
\quad
R \equiv |{\bf x}-{\bf x}'|,
\quad
\partial'_{\mu} \equiv \partial/\partial{{x'}^{\m}},
\EEq

We first write down 
(\ref{eq:covDerOf_lmn_generalResultProof}), which is a general result
for any symmetric tensor expressed in conformal coordinates,
\BEq
\label{eq:covDerOf_lmn_generalResult}
{l^{\m\n}}_{|\n}
=
\bar{g}^{\m\a}\bar{g}^{\n\b}l_{\a\b,\n}
- {\cal H}l\bar{g}^{\m0}+2{\cal H}{l}^{\m0}.
\EEq
Next, for our retarded solution \cite{AGSMK2016}, $l^{\m\n}$, as found in (\ref{eq:l_mn_retarded_solution}),
\BEq
\label{eq:STPerturbationEquationSolution_simpler}
 l_{\m\n}(\eta,{\bf x}) = 
\underbrace{
b(\eta)\int 
\frac{{\cal S}_{\m\n}(\eta',{\bf x}')}{|{\bf x}-{\bf x}'|}d^3x'
}_{\equiv Q_{\m\n}}
-\bar{g}_{\m\n}\frac{\varphi(\eta,{\bf x})}{\bar{\phi}(\eta)},
\quad
{\cal S}_{\m\n} = 
\frac{-4a^2\Lambda_{\m\n}}
{b\bar{\phi}},
\quad
{\cal B} \equiv \frac{\dot{b}}{b} = {\cal H}-\frac{{\cal F}}{2},
\EEq
we get 
\BEqA
{l^{\m\n}}_{|\n}
&=&
\bar{g}^{\m\a}\bar{g}^{\n\b}l_{\a\b,\n}
- {\cal H}l\bar{g}^{\m0}+2{\cal H}{l}^{\m0}
\nonumber \\
&=&
\bar{g}^{\m\a}\bar{g}^{\n\b}
\partial_{\n}
\left(
b
\int 
\frac{{\cal S}'_{\a\b}}{R}d^3x'
-\bar{g}_{\a\b}\frac{\varphi}{\bar{\phi}}
\right)
- {\cal H}l\bar{g}^{\m0}+2{\cal H}{l}^{\m0}
\nonumber \\
&=&
\bar{g}^{\m\a}\bar{g}^{\n\b}
(\partial_{\n}b)
\left(
\int 
\frac{{\cal S}'_{\a\b}}{R}d^3x'
\right)
+\bar{g}^{\m\a}\bar{g}^{\n\b}
b\int 
\partial_{\n}
\left(
\frac{{\cal S}'_{\a\b}}{R}\right)
d^3x'
- \bar{g}^{\m\a}\bar{g}^{\n\b}
\partial_{\n}
\left(
\bar{g}_{\a\b}\frac{\varphi}{\bar{\phi}}
\right)
- {\cal H}l\bar{g}^{\m0}+2{\cal H}{l}^{\m0}
\nonumber \\
%%%%%%%%%
&=&
\bar{g}^{\m\a}\bar{g}^{0\b}{\cal B}{Q}_{\a\b}
+\bar{g}^{\m\a}\bar{g}^{\n\b}
b\int 
\partial_{\n}
\left(
\frac{{\cal S}'_{\a\b}}{R}\right)
d^3x'
- \bar{g}^{\m\a}\bar{g}^{\n\b}
\partial_{\n}
\left(
\bar{g}_{\a\b}\frac{\varphi}{\bar{\phi}}
\right)
- {\cal H}l\bar{g}^{\m0}+2{\cal H}{l}^{\m0}
\nonumber \\
%%%%%%%%%
&=&
{\cal B}{Q}^{\m0}
+\bar{g}^{\m\a}\bar{g}^{\n\b}
b\int 
\partial_{\n}
\left(
\frac{{\cal S}'_{\a\b}}{R}\right)
d^3x'
- \bar{g}^{\m\a}\bar{g}^{\n\b}
\partial_{\n}
\left(
\bar{g}_{\a\b}\frac{\varphi}{\bar{\phi}}
\right)
- {\cal H}l\bar{g}^{\m0}+2{\cal H}{l}^{\m0}
%%%
\nonumber \\
%%%%%%%%%
&=&
{\cal B}\left({l}^{\m0}+\bar{g}^{\m0}\frac{\varphi}{\bar{\phi}}\right)
+\bar{g}^{\m\a}\bar{g}^{\n\b}
b\int 
\partial_{\n}
\left(
\frac{{\cal S}'_{\a\b}}{R}\right)
d^3x'
- \bar{g}^{\m\a}\bar{g}^{\n\b}
\partial_{\n}
\left(
\bar{g}_{\a\b}\frac{\varphi}{\bar{\phi}}
\right)
- {\cal H}l\bar{g}^{\m0}+2{\cal H}{l}^{\m0}
%%%%%%%%%%%%%%%%%%%%
\nonumber \\
&=&
{\cal B}{l}^{\m0}+{\cal B}\bar{g}^{\m0}\frac{\varphi}{\bar{\phi}}
+\bar{g}^{\m\a}\bar{g}^{\n\b}
b\int 
\partial_{\n}
\left(
\frac{{\cal S}'_{\a\b}}{R}\right)
d^3x'
- \bar{g}^{\m\a}\bar{g}^{\n\b}
\partial_{\n}
\left(
\bar{g}_{\a\b}\frac{\varphi}{\bar{\phi}}
\right)
- {\cal H}l\bar{g}^{\m0}+2{\cal H}{l}^{\m0}
%%%%%%%%%%%%%%%%%%%%
\nonumber \\
&=&
{\cal B}\bar{g}^{\m0}\frac{\varphi}{\bar{\phi}}
+\bar{g}^{\m\a}\bar{g}^{\n\b}
b\int 
\partial_{\n}
\left(
\frac{{\cal S}'_{\a\b}}{R}\right)
d^3x'
- \bar{g}^{\m\a}\bar{g}^{\n\b}
\partial_{\n}
\left(
\bar{g}_{\a\b}\frac{\varphi}{\bar{\phi}}
\right)
- {\cal H}l\bar{g}^{\m0}+\left(2{\cal H}+{\cal B}\right){l}^{\m0}
%%%
\nonumber \\
&=&
{\cal B}\bar{g}^{\m0}\frac{\varphi}{\bar{\phi}}
- \bar{g}^{\m\a}\bar{g}^{\n\b}
\partial_{\n}
\left(
\bar{g}_{\a\b}\frac{\varphi}{\bar{\phi}}
\right)
+\frac{1}{a^4} b
\int 
\partial_{\n}
\left(
\frac{f^{\m\a}f^{\n\b}{\cal S}'_{\a\b}}{R}\right)
d^3x'
- {\cal H}l\bar{g}^{\m0}+\left(2{\cal H}+{\cal B}\right){l}^{\m0}.
\EEqA
A few additional steps analogous to those in electrodynamics 
(see discussion following Eq.\ (\ref{eq:ED_gaugeCheck1})) give
\BEqA
{l^{\m\n}}_{|\n}&=&
{\cal B}\bar{g}^{\m0}\frac{\varphi}{\bar{\phi}}
- \bar{g}^{\m\a}\bar{g}^{\n\b}
\partial_{\n}
\left(
\bar{g}_{\a\b}\frac{\varphi}{\bar{\phi}}
\right)
- {\cal H}l\bar{g}^{\m0}+\left(2{\cal H}+{\cal B}\right){l}^{\m0}
\nonumber \\
&&
+\frac{1}{a^4} b
\int
\frac{1}{R}
\left\{
\partial'_{0}\left( f^{\m\a}f^{0\b}{\cal S}'_{\a\b} \right)
+
\left[\partial'_{k}\left(f^{\m\a}f^{k\b}{\cal S}'_{\a\b}\right)\right]_{t' = {\rm const}}
\right\}
d^3x'
-\frac{1}{a^4} b
\underbrace{
\int
\partial'_{k}\left(\frac{f^{\m\a}f^{k\b}{\cal S}'_{\a\b}}{R}\right)d^3x'
}_{= 0, {\rm \; divergence}}
\nonumber \\
&=&
{\cal B}\bar{g}^{\m0}\frac{\varphi}{\bar{\phi}}
- \bar{g}^{\m\a}\bar{g}^{\n\b}
\partial_{\n}
\left(
\bar{g}_{\a\b}\frac{\varphi}{\bar{\phi}}
\right)
- {\cal H}l\bar{g}^{\m0}+\left(2{\cal H}+{\cal B}\right){l}^{\m0}
\nonumber \\
&&+\frac{1}{a^4} b
\int
\frac{1}{R}
\underbrace{
\left\{
\partial'_{0}\left(f^{\m\a}f^{0\b}{\cal S}'_{\a\b} \right)
+
\left[\partial'_{k}\left(f^{\m\a}f^{k\b}{\cal S}'_{\a\b}\right)\right]_{t' = {\rm const}}
\right\}
}_{\equiv I'^{\mu}}
d^3x'.
\nonumber \\
\EEqA
Notice that
\BEq
\bar{g}^{\m\a}\bar{g}^{\n\b}
\partial_{\n}
\left(
\bar{g}_{\a\b}\frac{\varphi}{\bar{\phi}}
\right)
=
\bar{g}^{\m0}(2{\cal H}-{\cal F})\frac{\varphi}{\bar{\phi}}+\frac{\varphi^{|\m}}{\bar{\phi}},
\EEq
so
\BEqA
{\cal B}\bar{g}^{\m0}\frac{\varphi}{\bar{\phi}}
- \bar{g}^{\m\a}\bar{g}^{\n\b}
\partial_{\n}
\left(
\bar{g}_{\a\b}\frac{\varphi}{\bar{\phi}}
\right)
&=&
\bar{g}^{\m0}\left({\cal H}-\frac{\cal F}{2}\right)\frac{\varphi}{\bar{\phi}}
-
\bar{g}^{\m0}(2{\cal H}-{\cal F})\frac{\varphi}{\bar{\phi}}
-\frac{\varphi^{|\m}}{\bar{\phi}}
\nonumber \\
&=&
\left(-{\cal H}+\frac{\cal F}{2}\right)\bar{g}^{\m0}\frac{\varphi}{\bar{\phi}}
-\frac{\varphi^{|\m}}{\bar{\phi}},
\EEqA
and thus
\BEqA
\label{eq:covDivergence_l^mn}
{l^{\m\n}}_{|\n}
&=&
\left(-{\cal H}+\frac{\cal F}{2}\right)\bar{g}^{\m0}\frac{\varphi}{\bar{\phi}}
-\frac{\varphi^{|\m}}{\bar{\phi}}
- {\cal H}l\bar{g}^{\m0}+\left(2{\cal H}+{\cal B}\right){l}^{\m0}
\nonumber \\
&&+\frac{1}{a^4} b
\int
\frac{1}{R}
\underbrace{
\left\{
\partial'_{0}\left(f^{\m\a}f^{0\b}{\cal S}'_{\a\b} \right)
+
\left[\partial'_{k}\left(f^{\m\a}f^{k\b}{\cal S}'_{\a\b}\right)\right]_{t' = {\rm const}}
\right\}
}_{\equiv I'^{\mu}}
d^3x'.
\nonumber \\
\EEqA
For the $0$-th component of $I^{\mu}$ we get
\BEqA
I^{0}
&=&
\partial_{0}\left(f^{0\a}f^{0\b}{\cal S}_{\a\b}\right)
+\partial_{k}\left(f^{0\a}f^{k\b}{\cal S}_{\a\b}\right)
\nonumber \\
&=&
\partial_{0}\left(f^{00}f^{00}{\cal S}_{00}\right)
+\partial_{k}\left(f^{00}f^{km}{\cal S}_{0m}\right)
\nonumber \\
&=&
\partial_{0}\left({{\cal S}}_{00}\right)
-\partial_{k}\left({{\cal S}}_{{0}k}\right)
\nonumber \\
&=&
\partial_{0}\left({\frac{-4a^2\Lambda_{00}}
{b\bar{\phi}}}\right)
-\partial_{k}\left({\frac{-4a^2\Lambda_{0k}}
{b\bar{\phi}}}\right)
\nonumber \\
&=&
\partial_{0}\left({\frac{-4a \left(a\Lambda_{00}\right)}
{b\bar{\phi}}}\right)
-\partial_{k}\left({\frac{-4a \left(a\Lambda_{0k}\right)}
{b\bar{\phi}}}\right)
%%%%%%%%
\nonumber \\
&=&
\partial_{0}\left({\frac{-4a}
{b\bar{\phi}}}\right)\left(a\Lambda_{00}\right)
-
\underbrace{
\partial_{k}\left({\frac{-4a}
{b\bar{\phi}}}\right)
}_{=0}
\left(a\Lambda_{0k}\right)
+
{\frac{-4a}{b\bar{\phi}}} \partial_{0}\left(a\Lambda_{00}\right)
-
{\frac{-4a}{b\bar{\phi}}}
\partial_{k}\left(a\Lambda_{0k}\right)
%%%%%%%%
\nonumber \\
&=&
{\cal H} {\frac{-4a^2\Lambda_{00}}
{b\bar{\phi}}}
-{\cal B}{\frac{-4a^2\Lambda_{00}}
{b\bar{\phi}}}
-{\cal F} {\frac{-4a^2\Lambda_{00}}
{b\bar{\phi}}}
+\frac{-4a}{b\bar{\phi}}
\left\{
\partial_{0}\left(a\Lambda_{00}\right)
-\partial_{k}\left(a\Lambda_{0k}\right)
\right\}
\nonumber \\
&=&
\left({\cal H}-{\cal B} -{\cal F} \right){\cal S}_{00}
+\frac{-4a}{b\bar{\phi}}
\left\{
\partial_{0}\left[a\left(\Lambda_{00}+\Lambda^{\varphi}_{00}\right)\right]
-\partial_{k}\left[a\left(\Lambda_{0k}+\Lambda^{\varphi}_{0k}\right)\right]
\right\}
\nonumber \\
&=&
-\frac{{\cal F}}{2}{\cal S}_{00}
+\frac{-4a}{b\bar{\phi}}
\left\{
\partial_{0}\left(a\Lambda_{00}\right)
-\partial_{k}\left(a\Lambda_{0k}\right)
\right\}.
\EEqA
For the {\it i}-th component of $I^{\mu}$ we similarly get
\BEqA
I^{i}
&=&
\partial_{0}\left(f^{i\a}f^{0\b}{\cal S}_{\a\b}\right)
+\partial_{k}\left(f^{i\a}f^{k\b}{\cal S}_{\a\b}\right)
\nonumber \\
&=&
\partial_{0}\left(f^{ij}f^{00}{\cal S}_{j0}\right)
+\partial_{k}\left(f^{ij}f^{km}{\cal S}_{jm}\right)
\nonumber \\
&=&
-\partial_{0}\left({{\cal S}}_{i0}\right)
+\partial_{k}\left({{\cal S}}_{ik}\right)
\nonumber \\
&=&
-\partial_{0}\left({\frac{-4a^2\Lambda_{i0}}
{b\bar{\phi}}}\right)
+\partial_{k}\left({\frac{-4a^2\Lambda_{ik}}
{b\bar{\phi}}}\right)
\nonumber \\
&=&
-{\cal H} \left({\frac{-4a^2\Lambda_{i0}}
{b\bar{\phi}}}\right)
+{\cal B} \left({\frac{-4a^2\Lambda_{i0}}
{b\bar{\phi}}}\right)
+{\cal F} \left({\frac{-4a^2\Lambda_{i0}}
{b\bar{\phi}}}\right)
-\frac{-4a}{b\bar{\phi}}
\left\{
\partial_{0}\left(a\Lambda_{i0}\right)-\partial_{k}\left(a\Lambda_{ik}\right)
\right\}
\nonumber \\
&=&
-\left({\cal H}-{\cal B} -{\cal F} \right){\cal S}_{i0}
-\frac{-4a}{b\bar{\phi}}
\left\{
\left(a\Lambda_{i0}\right)-\partial_{k}\left(a\Lambda_{ik}\right)\right\}
\nonumber \\
&=&
+\frac{{\cal F}}{2}{\cal S}_{i0}
-\frac{-4a}{b\bar{\phi}}
\left\{
\left(a\Lambda_{i0}\right)-\partial_{k}\left(a\Lambda_{ik}\right)
\right\}.
\EEqA

To find out what $\partial_{0}\left(a\Lambda_{00}\right)
-\partial_{k}\left(a\Lambda_{0k}\right)$ and
$\partial_{0}\left(a\Lambda_{i0}\right)
-\partial_{k}\left(a\Lambda_{ik}\right)$ are
we need to do a few additional calculations.

First, by taking the covariant divergence of (\ref{eq:l_mn_perturbation_linearHubble1})
we get, in the linear Hubble approximation,
\BEq
8\pi{\Lambda^{\m\n}}_{|\n}
=
\frac{1}{2}\bar{\phi}^{|\m}
\underbrace{
\left(
{l^{\n\b}}_{|\n{\b}}+\frac{1}{2}{l^{|\n}}_{|\n}
-\frac{2\omega}{\bar{\phi}}{\varphi^{|\n}}_{|\n}
\right)
}_{={\cal O}({\cal H}), {\rm \; by\;}  (\ref{eq:trace_perturbation_linearHubble1})}
=0.
\EEq
On the other hand, for any symmetric tensor $s^{\m\n}$ we have
\BEq
\label{eq:LLformulaForSymmTensor}
{s^{\m\n}}_{|\n} 
= \bar{g}^{\m\a}
\left[
\frac{1}{\sqrt{-\bar{g}}}\partial_{\nu}\left(\sqrt{-\bar{g}}\bar{g}^{\b\n}s_{\a\b}\right)
-\frac{1}{2}\left(\partial_{\a}\bar{g}_{\b\n}\right) s^{\b\n}
\right],
\EEq
which in conformal coordinates takes the form
\BEqA
%\label{eq:covContinuityLambda_varphiOLD}
\partial_{0}\left(as_{00}\right) - \partial_{j}\left(as_{0j}\right)
&=&-{\cal H}\left(as_{kk}\right)
+a^5 {s^{0\n}}_{|\n},
\\
\partial_{0}\left(as_{i0}\right) - \partial_{j}\left(as_{ij}\right)
&=&-{\cal H}\left(as_{i0}\right)
- a^5 {s^{i\n}}_{|\n}.
\EEqA
Applying this to $\L^{\m\n}$ gives,
\BEqA
\label{eq:covContinuityLambda}
\partial_{0}\left(a\Lambda_{00}\right) - \partial_{j}\left(a\Lambda_{0j}\right)
&=&-{\cal H}\left(a\Lambda_{kk}\right),
\\
\partial_{0}\left(a\Lambda_{i0}\right) - \partial_{j}\left(a\Lambda_{ij}\right)
&=&-{\cal H}\left(a\Lambda_{i0}\right).
\EEqA
Therefore, for the $0$-th component of $I^{\m}$ we have
\BEqA
\label{eq:I^0}
I^{0}
&=&
-\frac{{\cal F}}{2}{\cal S}_{00}
+\frac{-4a}{b\bar{\phi}}
\left(
-{\cal H}a\Lambda_{kk}
\right)
\nonumber \\
&=&
-\frac{{\cal F}}{2}{\cal S}_{00}
-{\cal H}{\cal S}_{kk}.
\EEqA
%%%%%%%%%%%%%
Finally, plugging (\ref{eq:I^0}) in (\ref{eq:covDivergence_l^mn}) gives
\BEqA
\label{eq:covDivergence_l^0n}
{l^{0\n}}_{|\n} 
&=&
\left(-{\cal H}+\frac{\cal F}{2}\right)\bar{g}^{00}\frac{\varphi}{\bar{\phi}}
-\frac{\varphi^{|0}}{\bar{\phi}}
- {\cal H}l\bar{g}^{00}+\left(2{\cal H}+{\cal B}\right){l}^{00}
-\frac{1}{a^4}
b
\underbrace{
\int
\frac{1}{R}
\left(\frac{{\cal F}'}{2}{\cal S}'_{00}+{\cal H}'f^{ik}{\cal S}'_{ik}
\right)
d^3x'
}_{{\rm treat\;}{\cal H}'\;{\rm and\;}{\cal F}'{\rm \; as\; constants}}
\nonumber \\
%%%%%%
&=&
\left(-{\cal H}+\frac{\cal F}{2}\right)\bar{g}^{00}\frac{\varphi}{\bar{\phi}}
-\frac{\varphi^{|0}}{\bar{\phi}}
- {\cal H}l\bar{g}^{00}+\left(3{\cal H}-\frac{{\cal F}}{2}\right){l}^{00}
-\frac{{\cal F}}{2}{Q}^{00}
-{\cal H}\frac{\bar{g}^{ik}}{a^2}Q_{ik}
\nonumber \\
%%%%%
&=&
\left(-{\cal H}+\frac{\cal F}{2}\right)\bar{g}^{00}\frac{\varphi}{\bar{\phi}}
-\frac{\varphi^{|0}}{\bar{\phi}}
- {\cal H}l\bar{g}^{00}+\left(3{\cal H}-\frac{{\cal F}}{2}\right){l}^{00}
-\frac{{\cal F}}{2}
\left(
{l}^{00}
+
\bar{g}^{00}\frac{\varphi}{\bar{\phi}}
\right)
-{\cal H}\frac{\bar{g}^{ik}}{a^2}
\left(
{l}_{ik}
+
\bar{g}_{ik}\frac{\varphi}{\bar{\phi}}
\right)
\nonumber \\
%%%%%%
%%%%%%%
&=&
\left(-{\cal H}+\frac{\cal F}{2}\right)\bar{g}^{00}\frac{\varphi}{\bar{\phi}}
-\frac{\varphi^{|0}}{\bar{\phi}}
- {\cal H}l\bar{g}^{00}+\left(3{\cal H}-{\cal F}\right){l}^{00}
-\frac{{\cal F}}{2}\bar{g}^{00}\frac{\varphi}{\bar{\phi}}
-{\cal H}\frac{\bar{g}^{ik}}{a^2}
\left(
{l}_{ik}
+
\bar{g}_{ik}\frac{\varphi}{\bar{\phi}}
\right)
\nonumber \\
%%%%%%%
&=&
\left(-{\cal H}+\frac{\cal F}{2}\right)\bar{g}^{00}\frac{\varphi}{\bar{\phi}}
-\frac{\varphi^{|0}}{\bar{\phi}}
- {\cal H}l\bar{g}^{00}+\left(3{\cal H}-{\cal F}\right){l}^{00}
-\frac{{\cal F}}{2}\bar{g}^{00}\frac{\varphi}{\bar{\phi}}
-{\cal H}\frac{\bar{g}^{ik}}{a^2}{l}_{ik}
-{\cal H}\frac{\bar{g}^{ik}}{a^2}
\bar{g}_{ik}\frac{\varphi}{\bar{\phi}}
\nonumber\\
&&
+\underbrace{
\left(
{\cal H}\bar{g}_{00}{l}^{00}\bar{g}^{00}
-{\cal H}\bar{g}_{00}{l}^{00}\bar{g}^{00}
\right)
}_{\rm added\;and\;subtracted\;same\;thing}
\nonumber \\
%%%%%%%
&=&
\left(-{\cal H}+\frac{\cal F}{2}\right)\bar{g}^{00}\frac{\varphi}{\bar{\phi}}
-\frac{\varphi^{|0}}{\bar{\phi}}
\underbrace{
- {\cal H}l\bar{g}^{00}
}_{1}
+
\underbrace{
\left(3{\cal H}-{\cal F}\right){l}^{00}
}_{2}
-\frac{{\cal F}}{2}\bar{g}^{00}\frac{\varphi}{\bar{\phi}}
\underbrace{
-{\cal H}\frac{\bar{g}^{ik}}{a^2}{l}_{ik}
}_{1}
-{\cal H}\frac{\bar{g}^{ik}}{a^2}
\bar{g}_{ik}\frac{\varphi}{\bar{\phi}}
\nonumber\\
&&
+
\left(
\underbrace{
{\cal H}\bar{g}_{00}{l}^{00}\bar{g}^{00}
}_{1}
-
\underbrace{
{\cal H}\bar{g}_{00}{l}^{00}\bar{g}^{00}
}_{2}
\right)
%%%%%%
\nonumber \\
&=&
\left(-{\cal H}+\frac{\cal F}{2}\right)\bar{g}^{00}\frac{\varphi}{\bar{\phi}}
-\frac{\varphi^{|0}}{\bar{\phi}}
+\left(2{\cal H}-{\cal F}\right){l}^{00}
-\frac{{\cal F}}{2}\bar{g}^{00}\frac{\varphi}{\bar{\phi}}
-{\cal H}\frac{\bar{g}^{ik}}{a^2}
\bar{g}_{ik}\frac{\varphi}{\bar{\phi}}
\nonumber \\
%%%%%
\nonumber \\
&=&
\left(-{\cal H}+\frac{\cal F}{2}\right)\bar{g}^{00}\frac{\varphi}{\bar{\phi}}
-\frac{\varphi^{|0}}{\bar{\phi}}
+\left(2{\cal H}-{\cal F}\right){l}^{00}
-\frac{{\cal F}}{2}\bar{g}^{00}\frac{\varphi}{\bar{\phi}}
+{\cal H}\bar{g}^{00}\bar{g}^{ik}
\bar{g}_{ik}\frac{\varphi}{\bar{\phi}}
\nonumber \\
%%%%%
\nonumber \\
&=&
\left(-{\cal H}+\frac{\cal F}{2}\right)\bar{g}^{00}\frac{\varphi}{\bar{\phi}}
-\frac{\varphi^{|0}}{\bar{\phi}}
+\left(2{\cal H}-{\cal F}\right){l}^{00}
-\frac{{\cal F}}{2}\bar{g}^{00}\frac{\varphi}{\bar{\phi}}
+3{\cal H}\bar{g}^{00}\frac{\varphi}{\bar{\phi}}
\nonumber \\
%%%%%
\nonumber \\
&=&
\left(-{\cal H}+\frac{\cal F}{2}\right)\bar{g}^{00}\frac{\varphi}{\bar{\phi}}
-\frac{\varphi^{|0}}{\bar{\phi}}
+\left(2{\cal H}-{\cal F}\right){l}^{00}
-\frac{{\cal F}}{2}\bar{g}^{00}\frac{\varphi}{\bar{\phi}}
+3{\cal H}\bar{g}^{00}\frac{\varphi}{\bar{\phi}}
\nonumber \\
%%%%%
\nonumber \\
&=&
\left(-{\cal H}+\frac{\cal F}{2}\right)\bar{g}^{00}\frac{\varphi}{\bar{\phi}}
-\frac{\varphi^{|0}}{\bar{\phi}}
+\left(2{\cal H}-{\cal F}\right){l}^{00}
-\frac{{\cal F}}{2}\bar{g}^{00}\frac{\varphi}{\bar{\phi}}
+3{\cal H}\bar{g}^{00}\frac{\varphi}{\bar{\phi}}
\nonumber \\
%%%%%
\nonumber \\
&=&
\left(2{\cal H}-{\cal F}\right){l}^{00}
+2{\cal H}\bar{g}^{00}\frac{\varphi}{\bar{\phi}}
-\frac{\varphi^{|0}}{\bar{\phi}},
\EEqA
in agreement with (\ref{eq:generalizedGaugeActual}).

Now, for the {\it i}-th component,
\BEqA
I^{i}
&=&
+\frac{{\cal F}}{2}{\cal S}_{i0}
-\frac{-4a}{b\bar{\phi}}
\left(
-{\cal H}a\Lambda_{i0}
\right)
\nonumber \\
&=&
+\frac{{\cal F}}{2}{\cal S}_{i0}
+{\cal H}{\cal S}_{i0},
\EEqA
and thus,
\BEqA
\label{eq:covDivergence_l^in}
{l^{i\n}}_{|\n}
&=&
-
\frac{\varphi^{|i}}{\bar{\phi}}
+
\left(2{\cal H}+{\cal B}\right){l}^{i0}
+\frac{1}{a^4}b
\underbrace{
\int
\left({\cal H}'+\frac{{\cal F}'}{2} \right)\frac{{\cal S}'_{i0}}{R}
d^3x'
}_{{\rm treat\;}{\cal H}'\;{\rm and\;}{\cal F}'{\rm \; as\; constants}}
\nonumber \\
&\approx&
-
\frac{\varphi^{|i}}{\bar{\phi}}
+
\left(3{\cal H}-\frac{{\cal F}}{2}\right){l}^{i0}
+\left({\cal H}+\frac{{\cal F}}{2}\right)
\frac{1}{a^4} 
b
\int
\frac{{\cal S}'_{i0}}{R}
d^3x'
\nonumber \\
&=&
-
\frac{\varphi^{|i}}{\bar{\phi}}
+
\left(3{\cal H}-\frac{{\cal F}}{2}\right){l}^{i0}
+\left({\cal H}+\frac{{\cal F}}{2}\right)\frac{1}{a^4}{l}_{i0}
\nonumber \\
&=&
-
\frac{\varphi^{|i}}{\bar{\phi}}
+
\left(3{\cal H}-\frac{{\cal F}}{2}\right){l}^{i0}
+\left({\cal H}+\frac{{\cal F}}{2}\right)\frac{1}{a^4}\bar{g}_{i\m}\bar{g}_{0\n}{l}^{\m\n}
\nonumber \\
&=&
-
\frac{\varphi^{|i}}{\bar{\phi}}
+
\left(3{\cal H}-\frac{{\cal F}}{2}\right){l}^{i0}
+\left({\cal H}+\frac{{\cal F}}{2}\right)f_{i\m}f_{0\n}{l}^{\m\n}
\nonumber \\
&=&
-
\frac{\varphi^{|i}}{\bar{\phi}}
+
\left(3{\cal H}-\frac{{\cal F}}{2}\right){l}^{i0}
+\left({\cal H}+\frac{{\cal F}}{2}\right)f_{ik}f_{00}{l}^{k0}
\nonumber \\
&=&
-
\frac{\varphi^{|i}}{\bar{\phi}}
+
\left(3{\cal H}-\frac{{\cal F}}{2}\right){l}^{i0}
-\left({\cal H}+\frac{{\cal F}}{2}\right){l}^{i0}
\nonumber \\
&=&\left(2{\cal H}-{\cal F}\right){l}^{i0} -
\frac{\varphi^{|i}}{\bar{\phi}},
\EEqA
as required.

%%%%%%%%%%%%%%%%%%%%%%%%%%%%%

%\newpage

\section{APPENDIX: Background Friedman cosmology}
\label{sec:BackgroundFriedmanCosmology}

Here, for convenience, we list a few results related to the background Friedman 
cosmology.

\subsection{General considerations}
\label{eq:BackgroundCosmologyGeneralConsiderations}

In this Subsection, the derivatives with respect to the coordinate time $x^0$, 
cosmic time $T$, conformal time $\eta$, and the scalar field $\phi$ will be denoted by
\BEq
\frac{\partial F(x^0,x^i)}{\partial x^0}\equiv F_{,0},
\quad
\frac{dF(T)}{dT}\equiv F_{,T},
\quad
\frac{dF(\eta)}{d\eta}\equiv \dot{F},
\quad
\frac{dF(\phi)}{d\phi}\equiv F'.
\EEq
We work with the conformally {\it flat} background FLRW metric,
\BEq
d\bar{g}^2 = a^2(\eta)\left(-d\eta^2 + \d_{ij}dx^idx^j\right),
\quad
\bar{g}_{\a\b} = a^2(\eta){f}_{\a\b},
\quad 
{f}_{\a\b}={\rm diag}\left(-1,1,1,1\right),
\EEq
so that
\BEq
\label{eq:backgroundConnection}
\bar{\Gamma}^{\a}_{\b\g}=
-\frac{\cal{H}}{a}
\left(
\d^{\a}_{\b}\bar{u}_{\g} + \d^{\a}_{\g}\bar{u}_{\b} 
- \bar{u}^{\a}\bar{g}_{\b\g}
\right),
\quad
\bar{\Gamma}^{0}_{00}
=\bar{\Gamma}^{1}_{10}=\bar{\Gamma}^{2}_{20}=\bar{\Gamma}^{3}_{30}
=\bar{\Gamma}^{0}_{11}=\bar{\Gamma}^{0}_{22}=\bar{\Gamma}^{0}_{33}
=\cal{H},
\EEq
\BEq
\label{eq:backgroundRicci}
\bar{R}_{\a\b}
=\frac{1}{a^2}
\left[
\dot{\cal H}
\left(
\bar{g}_{\a\b}-2\bar{u}_{\a}\bar{u}_{\b}
\right)
+2{\cal H}^2\left(\bar{g}_{\a\b}+\bar{u}_{\a}\bar{u}_{\b}\right)
\right],
\quad
\bar{R}_{00}=-3\dot{\cal H}, 
\quad
\bar{R}_{11}=\bar{R}_{22}=\bar{R}_{33}=\dot{\cal H}+2{\cal H}^2,
\EEq
\BEq
\label{eq:backgroundScalarCurvature}
\bar{R}=\frac{6}{a^2}\left(\dot{\cal H}+{\cal H}^2\right),
\EEq
where the velocity of the Hubble flow is
\BEq
\bar{u}^{\m}= (1/a)\d^{\m}_{0}=\left(1/a,0,0,0\right), 
\quad 
\bar{u}_{\m}=-a\d^0_{\mu}=(-a,0,0,0),
\quad
\bar{u}_{\m}\bar{u}^{\m}=-1,
\EEq
and the (conformal) Hubble parameter is
\BEq
{\cal H}\equiv \dot{a}/a.
\EEq
Then, in the isotropic conformal coordinates,
\BEqA
&&
{\bar{\phi}_{|00}}=\ddot{\bar{\phi}}-{\cal H}\dot{\bar{\phi}},
\quad
{\bar{\phi}_{|0i}}=0,
\quad
{\bar{\phi}_{|ij}}=-{\cal H}\dot{\bar{\phi}}f_{ij},
\\
&&
{\bar{\phi}^{|\a}}_{|\a} 
=-\frac{1}{a^2}\left(\ddot{\bar{\phi}}+2{\cal H}\dot{\bar{\phi}}\right),
\\
&&
\bar{\phi}^{|\a}\bar{\phi}_{|\a} 
= -\frac{1}{a^2}\dot{\bar{\phi}}^2,
\EEqA
and
\BEqA
\bar{T}^{M}_{\m\n} 
&=& \left(\bar{\epsilon}+\bar{p}\right) \bar{u}_{\m}\bar{u}_{\n}+\bar{p}\bar{g}_{\m\n}
={\rm diag}(a^2\bar{\epsilon},a^2\bar{p},a^2\bar{p},a^2\bar{p}),
\\
\bar{T}_{M}^{\m\n} 
&=& \left(\bar{\epsilon}+\bar{p}\right) \bar{u}^{\m}\bar{u}^{\n}+\bar{p}\bar{g}^{\m\n}
={\rm diag}(\bar{\epsilon}/a^2,\bar{p}/a^2,\bar{p}/a^2,\bar{p}/a^2),
\\
\bar{T}^{M} &=& \bar{g}^{\m\n} \bar{T}^{M}_{\m\n} 
= -\bar{\epsilon}+3\bar{p}.
\EEqA

\subsection{Background Friedman equations with conformal time}

Combining the results of Subsection \ref{eq:BackgroundCosmologyGeneralConsiderations}
with Eqs.\ (\ref{eq:backgroundFieldEqs21}),
(\ref{eq:backgroundFieldEqs22}), and (\ref{eq:backgroundFieldEqs23}),
and calculating the sums
\[
\bar{R}_{00}+\frac{1}{2}a^2\bar{R}-\frac{a^2}{\bar{\phi}}{\bar{\phi}^{|\a}}_{|\a}
\]
and
\[
\frac{1}{2}a^2\bar{R} - 3{\cal H}^2,
\]
we get, in the isotropic conformal coordinates with conformal time $\eta$, the 
background Friedman equations,
\BEqA
\label{eq:axil1general}
3{\cal H}^2
&=& \frac{8\pi a^2}{\bar{\phi}}\bar{\epsilon} 
- 3{\cal H}\frac{\dot{\bar{\phi}}}{\bar{\phi}}
+\frac{\omega}{2}\left(\frac{\dot{\bar{\phi}}}{\bar{\phi}}\right)^2
+a^2 \lambda,
\\
\label{eq:axil3general}
3\dot{\cal H}
&=& -\frac{8\pi a^2}{\bar{\phi}}
\left[
\frac{3+\omega}{3+2\omega}\bar{\epsilon} 
+
\frac{\omega}{3+2\omega}(3\bar{p})
\right]
+ 3{\cal H}\frac{\dot{\bar{\phi}}}{\bar{\phi}}
-\left(\omega-\frac{3}{2}\frac{\omega' \bar{\phi}}{3+2\omega}\right)
\left(\frac{\dot{\bar{\phi}}}{\bar{\phi}}\right)^2
+\frac{a^2}{3+2\omega}\left(2\omega\lambda +3\lambda'\bar{\phi}\right),
\\
\label{eq:axil2general}
\ddot{\bar{\phi}}+ 2{\cal H}\dot{\bar{\phi}}
&=& \frac{8\pi a^2}{3+2\omega}\left(\bar{\epsilon}-3\bar{p}\right)
-\frac{\omega'}{3+2\omega}\dot{\bar{\phi}}^2.
\EEqA

\subsection{Background Friedman equations with Hubble time}
\label{sec:BD_background}

In this Subsection we work with the usual Hubble time, $t$, defined by
\BEq
\label{eq:axil_from_eta_to_t}
\frac{dt}{d\eta}\equiv a, \quad 
{\cal H}=Ha, \quad
\frac{d{\cal H}}{d\eta} =  \left(H^2 + \frac{d{H}}{dt}\right)a^2.
\EEq
Then,
\BEq
\label{eq:axil2_LeftSide_from_eta_to_t}
\frac{d^2 \bar{\phi}}{d\eta^2}+2{\cal H}\frac{d\bar{\phi}}{d\eta}
= \left(\frac{d^2 \bar{\phi}}{dt^2}+3H\frac{d\bar{\phi}}{dt}\right)a^2.
\EEq
For the remainder of this subsection the derivative with respect to $t$ will be denoted with an 
overdot.

\subsubsection{Friedman equations in general background cosmology}

The background ``matter'' is assumed to be a fluid described by the equation of state,
\BEq
{\bar p} = {\a} \bar{\epsilon},
\EEq
and obeying the law of conservation,
\BEq
\label{eq:axil0}
\bar{\epsilon}a^{3(1+{\a})}=\bar{\epsilon}_{0}a_{0}^{3(1+{\a})}.
\EEq
From (\ref{eq:axil1general}), (\ref{eq:axil3general}), (\ref{eq:axil2general}),  
(\ref{eq:axil_from_eta_to_t}), (\ref{eq:axil2_LeftSide_from_eta_to_t}), 
we get the background Friedmann equations,
\BEqA
\label{eq:axil1}
\left({H}+\frac{1}{2} \frac{\dot{\bar{\phi}}}{\bar{\phi}}\right)^2
&=& \frac{3+2\omega}{12}
\left(\frac{\dot{\bar{\phi}}}{\bar{\phi}}\right)^2
+ \frac{8\pi\bar{\epsilon}}{3\bar{\phi}} +\frac{\lambda}{3},
\\
\label{eq:axil3}
H^2+\dot{ H}
&=& - \frac{8\pi \bar{\epsilon} }{3\bar{\phi}}
\left[\frac{3+(1+3\a)\omega}{3+2\omega}\right]
+ {H}\frac{\dot{\bar{\phi}}}{\bar{\phi}}
-\left(\omega-\frac{3}{2}\frac{\omega' \bar{\phi}}{3+2\omega}\right)
\left(\frac{\dot{\bar{\phi}}}{\bar{\phi}}\right)^2
+\frac{1}{3+2\omega}\left(2\omega\lambda +3\lambda'\bar{\phi}\right),
\\
\label{eq:axil2}
\ddot{\bar{\phi}}+3\dot{\bar{\phi}} H
&=& \frac{8\pi }{3+2\omega}\left(1-3\a\right){\bar{\epsilon}}
-\frac{\omega'}{3+2\omega}\dot{\bar{\phi}}^2.
\EEqA

\subsubsection{Friedman equations in standard Brans-Dicke background cosmology}

\label{sec:Standard_Brans-Dicke_background_cosmology}

The Brans-Dicke theory is recovered by setting 
\BEq
\label{eq:simplifiedCosmology_BD}
\lambda = 0, \quad \omega' = 0.
\EEq
One possible solution maybe found in the power-law form by using the {\it Ansatz},
\BEq
\label{eq:ansatz_matterGeneral}
a=a_0\left(\frac{t}{t_0}\right)^q,
\quad
\bar{\phi}=\bar{\phi}_0\left(\frac{t}{t_0}\right)^r,
\quad
\bar{\e}=\bar{\e}_0\left(\frac{t}{t_0}\right)^s,
\EEq
so that
\BEq
\label{eq:ansatz_matterGeneral_derivatives}
\dot{a}=a_0q\frac{t^{q-1}}{t_0^q},
\quad
\dot{\bar{\phi}}=\bar{\phi}_0r\frac{t^{r-1}}{t_0^r},
\quad
\ddot{\bar{\phi}}=\bar{\phi}_0r(r-1)\frac{t^{r-2}}{t_0^r},
\quad
H=\frac{1}{a}\frac{da}{dt} = \frac{q}{t},
\quad
\frac{1}{\bar{\phi}}\frac{d{\bar{\phi}}}{dt} = \frac{r}{t},
\quad
\frac{\bar{\e}}{\bar{\phi}} = \frac{\bar{\e}_0}{\bar{\phi}_0}\left(\frac{t}{t_0}\right)^{s-r}.
\EEq
Using (\ref{eq:simplifiedCosmology_BD}) and (\ref{eq:ansatz_matterGeneral_derivatives}) 
in (\ref{eq:axil0}), (\ref{eq:axil1}) and (\ref{eq:axil2}), we get the system,
\BEqA
\label{eq:axil01}
s+3q(1+\a)&=&0,
\\
\label{eq:axil11}
(2q+r)^2
&=& \left(1+\frac{2}{3}\omega\right)r^2+\frac{4(3+2\omega)r(r-1+3q)}{3(1-3\a)},
\\
\label{eq:axil21}
s-r+2&=&0,
\\
\label{eq:axil22}
{\bar{\phi}}_0&=&\frac{8\pi }{3+2\omega}
\frac{(1-3\a)}{r(r-1+3q)}\bar{\epsilon}_0t_0^{2},
\EEqA
whose solution (which in the limit $\omega \rightarrow \infty$ correctly reproduces
the standard Friedmann cosmology) is
\BEqA
q&=&\frac{2[1+(1-\a)\omega]}{4+3(1-\a^2)\omega},\\
r&=&\frac{2(1-3\a)}{4+3(1-\a^2)\omega},\\
s&=&\frac{6(1+\a)[1+(1-\a)\omega]}{4+3(1-\a^2)\omega}.
\EEqA

It is interesting to notice that 
\BEq
\frac{1}{H}\frac{\dot{{\bar{\phi}}}}{\bar{\phi}} 
= \frac{(1-3\a)}{1+(1-\a)\omega},
\EEq
and thus there are two small parameters in our theory,
\BEq
\chi_1 = {H}T_0, \quad 
\chi_2 = \frac{\dot{{\bar{\phi}}}}{\bar{\phi}}T_0 
= \frac{(1-3\a)}{1+(1-\a)\omega}HT_0,
\EEq
where $T_0$ is the characteristic time of the dynamical evolution
of the system, say, its orbital or rotational period.  If 
\BEq
\a\neq 1, \quad \omega \gg 1,
\EEq
which is a typical situation (currently accepted value is $\omega \simeq 4\times 10^4$), then
\BEq
\label{eq:twoSmallParametersInequality}
\chi_1 \gg \chi_2.
\EEq

%%%%%%%%%%%%%%%%%

%%%%%%%%%% BEGIN FIG. 1
%\begin{figure}[!ht]
%\includegraphics[angle=0,width=1.00\linewidth]{fig1}
%\caption{ \label{fig:1} 
%(color online). 
%
%}
%\end{figure}
%%%%%%%%%% END FIG. 1

%\begin{acknowledgments}
%
%MORE...
%
%\end{acknowledgments}
%